\documentclass[12pt]{article}
\usepackage{graphicx}
\usepackage{epsf}
\usepackage{amsmath}
\usepackage{here}
\usepackage{setspace}
\usepackage{natbib}
\usepackage{xcolor}
\usepackage{varioref}
 \usepackage{wrapfig}
 \usepackage{threeparttable}
 \usepackage{dcolumn}
  \newcolumntype{d}{D{.}{.}{-1}}
 \usepackage{nomencl}
  \makeglossary
  \usepackage{subfigure}
 \usepackage{subfigmat}
 \usepackage{fancyvrb}
  \fvset{fontsize=\footnotesize,xleftmargin=2em}
 \usepackage{lettrine}

\usepackage{caption}
\usepackage{epstopdf}

\begin{document}
   \begin{center}
{\bf Mixing and Combustion in a Laminar Shear Layer with Imposed Counterflow}

{\bf William A. Sirignano}\\
{\normalsize\itshape  Department of Mechanical and Aerospace Engineering }\\
   {\normalsize\itshape  University of California, Irvine, CA 92697}  \\
   {\normalsize\itshape  May 26 , 2020 }
\end{center}

\begin{abstract}
 Three-dimensional laminar flow structures with mixing, chemical reaction, normal strain, and shear strain qualitatively representative of turbulent combustion at the small scales are analyzed.   A mixing layer is subjected to counterflow in the transverse y- and z-directions. Both non-reactive and reactive flows are examined.   Reduction of the three-dimensional boundary-layer equations to a one-dimensional similar form is obtained allowing for heat and mass diffusion with variations in density and properties.   In steady configurations, a set of ODEs governs the three velocity components as well as the scalar-field variables.  The transverse velocity is determined as a functional of the similarity coordinate. A generalization is found extending the Crocco integral for non-unitary Prandtl number and for imposed normal strain.   A flamelet model for individual diffusion flames with combined shear and normal strain is developed. Another  model with solution in similar form is obtained for a configuration with a dominant diffusion flame and a weaker fuel-rich premixed flame. Results for the velocity and scalar fields are found for ranges of Damk\"{o}hler number $Da$,  normal strain rate due to the counterflow, streamwise-velocity ratio across the mixing layer, Prandtl number, and Mach number.  For the flamelet model, a conserved scalar is cast as the independent variable to give an alternative description of the results.    The imposed normal strain decreases mixing-layer thickness and increases scalar gradients and transport rates.  There is indication of diffusion control for partially premixed flames in the multi-branched flame situation. The enhancement of the mixing and combustion rates by imposed normal strain on a shear layer can be very substantial. Also, the imposition of shear strain and thereby vorticity on the counterflow can be substantial indicating the need for flamelet models with both shear strain and normal strain.
\end{abstract}



\newcommand{\hs}{\mbox{\hspace{0.10in}}}
\newcommand{\hsm}{\mbox{\hspace{-0.06in}}}
\newcommand{\be}{\begin{equation}}
\newcommand{\ee}{\end{equation}}
\newcommand{\bea}{\begin{eqnarray}}
\newcommand{\eea}{\end{eqnarray}}

 \newcommand{\eqnref}[1]{(\ref{#1})}
 \newcommand{\class}[1]{\texttt{#1}}
 \newcommand{\package}[1]{\texttt{#1}}
 \newcommand{\file}[1]{\texttt{#1}}
 \newcommand{\BibTeX}{\textsc{Bib}\TeX}

\renewcommand{\figurename}{Fig.}
\captionsetup[figure]{labelsep=quad,textfont={small,bf},labelfont={small,bf}}

\linespread{1.5}

\section*{Nomenclature}
{\begin{tabular}{@{}ll}
$c_p$ & Specific heat under constant pressure   $J/(^oK kg)$ \\
$D$ & Mass diffusivity  $m^2/s$ \\
$Da$ & Damk\"{o}hler number     \\
$E$   &   Integral effect of normal strain in $z$-direction  \\
$f$  &   Normalized  pseudo stream function  \\
$G$  &  Normalized rate of expansive normal strain    \\
$g$  &  Weighting function for similarity    $kg/m^2$     \\
$h$ & Specific enthalpy $J/kg$ \\
$h_{f, m}$ & Heat of formation for species $m$ $J/kg$ \\
$K$  &  Ratio for Damk\"{o}hler number   \\
$Le$   &    Lewis number        \\
$M$   &  Mach number      \\
$N$ & Number of species \\
$p$ & Pressure $N /m^2$  \\
$Pr$ & Prandtl number     \\
$Q$    &   Fuel heating value $J/kg$   \\
$R$ & Specific gas constant $J/(^oK kg)$ \\
$Sc$  &  Schmidt number    \\
$u, v, w$ & Velocity components  $m/s$ \\
$x, y, z$ & Cartesian coordinate (m) \\
$Y_{m}$ &  Mass fraction of species $m$\\
$\alpha, \beta$   & Shvab-Zel'dovich conserved scalars    \\
$\gamma$  & Ratio of specific heats    \\
$\zeta$  &    Dummy variable for integration    \\
$\eta$   &   Density-weighted similarity coordinate   \\
$\theta$  &    Dummy variable for integration    \\
$\kappa$   &  Coefficient of $z$ for $w$ component of velocity  $s^{-1}$  \\
$\lambda$  &    Thermal conductivity $J/(s m^2)$  \\
$\mu$     &    Coefficient of viscosity   $N s/m^2$    \\
$\nu$   &  Mass stoichiometric ratio     \\
$\rho$ & Density $kg/m^3$ \\
$\Sigma$  &  Normalized conserved scalar       \\
$\Psi$  &  Pseudo stream function   $kg/(m s)$  \\
$\Omega$  &   Normalized shear strain rate       \\
$\omega_m$ & Reaction rate  for species $s^{-1}$\\
\end{tabular}}
\newpage
{\begin{tabular}{@{}ll}
Superscripts \\
${ }^*$ & Dimensional values \\
${ }'$ & Ordinary derivative with respect to $\eta$ \\
Subscripts \\\
$m$ & Integer for species designation \\
$\infty$ & Conditions at positive  infinite $y$  \\
$-\infty$ & Conditions at negative  infinite  $y$
\end{tabular}}

\section{Introduction}

There is need to understand the laminar mixing and combustion that commonly occurs within turbulent eddies. These  laminar flamelet sub-domains  experience significant strain. Some important work has been done here but typically for counerflows or simple vortex structures in two-dimensions or axisymmetry and often with a constant-density approximation. See  \cite{Linan},  \cite{Marble},  \cite{Karagozian},  \cite{Cetegen1, Cetegen2},  \cite{Peters}, and  \cite{Pierce}. Linan and  Peters focused on the counterflow configuration. Karagozian and Marble examined a three-dimensional  flow  with radial inward velocity, axial jetting, and a vortex  centered on the axis. The flame sheet wrapped around the axis due to the vorticity. Pierce and Moin modified the counterflow configuration by fixing domain size and forcing flux to zero at the boundaries.

These models are built around the postulate that the flamelets are always nonpremixed (i.e., diffusion) flames and subject to flow strain.  \cite{Tuan1} and  \cite{Tuan2} employed the Pierce-Moin flamelet approach in the simulation of a single-injector rocket engine. They showed the importance of flamelets subject to high strain rates. However, contradictions occurred in that both premixed flames and nonpremixed flames appeared in the predictions.  In fact, they report multi-branched flames; in particular, the combination is often seen of a fuel-lean premixed-flame branch with a branch consisting of a merged diffusion flame and  fuel-rich  premixed flame.

Experiments and asymptotic analysis  \citep{Seshadri} showed that a partially premixed fuel-lean flame and a diffusion flame may co-exist in a counterflow with opposing streams of heptane vapor and methane-oxygen-nitrogen mixture.   Thus, a need exists for flamelet theory to address both premixed and non-premixed flames. Recently,  \cite{Rajamanickam} has provided an interesting three-dimensional triple-flame analysis. \cite{Sirignano2019a} has provided  a counterflow analysis with three-dimensional strain and shown the possibility for a variety of flame configurations to exist depending on the compositions of the inflowing streams: (i) three flames including fuel-lean partially premixed, nonpremixed (i.e., diffusion-controlled), and fuel-rich partially premixed; (ii) nonpremixed and fuel-rich partially premixed; (iii) fuel-lean partially premixed and nonpremixed; (iv) nonpremixed; and (v) premixed. \cite{Lopez2019} has extended the counterflow analysis  to consider detailed kinetics for methane-oxygen detailed chemical kinetics.

\begin{figure}
\centering
{\includegraphics[height =5.0cm, width=0.45\linewidth]{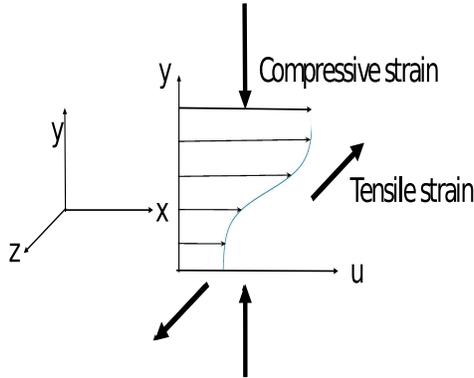}}
\caption{Sketch of mixing-layer flow with imposed counterflow.}
\label{Sketch}
\end{figure}

There is a strong need to study mixing and combustion in three-dimensional flows with both imposed normal strain and shear strain (and therein imposed vorticity with global circulation).    It is well known that the vorticity vector will tend to align with the direction of tensile (i.e., extensional) normal strain in a flow which leads us to choose a certain three-dimensional configuration that combines the mixing layer and the counterflow. We also expect a material interface to align to be normal to the direction of the compressive normal strain. See \cite{Nomura1992, Nomura1993, Boratav1996} and \cite{Boratav1998}.       In this work, we  extend flamelet theory in two significant aspects: the inclusion of both premixed and non-premixed flame structures and the extension to three-dimensional fields with both shear and normal strains. We address a steady three-dimensional mixing layer flow with primary flow component $u$ in the $x$-direction and an imposed counterflow with $v$ and $w$ velocity components in the $y$ and $z$ directions, respectively.  
The particular flow configuration considered here is sketched in Figure \ref{Sketch}.  The monotonic profile of $u(y)$ at fixed $x$ and $z$ is shown with the features of a traditional mixing layer. It also shows the imposed compressive normal strain is in the $y$-direction  and the commensurate expansive normal strain is in the $z$-direction. So, convergence of streamline projections occurs in the $x-y$ plane with divergence of streamline projections in the other two planes.

Combustion, variable density, and variable properties are examined.  The classical counterflow treatment by \cite{Peters} has two opposing streams, one of fuel or fuel plus a chemically inert gas and the other of oxidizer or oxidizer plus an inert gas. Our computations address that situation where a single diffusion flame exists. We also provide here  some background analytical considerations for  situations where  the inflowing streams from $y(\infty)$ and $y(-\infty)$ (each of which also now has a parallel component of forced velocity in the $x$-direction) may consist of either only one reactant or a combustible mixture of fuel and oxidizer, thereby allowing another flame besides the simple diffusion flame to co-exist.  Propane and oxygen are specifically considered with one-step,  \cite{Westbrook_Dryer:1984} kinetics; however, the qualitative conclusions are expected to be more general.  Those one-step kinetic relations were obtained by fitting to experiments for premixed flames and are expected to be less accurate for diffusion flames. Nevertheless, we may accept some error here because diffusion rather than kinetics is rate controlling for diffusion flames. The approach here expands on recent work  \citep{Sirignano2019b} which used infinite kinetics for three-dimensional counterflow diffusion flames and another work \citep{Sirignano2019a} which used one-step kinetics.

In Section \ref{analysis}, the analysis is presented. In sequence, the three-dimensional problem is reduced to a two-dimensional form and then, for the downstream mixing-layer flow, to a one-dimensional similar form.  The system of ordinary differential equations (ODEs) is presented for the thermo-chemical variables and the  velocity components. An analysis is given in Subsection \ref{Croccosec} to relate a conserved scalar to the velocity solution. The validity of the similar solution form for mixing layers with  certain thin reaction zones are discussed using concepts from singular perturbation theory. The chemical kinetic model is described in a form to be used as a source term for an ODE. In  Subsection \ref{sigma}, the relevance of these new findings for flamelet theory is discussed. Then, in Section \ref{results}, the findings from calculations  for the  mixing layer with imposed counterflow are presented for both the non-reacting flow and the flow with the diffusion flame.  Conclusions are presented in Section \ref{conclusions}.

\section{Analysis} \label{analysis}

The planar mixing-layer has been widely used in combustion studies for development of flamelet models. In particular, a similar solution can be produced, offering the convenience of  reduction to a system of ordinary differential equations to describe a multi-dimensional configuration. Here, we develop a similar solution for a specific three-dimensional configuration. Consider a mixing layer with primary flow in the $x$-direction and diffusion primarily in the $y$-direction. An oxidizer rich gas enters the mixing-layer domain at negative $y$ values while a fuel-rich gas enters at positive $y$ values. Shear strain and associated vorticity with a vector in the $z$-direction are obviously introduced here. Superimposed on this otherwise-planar flow is a counterflow with an incoming stream with compressive (i.e., negative) normal strain in the $y$-direction and an outgoing stream with expansive (i.e., positive) normal strain in the $z$-direction.   The velocity $\vec{u}$ has the components $u, v, $ and $w$ in the $x,y,$ and $z$ directions, respectively. The interface of the two incoming, opposing transverse flows is specified at $y$ = 0. The chosen configuration here has no $x$-component to the pressure gradient. The classical boundary-layer approximation is made, rendering the $y$-component of the pressure gradient to be negligible compared to other terms. If the approaching streams have the same pressure at a distance from the interface and its viscous layer, we expect that, in a frame of reference attached to the interface, momentum balance for steady flow yields $\rho_{-\infty} v_{-\infty}^2 = \rho_{\infty} v_{\infty}^2$.   The $y$-directed inflowing streams in all cases  bring together fluids of differing temperature and / or composition; so, heat diffusion and mass diffusion are  in the $y$-direction.

The two streams will generally have different upstream values for velocity $v$, temperature $T$, enthalpy $h$, density $\rho$, or composition reflected through mass fraction $Y_m$ for chemical species $m$. Pressure $p$ will be given the same upstream values for the two streams. Fickian mass diffusion and Fourier heat conduction are considered so that all fluid properties are continuous across the interface.  The Prandtl number ($Pr$) and the Schmidt number ($Sc$)  will be assumed to have the same constant  value. Thus, the Lewis number ($Le = Sc/Pr$) has unitary value. Radiation and  gravity are neglected. A Newtonian fluid with the Stokes hypothesis is examined. In the reactive case,  kinetic energy and viscous dissipation will be neglected in the energy consideration because of a focus on low-Mach-number flow. However, those terms will be kept in the reactive case.

\subsection{Three-dimensional Formulation}  \label{3D}

The governing equations for steady 3D flow with the boundary-layer approximation are given as
\begin{eqnarray}
 \frac{\partial (\rho u)}{\partial x} + \frac{\partial (\rho v)}{\partial y} +  \frac{\partial (\rho w)}{\partial z} =  0
\label{cont}
\end{eqnarray}
\begin{eqnarray}
 \rho u\frac{\partial u}{\partial x} + \rho v\frac{\partial u}{\partial y}  +
 \rho w\frac{\partial u}{\partial z} = \frac{\partial}{\partial y} \Big(\mu\frac{\partial u}{\partial y}\Big)
\label{x-momentum}
\end{eqnarray}
\begin{eqnarray}
 \rho u\frac{\partial w}{\partial x} + \rho v\frac{\partial w}{\partial y}  +
 \rho w\frac{\partial w}{\partial z}+\frac{\partial p}{\partial z} = \frac{\partial}{\partial y} \Big(\mu\frac{\partial w}{\partial y}\Big)
\label{z-momentum}
\end{eqnarray}
\begin{eqnarray}
 \rho u\frac{\partial h}{\partial x} + \rho v\frac{\partial h}{\partial y} +
   \rho w\frac{\partial h}{\partial z} =   \frac{1}{Pr}\frac{\partial}{\partial y} \Big( \mu \frac{\partial h}{\partial y}  \Big)
 -\rho \Sigma_{m=1}^N h_{f,m} \omega_m
\label{energy}
\end{eqnarray}
\begin{eqnarray}
 \rho u\frac{\partial Y_m}{\partial x} + \rho v\frac{\partial Y_m}{\partial y} +
   \rho w\frac{\partial Y_m}{\partial z} = \frac{1}{Pr}\frac{\partial}{\partial y} \Big( \mu \frac{\partial Y_m}{\partial y}  \Big) + \rho \omega_m   \;\;;\;\; m=1, 2, ...., N
\label{species}
\end{eqnarray}
The equation for the $y$-component of momentum has been replaced by the classical boundary-layer approximation that no variation of pressure in the $y$-direction through the mixing layer exists. Note that pressure will have a gradient in the y-direction outside of the boundary layer to create the counterflow. However, the gradient must change direction in the mixing layer passing through the zero value. Thus, the classical zero-gradient assumption using the boundary-layer approximation for a thin mixing layer is valid.

\subsection{Reduction to Two Dimensions and Conserved Scalars} \label{2D}

Following \cite{Rajamanickam} , we consider $w = \kappa z$, differing in that we allow $\kappa$ to be a function of $x$ and $y$ rather than constant. Accordingly, $p$ will also vary with $z$. In particular, since $p$ will not vary with $x$ or $y$, we have $p=p_o - (1/2)\rho_{\infty}w_{\infty}^2 = p_o - 1/2)\rho_{\infty}\kappa_{\infty}^2z^2$ where $p_o$ is constant.   All other variables ($u, v, \rho, h, Y_m $) will depend only on $x$ and $y$. The resulting two-dimensional system of equations follows.
\begin{eqnarray}
 \frac{\partial (\rho u)}{\partial x} + \frac{\partial (\rho v)}{\partial y} +
 \rho \kappa =  0
\label{cont2}
\end{eqnarray}
\begin{eqnarray}
 \rho u\frac{\partial u}{\partial x} + \rho v\frac{\partial u}{\partial y}  = \frac{\partial}{\partial y} \Big(\mu\frac{\partial u}{\partial y}\Big)
\label{x-momentum2}
\end{eqnarray}
\begin{eqnarray}
 \rho u\frac{\partial \kappa}{\partial x} + \rho v\frac{\partial \kappa}{\partial y}  +
 \rho \kappa^2  - \rho_{\infty} \kappa_{\infty}^2 = \frac{\partial}{\partial y} \Big(\mu\frac{\partial \kappa}{\partial y}\Big)
\label{z-momentum2}
\end{eqnarray}
Since the pressure does not vary with $y$, we must have $ \rho_{\infty} \kappa_{\infty}^2  =   \rho_{-\infty} \kappa_{-\infty}^2$. Thus, the derivates of $\kappa$ will go to zero at plus and minus infinity for the $y$ value.
\begin{eqnarray}
 \rho u\frac{\partial h}{\partial x} + \rho v\frac{\partial h}{\partial y}  = \frac{1}{Pr}\frac{\partial}{\partial y} \Big( \mu \frac{\partial h}{\partial y}  \Big)
 -\rho \Sigma^N_{m=1}h_{f,m} \omega_m
\label{energy2}
\end{eqnarray}
\begin{eqnarray}
 \rho u\frac{\partial Y_m}{\partial x} + \rho v\frac{\partial Y_m}{\partial y} = \frac{1}{Pr}\frac{\partial}{\partial y} \Big( \mu \frac{\partial Y_m}{\partial y}  \Big) + \rho \omega_m   \;\;;\;\; m=1, 2, ...., N
\label{species2}
\end{eqnarray}

Here, the sensible enthalpy $h =c_pT$ is based on the assumption of a calorically perfect gas. When normalized by ambient conditions, the non-dimensional values of enthalpy and temperature are identical. For simplification, we neglect the effect of species composition on specific heats and the specific gas constant. For the one-step kinetics considered here, the last term in Equation (\ref{energy2}) may be replaced by $- \rho Q \omega_F$ where $Q$ is the heating value (energy/mass) of the fuel and $\omega_F < 0$ is the chemical oxidation rate of the fuel. Then, $Y_F$ and $Y_O$ are the mass fractions of propane and oxygen, respectively. $\nu =0.275$ is the stoichiometric  ratio of propane mass to oxygen mass.

The boundary conditions on Equations (\ref{x-momentum2}) through (\ref{species2}) involve specifications of the dependent variables at $y=\infty$ and $y= - \infty$ as well as their values at an upstream value of the coordinate $x$. From these primitive equations for $h$ and $Y_m$, we may form equations for conserved scalars. We define two conserved scalars as
\begin{eqnarray}
\alpha \equiv  Y_F - \nu Y_O \nonumber  \\
\beta \equiv    h + \nu Y_O Q    \nonumber  \\
\label{consscalar}
\end{eqnarray}
to obtain
\begin{eqnarray}
\rho u\frac{\partial \alpha}{\partial x} + \rho v\frac{\partial \alpha}{\partial y} = \frac{1}{Pr}\frac{\partial}{\partial y} \Big( \mu \frac{\partial \alpha}{\partial y}  \Big)
\nonumber \\
\rho u\frac{\partial \beta}{\partial x} + \rho v\frac{\partial \beta}{\partial y} = \frac{1}{Pr}\frac{\partial}{\partial y} \Big( \mu \frac{\partial \beta}{\partial y}  \Big)
\label{scalarPDE}
\end{eqnarray}
Note that, for the non-reacting case, $h$ and $Y_m$ are conserved scalars satisfying this same partial differential equation.

The boundary  conditions for $u, h, Y_m, \alpha,$ and $\beta$ at $y = \infty$ and at $y= - \infty$ remain constant as both $x$ and $z$ vary. The boundary values $\kappa(\infty) = \kappa_{\infty}$ and $ \kappa(- \infty) = \kappa_{-\infty} $ will vary with $x$.  As downstream distance $x$ increases, the mixing-layer solution becomes less dependent on upstream inflow profiles; in fact, it is known to become independent asymptotically, depending only on the boundary conditions.

When $Pr = 1$, the solutions for $\alpha, \beta,$ and, in the nonreacting case, $h$ and $Y_m$ become linear functions of $u$ as known through the classical Crocco integral \citep{Crocco_1932}. In the nonreacting case where terms of order of the square of Mach number are retained and $Pr =1$, we repeat Crocco's finding that $h + u^2/2$ is linear in $u$.

\subsection{Similarity in One-dimensional Form}  \label{1D}

Our development of the similarity follows a pattern originally developed for compressible boundary layer flow. However, the appearance of the $z$-momentum equation causes a need for new features.

\cite{Howarth1948} assumed a perfect gas with dynamic viscosity  $\mu$ directly proportional to temperature $T$ and used  a transformation of variable  $\bar{y} \;\;\alpha \;\; \int  (\nu)^{-1/2} dy$ where $\nu$ is the kinematic viscosity. This leads to
$\bar{y} \;\;\alpha \;\; p^{1/2} \int (1/T) dy$.  \cite{Stewartson} simply stated $\bar{y}\;\; \alpha \;\;\int \rho dy$  which does not require an assumption about the relation between temperature and viscosity. \cite{Dorodnitsyn} did parallel work earlier.  \cite{Lees} generalized the transformation for situation where free stream velocity and pressure varied in the streamwise direction.
The density-weighted  transformation is used here to replace $y$ with $\bar{y}  \equiv \int^y_0 \rho (y') dy'$ .
 The product $\rho \mu$ is assumed to remain constant and equal to $\rho_{\infty} \mu_{\infty}$ throughout the mixing layer. (This is convenient but not necessary to obtain a similar solution.)  The similarity variable $\eta$ and other variables are defined.
\begin{eqnarray}
\eta \equiv \frac{\bar{y}}{g(x)}  \nonumber \\
g(x)   \equiv   \sqrt{\frac{2 \rho_{\infty} \mu_{\infty} x}{u_{\infty}} }  \nonumber  \\
G(\eta) \equiv \frac{\kappa g^2}{\rho_{\infty} \mu_{\infty}} = \frac{2 \kappa x} {u_{\infty}}  =\frac{2 w x}{z u_{\infty}}   \nonumber  \\
E \equiv  \int_0^{\eta}  G(\eta') d \eta'
\label{definitions}
\end{eqnarray}
Note the implication that $\kappa$ will vary as the reciprocal of $x$. $\eta, G,$ and $E$ are non-dimensional here.  $G$ is both the normalized $z$-component of velocity and the indicator of the imposed normal strain.

In standard fashion, a pseudo stream function $\Psi$ and a pseudo $y$-component of velocity $\tilde{v}$ are created mimicking an incompressible flow.
\begin{eqnarray}
\Psi &=& f(\eta)g(x) u_{\infty}  \nonumber  \\
u &=& \frac{\partial \Psi}{\partial \bar{y}}  = u_{\infty} \frac{df}{d\eta} \nonumber  \\
\rho_{\infty}\mu_{\infty}\tilde{v} &=& -  \frac{\partial \Psi}{\partial x}  = \rho v + u\int_0^y\frac{\partial \rho}{\partial x}dy'
+ \sqrt{\frac{\rho_{\infty}\mu_{\infty}u_{\infty} }{2x}}E
\label{v}
\end{eqnarray}
It follows that
\begin{eqnarray}
\tilde{v} g = \eta \frac{df}{d \eta} - f
\label{tildev}
\end{eqnarray}
Thereby, $\tilde{v}$  will vary as a function of the similarity variable multiplied by the reciprocal of the square root of x. So,  $\tilde{v}g$ becomes a function only of the similarity variable $\eta$.

Generally, for two-dimensional mixing layers and boundary layers with variable density, the interest in the precise determination of $v$ beyond Equation (\ref{v}) has not been high because $v^2  << u^2$. Here, because of the imposed counterflow, we have interest in the determination of $vg$ as a function of $\eta$.
The second term on the right side of the second line of equation (\ref{v}) requires attention in order to determine $v(\eta, x)$. In that term, a derivative with respect to $x$ is taken of an integral over $y$ space. We may write
\begin{eqnarray}
y = \int^{\bar{y}}_0 \frac{1}{\rho} d\bar{y}' =g(x) \tilde{I}(\eta)
\end{eqnarray}
where $\tilde{I}(\eta) \equiv   \int^{\eta}_0 (1/\rho(\eta')) d \eta'$.
Furthermore, taking the derivative at constant $y$,
\begin{eqnarray}
\frac{dy}{dx} &=&  0 = \tilde{I}\frac{dg}{dx} +g\frac{d\tilde{I}}{dx} = \tilde{I}\frac{dg}{dx} +\frac{g}{\rho}\frac{d\eta}{dx}|_{y = constant}  \nonumber \\
\frac{d\bar{y}}{dx} &=&  \eta \frac{dg}{dx} + g\frac{d\eta}{dx} = (\eta - \tilde{I} \rho)\frac{dg}{dx}
\end{eqnarray}
Now, from Equations (\ref{v}) and (\ref{tildev}),
\begin{eqnarray}
 \rho v g  &=& \rho_{\infty}\mu_{\infty}(\eta \frac{df}{d \eta} - f)  - u(\eta - \tilde{I} \rho)g \frac{dg}{dx}
- g\sqrt{\frac{\rho_{\infty}\mu_{\infty}u_{\infty} }{2x}}E  \nonumber  \\
    & = &  \rho_{\infty}\mu_{\infty} \Big[ \tilde{I}\rho\frac{df}{d \eta}- f -E \Big]
    \label{v2}
\end{eqnarray}
\cite{Libby1968} produced the parallel analytical result for the two-dimensional boundary-layer flow but never computed $v$. \cite{Pruett} made $v$ calculations for a wall-layer similar solution. \cite{Kennedy} made $v$ calculations for a mixing layer similar solution.
Those studies all considered a two-dimensional shear layer without imposed normal strain, i.e., $E =0$.

Using the perfect-gas relation and neglecting terms of order Mach number squared, we find
\begin{eqnarray}
\Big(\frac{\rho}{\rho_{\infty}} \Big)^{-1}= \frac{T}{T_{\infty}} = \frac{h}{h_{\infty}} = \frac{\mu}{\mu_{\infty}}
\end{eqnarray}
For the perfect gas, $\tilde{I}$ may be written with $\tilde{h}$ as the integrand and Equation (\ref{v2})  for $v$ can be modified.
\begin{eqnarray}
 \rho v g  =  \rho_{\infty}\mu_{\infty} \Big[ \frac{\int_0^{\eta}\tilde{h}(\eta')
 d \eta'}{\tilde{h}}\frac{df}{d \eta}- f -E \Big]
\end{eqnarray}

\subsection{Formulation of Ordinary Differential Equations}\label{ODEs}

We define  $( )' \equiv d( )/d\eta, \tilde{h} \equiv h/ h_{\infty}= \rho_{\infty}/\rho$, and $\tilde{\beta} \equiv \beta/h_{\infty}$. Now, the following ordinary differential equations (ODEs) and boundary conditions do follow.
\begin{eqnarray}
f''' + (f + E) f''  = 0   \; ; \;
f'(-\infty)=\frac{u_{-\infty}}{u_{\infty}} \; ; \; f(0) = 0 \;  ; \; f'(\infty)= 1
\label{Blasius}
\end{eqnarray}
\begin{eqnarray}
G'' + (f + E) G'  - G^2  +  G_{\infty}^2 \tilde{h} =0  \; ; \;
G(-\infty)= G_{-\infty}\; ; \;  G(\infty)= G_{\infty}
\label{G}
\end{eqnarray}
Equation (\ref{Blasius}) differs from the classical Blasius equation because of the presence of $E$ which couples it to Equation (\ref{G}). That equation may actually be interpreted as a third-order ODE since $G = dE/d\eta.$  In the coefficient of the first derivatives in the above and the following equations, $f$ describes the contribution to transverse transport due to the shear strain while $E$ represents the contribution to transverse transport due to the imposed normal strain. The boundary conditions on $G$ are chosen so that pressure will be the same with variation for $z$ within the two free streams. Thus, $\rho w^2$ will be same in the two streams for a given $x$ and $z$. If the two free-streams have the same temperature (and therefore the same density), $G_{-\infty}= G_{\infty}$; otherwise, they must differ.
\begin{eqnarray}
\tilde{h}'' + Pr(f + E) \tilde{h}'  =  2Pr\frac{\omega_F x}{u_{\infty}}\frac{Q}{h_{\infty}}     \; ; \;
\tilde{h}(-\infty)=\frac{h_{-\infty}}{h_{\infty}}\; ; \; \tilde{h}(\infty)= 1
\label{h}
\end{eqnarray}
\begin{eqnarray}
Y_F'' + Pr(f + E) Y_F' =- 2Pr\frac{\omega_F x}{u_{\infty}}   \; ; \;
Y_F(-\infty)=Y_{F, -\infty}\; ; \; Y_F(\infty)= Y_{F,\infty}
\label{YF}
\end{eqnarray}
As written, $\omega_F$ is to be taken as negative when fuel is being consumed.
\begin{eqnarray}
\alpha'' + Pr(f + E) \alpha'  = 0   \;\;   ; \;\;
\alpha(-\infty)=\alpha_{-\infty}\;\; ; \;\; \alpha(\infty)= \alpha_{\infty}
\label{alpha}
\end{eqnarray}
\begin{eqnarray}
\tilde{\beta}'' + Pr(f + E) \tilde{\beta}' &=& 0    \; ; \; \nonumber \\
\tilde{\beta}(-\infty)=\tilde{h}_{-\infty}&+& \frac{Q\nu Y_{O,-\infty}}{h_{\infty}}\; ; \;
\tilde{\beta}(\infty)= 1 + \frac{Q\nu Y_{O,\infty}}{h_{\infty}}
\label{betatilde}
\end{eqnarray}

Several important parameters can be identified in the equations. A Damk\"{o}hler number $Da$ will be embedded in the chemical-rate function $\omega_F$ to be discussed later. Of course, the composition and temperature at $\eta = \infty$ and $\eta = - \infty$ will be influential. $Pr$ affects the mass and energy diffusion rate as compared to the diffusion of momentum due to viscosity. $G_{\infty}$  is the normalized magnitude of the imposed normal strain rate $\kappa$ using  $u_{\infty}/x $  (the reciprocal of a residence time) as the normalizing factor. The shear strain rate is estimated by $(u_{\infty} - u_{-\infty})/ \delta(x)$ where $\delta(x)$ is the mixing-layer thickness. Using the same normalization factor, the normalized shear strain is estimated as $(x/\delta)(1 - u_{-\infty}/u_{\infty})$.  Clearly, the velocity ratio $ u_{-\infty}/u_{\infty} $  is important. $\delta$ will be affected by both the shear strain and the compressive strain. For the pure counterflow without shear, $\delta_C = O((\nu/\kappa)^{1/2})$ where $\nu$ is the kinematic viscosity. For the pure shear layer without compressive normal strain, $\delta_S = O( x/Re_x^{1/2} ) = O((x\nu/u_{\infty})^{1/2})$. Thus, $\delta_S/\delta_C = O((\kappa x/u_{\infty})^{1/2}  )$ and $G = O(( \delta_S/\delta_C  )^2)$. So, high (low) values of $G$ will imply that the compressive (shear) strain plays a dominant role in determining layer thickness.

Equations (\ref{Blasius}) through (\ref{betatilde}) could  be made more general by considering variation of $\rho \mu$ through the mixing layer.  Terms with the first derivative in $\eta$ space of the normalized product $\rho \mu / (\rho_{\infty}\mu_{\infty}) $ would appear in the equations. This secondary effect will be neglected here. In the classical two-dimensional mixing layer without imposed normal strain (i.e., $\kappa = 0, G =0, $ and $E=0$), Equation (\ref{G}) disappears while Equations (\ref{Blasius}) and (\ref{h}) through (\ref{betatilde}) simplify to the well known forms. For the case where no counterflow is applied (i.e., $E = 0$), Equation (\ref{Blasius}) has been extended to cases where
$d(\rho \mu)/d\eta \neq 0$.  \cite{poblador2020} have examined the solution of Equation (\ref{v2}) where $d(\rho \mu)/d\eta \neq 0$  and $E=0$.  Also, they show  favorable comparisons between similar solutions and two-dimensional computational results for $v$.

\subsection{Generalized Crocco Integral}\label{Croccosec}

\cite{Crocco_1932} developed his integral solution for a compressible wall boundary layer at significant values of free stream Mach number. Assuming $Pr = 1$, he showed that, knowing the solution for velocity $u$ and the boundary conditions for the conserved scalar, a similar solution existed with the conserved scalar $h_o \equiv h + u^2/2$ linear in $u$. Here, we show that, for  $Pr \neq  1$, a conserved scalar solution may still be found as a integral function of the velocity derivative. In a non-reacting case at  Mach number $M <<1$, $h_o \approx h$; thus,  the static enthalpy $h$ may be approximated as a conserved scalar because viscous dissipation is negligible. We also show that the enthalpy, with account for viscous dissipation at any subsonic $M$ value and any $Pr$ value, can be described as an integral function of the velocity derivative.

Visualize Equation (\ref{Blasius}) as a first-order ODE governing $f'' = du/d\eta$ and Equations (\ref{h}) through (\ref{betatilde}) as first-order ODEs governing the first derivative of the dependent variable. In the case without imposed normal strain (i.e., $G = 0$), the solutions for $f(\eta)$ and $u(\eta)$ will be independent of $Pr$ and $M$. (Realize that a coupling actually applies through $\eta(x,y)$.) $G$ and thus $E$ are coupled with enthalpy and thereby have dependencies on both $Pr$ and $M$. In that case, $f$ and $u$ will also experience the dependencies. Consider the conserved scalars $h$ and $Y_m$ in the non-reacting case and $\alpha$ and $\beta$ in any case. Clearly, a solution exists where the first derivative of the scalar is linear in $(du/d\eta)^{Pr}$. In fact, since all derivatives go to zero at plus and minus infinity, we have a direct proportionality.
Subsequently, we obtain the generalized Crocco integral for any conserved scalar $CS$.
\begin{eqnarray}
\frac{CS(\eta) - CS_{-\infty}}{CS_{\infty} - CS_{-\infty}}  = \frac{\int_{-\infty}^{\eta}  \Big(\frac{d u}{d\eta}\Big)^{Pr} d \zeta}{\int_{-\infty}^{\infty}  \Big(\frac{d u}{d\eta}\Big)^{Pr} d \eta}
 =  \frac{\int_{-\infty}^{\eta}  \big(f''(\zeta)\big)^{Pr} d \zeta}{\int_{-\infty}^{\infty}   \big(f''(\eta)\big)^{Pr}d \eta}
= \frac{J(\eta)}{J(\infty)}
\label{CS}
\end{eqnarray}
where solution of Equation (\ref{Blasius}) as a first-order ODE for $f''$ yields
\begin{eqnarray}
J(\eta)  \equiv  \int^{\eta}_{-\infty} \big(f''(\zeta)\big)^{Pr} d\zeta =\int_{-\infty}^{\eta} e^{-Pr I(\zeta)}d\zeta  \;\;  ; \;\;
I(\eta) &\equiv&
\int_{0}^{\eta} [f + E ]d\zeta
\label{Crocco}
\end{eqnarray}
Equations (\ref{Blasius}) and (\ref{G}) must still be integrated to determine the functions $f$ and $E$ which appear in the integrand of Equation (\ref{Crocco}). The integration will couple with other equations since $\tilde{h}$ appears in (\ref{G}). For the nonreacting case, the functions $\tilde{h}, Y_F, \alpha, $ and $\beta$ may be determined using the generalized Crocco integral in Equation (\ref{CS}). For the reacting case, the integral may be used to evaluate $\alpha$ and $\beta$ but either $\tilde{h}$ or $Y_F$ must be solved from (\ref{h}) or (\ref{YF}).

Obviously, in cases where $Pr$ and Schmidt number $Sc$ differ, $Sc$ should appear in the exponent for mass fraction (for the non-reacting case only) and $\alpha$ solutions. In such a case, Equation (\ref{h}) must be re-formulated since it is now based on $Pr = Sc$.

It is possible to generalize the Crocco integral also in the case where Mach number squared terms are not neglected. Here, under the boundary-layer approximation the dissipation term $\mu(\partial u/\partial y)^2$ is added to the right sides of Equations (\ref{energy}) and (\ref{energy2}). Consequently, the term
$-Pr (u_{\infty}^2/h_{\infty})(f'')^2$ must be added to the right side of Equation (\ref{h}). The non-reacting case will be considered so that $\omega_F =0$ in that equation.  Again, the second-order ODEs will be treated as first-order for determining the first derivative. Then, a simple integration follows for determining $\tilde{h}$ from knowledge of its first derivative. The result is
\begin{eqnarray}
\tilde{h} &=& \frac{h_{-\infty}}{h_{\infty}} + C\frac{\int_{-\infty}^{\eta}(f'')^{Pr} d\zeta}{ \int_{-\infty}^{\infty}(f'')^{Pr} d\zeta   }
- Pr\frac{u_{\infty}^2}{h_{\infty}}\int_{-\infty}^{\eta}(f'')^{Pr}\Big[\int_{-\infty}^{\zeta} (f'')^{2 - Pr} d \theta\Big]d \zeta
\nonumber \\
&=& \frac{h_{-\infty}}{h_{\infty}} + C \frac{J(\eta)}{J(\infty)} - Pr \Omega^2\frac{u_{\infty}^2}{h_{\infty}}\int_{-\infty}^{\eta}\int_{-\infty}^{\zeta}
e^{-PrI(\zeta) -(2-Pr)I(\theta)} d\theta d\zeta
\nonumber \\
C &\equiv& 1 - h_{-\infty}/h_{\infty}  +(Pr u_{\infty}^2/h_{\infty})\int_{-\infty}^{\infty}(f'')^{Pr} \Big[\int_{-\infty}^{\eta}(f'')^{2 - Pr} d\zeta   \Big] d\eta
   \nonumber \\
   &=& 1 - h_{-\infty}/h_{\infty}  + (Pr \Omega^2u_{\infty}^2/h_{\infty})\int_{-\infty}^{\infty}  \int_{-\infty}^{\eta}
   e^{-Pr I(\eta)-(2 - Pr)I(\zeta)} d\zeta    d\eta
   \nonumber \\
    \Omega& \equiv  & f''_{max}    = \frac{f'(\infty) - f'(-\infty)}{\int_{-\infty}^{\infty} e^{-I(\eta)} d\eta }
\label{Crocco2}
\end{eqnarray}
$\Omega$  is a non-dimensional indicator of the magnitude of the maximum shear strain rate.
Upon neglect of terms of $O(u_{\infty}^2/h_{\infty})$, Equation (\ref{Crocco2}) reduces to Equation (\ref{CS}). When $Pr =1$, the integrals can be evaluated analytically yielding the original Crocco integral relation where $h + u^2/2$ becomes linear in $u = u_{\infty} f'$.  For the perfect gas, the parameter $u_{\infty}^2/h_{\infty} = (\gamma -1)M^2$. Here, $M$ is the Mach number of the free stream at $y =\infty$ which is generally the faster stream in the calculations presented.  \cite{Illingworth} in a flat-plate boundary-layer analysis produced an integral similar to the form of the second line of Equation (\ref{Crocco2}) to evaluate enthalpy and presented some calculations for $Pr =0.725$. However, the connection with the velocity derivative and the nature of the result as a generalized Crocco integral was not mentioned.

The analytical results in this subsection apply with or without imposed normal strain. They can be applied to a wall boundary layer by simply replacing boundary conditions at minus infinity by the wall boundary conditions.  Therefore, we have a general analytical relation for the scalar variable in terms of  the Blasius solution.

\subsection{Non-similar Behavior of the Reaction Zone}\label{innerouter}

There is a challenge in defending the claim that Equations (\ref{h}) and (\ref{YF}) are in similar form because of the appearance of the factor $\omega_F x/ u_{\infty}$ in the source and sink terms. $\omega_F$ will depend on pressure, temperature, and mass fractions which are expected to be similar (or at least near-similar) while $x$ is clearly non-similar. We cannot expect that $\omega_F \sim 1/x$. A suitable explanation can be made using the concept of inner and outer solutions from singular perturbation theory \citep{Cole, VanDyke}.

The reaction rates are negligible outside of thin reaction zones within the mixing layer.
Within those zones, second derivatives become notably larger than first derivatives and Equations (\ref{h}) and (\ref{YF}) become
\begin{eqnarray}
\tilde{h}''  \approx 2Pr\frac{\omega_F x}{u_{\infty}}\frac{Q}{h_{\infty}}  \; ; \nonumber \\
Y_F''  \approx - 2Pr\frac{\omega_F x}{u_{\infty}}
\label{inner}
\end{eqnarray}
These relations determine our inner solutions which must be matched to the outer solutions which apply to the much larger portion of the mixing layer where reaction rate is negligible. Integration of the inner solutions over the reaction zone yields
\begin{eqnarray}
\tilde{h}'|_+ - \tilde{h}'|_-  \approx  2 Pr \frac{Q}{h_{\infty}} \int_-^+\frac{\omega_F x}{u_{\infty}} d\eta \; ; \nonumber \\
Y_F'|_+ - Y_F'|_-  \approx - 2 Pr \int_-^+ \frac{\omega_F x}{u_{\infty}} d\eta
\label{innerint}
\end{eqnarray}
The ``plus" and ``minus" subscripts and integral limits denote the far edges of the inner zone: namely, the infinity limits on the small inner scale in singular perturbation theory. These results apply for both diffusion flames and premixed flames.

First, let us discuss diffusion flames.  Physically, we have a diffusion-controlled situation where peak temperature in the reaction zone varies weakly. Thus, the peak value of the reaction rate per unit volume $\omega_F$ varies weakly with downstream position and the reaction-zone thickness adjusts to have production and consumption rates match the diffusion rate. The gradients of the scalars in the $y$ space will decrease as $1/\sqrt{x}$  which implies no change with $x$ for the derivatives in $\eta$ space.  The important point is that the width of the reaction zone measured in $\eta$ space narrows with increasing downstream distance $x$. In particular, the measure $\Delta y$ of the zone width behaves as $1/\sqrt{x}$ which implies  that the ratio of reaction-zone thickness to mixing-layer thickness behaves as $1/x$. Consequently, $\Delta \eta$ for the reaction zone varies as $1/x$. Thus, the integrals with their integrands proportional to $x$  will actually not vary with $x$ so that  the jump in first-derivative values with respect to $\eta$  are not dependent on $x$. The jumps give the important quantities of heat production and fuel-mass consumption in the reaction zone and its impact on the remainder of the mixing layer. In other words, the rate of heat and mass diffusion in or out of the reaction zone depends solely on the variable $\eta$, giving similarity for the bulk of the mixing layer. An important point here is that, under the boundary-layer approximation, the communication between the reaction zone at any $x$ position and the rest of the mixing layer occurs only in the $y$ direction through transport and diffusion with the latter dominant; there is no direct transfer of information from the reaction zone at one $x$ value to the reaction zone at another $x$ value.  Note that the infinite-kinetics model yields consistent behavior with our results for the derivatives outside the reaction zone.

A premixed flame which is not excessively fuel lean or fuel rich will propagate in a wave-like manner relative to the combustible mixture. The rate of fuel consumption and heat production (and the flame speed) depends on the product of diffusion rate and chemical reaction rate. Flame speed (relative to the upstream incoming fluid velocity), flame thickness, and scalar gradients in $y$ space will not vary with $x$. The implication therefore is that flame speed, flame thickness, and scalar gradients in $\eta$ space vary as $1/\sqrt{x}, 1/\sqrt{x},$ and $\sqrt{x}$, respectively. The imposed counterflow velocity decreases as $1/\sqrt{x}$, and therefore the premixed flame transverse motion across the mixing layer could not be arrested at the same $\eta$ value for all $x$ values. Premixed flame position would change with $x$ position, moving away from $\eta=0$, in both $y$ space and $\eta$ space while diffusion flame position would not change in $\eta$ space. Thus, the similar solution that predicts  premixed or partially premixed flames of the classical form with wave-like propagation cannot be accurate and should only be trusted to support the plausibility of the premixed flame occurrence. The premixed  configuration deserves further examination with a two-dimensional analysis addressing Equations (\ref{cont2}) through (\ref{scalarPDE}). In this work, premixed flames with mixture ratios far from stoichiometric will be examined; they are more likely to be diffusion controlled with weak influence of chemical kinetics in determining a flame speed.

For reacting counterflow calculations with one-step propane-oxygen kinetics, \cite{Sirignano2019a}  found that, in a configuration with a diffusion flame and a fuel-rich flame, the latter flame did not show a tendency to propagate at speed proportional to the square root of the integrated reaction rate through the reaction zone. The change in flame speed with increasing Damk\"{o}hler number ($Da$) was notably weaker than the expected square root dependence for a premixed laminar flame. The situation was closer to one where the width of the reaction zone increased with the reciprocal of the reaction rate; thus, there was little variation in the integrated reaction rate or total consumption rate for fuel over the zone. Accordingly, the mean temperature and jumps in enthalpy gradient and mass fraction gradient across the reaction zone did not differ much depending on $Da$. It appears that the flame was a zone whose volume adjusted (at constant reaction rate) to accommodate the mass flux rate passing through the fuel-rich flame and moving towards the diffusion flame.

\subsection{Chemical Kinetic Model}\label{kinetics}

 In a one-step chemical reaction, each species is consumed or produced at a rate in direct proportion to the rate of some other species that is produced or consumed. We will focus on propane-oxygen flows with one-step kinetics. However, results are expected to be qualitatively more general, applying to situations with more detailed kinetics and to other  hydrocarbon /oxygen-or-air combination.  \cite{Westbrook_Dryer:1984} kinetics are used; they were developed for premixed flames but any error for nonpremixed flames is viewed as tolerable here because diffusion would be rate-controlling.  The reaction rate (rate of change of fuel mass fraction) in units of reciprocal seconds is given as
\begin{eqnarray}
\omega_F = - A {\rho}^{0.75} Y_F^{0.1} Y_O^{1.65} e^{-50.237/\tilde{h}}
\label{onestep}
\end{eqnarray}
where the ambient reference temperature is set at 300 K and density $\rho$ is to be given in units of kilograms per cubic meter. Here, $A =4.788 \textrm{x} 10^8 (kg/m^3)^{-0.75}/s$. The dimensional reciprocal of residence time $u_{\infty}/x$ is used to normalize time and reaction rate.
In non-dimensional terms,
\begin{eqnarray}
\frac{2\omega_F x}{u_{\infty}}&=& - \frac{2 A {\rho_{\infty}}^{0.75}}{u_{\infty}/x}\tilde{h}^{-0.75} Y_F^{0.1} Y_O^{1.65} e^{-50.237/\tilde{h}}   \nonumber  \\
\frac{2\omega_F x}{u_{\infty}}&=& - \frac{Da}{\tilde{h}^{0.75}} Y_F^{0.1} Y_O^{1.65} e^{-50.237/\tilde{h}}
\label{Damkohler}
\end{eqnarray}
The above equation defines the Damk\"{o}hler number $Da$. Furthermore, we set $Da \equiv K Da_{ref}$ where
\begin{eqnarray}
Da_{ref} \equiv \frac{ 2A(10 kg/m^3)^{0.75}}{(20/s)} =2.693 \;\textrm{x}\; 10^6 \;\;  ;  \;\;
 K \equiv \Big[\frac{\rho_{\infty}}{10 kg/m^3}\Big]^{0.75}\frac{20/s}{u_{\infty}/x}
\end{eqnarray}
$\rho_{\infty} =$10$ kg/m^3$ and $u_{\infty}/x =$20$/s$ are arbitrarily chosen as reference values for density and reaction rate, respectively.    The reference value for density implies an elevated pressure. The $20/s$ reference value is in the middle of an interesting range for this chemical reaction. Clearly, there is no need to set pressure (or its proxy, density) and the strain rate separately for a one-step reaction. For propane and oxygen, the mass stoichiometric ratio $\nu =0.275$.

The non-dimensional parameter $K$ will increase (decrease) as $u_{\infty}/x$ decreases (increases) and/ or the pressure increases (decreases). $K =1$ is our base case and the range covered will include $10^{-2} \leq K\leq  3$.

\subsection{$\Sigma$ Space}  \label{sigma}

Flamelet theory \citep{Peters, Pierce}  has evolved with the use of a conserved scalar as the independent variable replacing the $y$ or $\eta$ coordinate. \cite{Bilger} has emphasized the use of element-based mass fractions which become conserved scalars because chemistry does not destroy atoms but only changes molecules. Bilger refers to it as a "mixture fraction". It only is useful as a replacement for $y$ if it remains monotonic in $y$. That will be always true for the steady-state, one-dimensional case; however, in the multidimensional case or in the one-dimensional unsteady case, monotonic behavior of the upstream boundary conditions or initial conditions becomes a requirement. Thereby, the use of the mixture  fraction  is not always optimal.

\cite{Sirignano2019b} argued that any conserved scalar that was monotonic varying in the direction normal to the flame surface would suffice. In fact, it need not have physical meaning.  A simple option is to use the solution $\Sigma$ of the following equation and boundary conditions:
\begin{eqnarray}
\Sigma'' +Pr (f + E) \Sigma'&=&  0 \nonumber \\
 \Sigma(-\infty) = 0 \;\; ; \;\; \Sigma(\infty) &=& 1
\end{eqnarray}
Using the generalized Crocco integral, the result is
\begin{eqnarray}
\Sigma(\eta) = \frac{J(\eta)}{J(\infty)}  \;\; ; \;\;
J(\eta)  \equiv  \int_{-\infty}^{\eta} e^{-I(\eta')}d\eta'  \;\; ; \;\;
I(\eta) \equiv      \int_{-\infty}^{\eta} Pr[f + E ]d\zeta
\label{Sigma}
\end{eqnarray}

$\Sigma$ equals a normalized steady-state conserved scalar. For two examples at any $Pr$ value,
\begin{eqnarray}
\Sigma = \frac{\alpha(\eta) - \alpha(-\infty)}{\alpha(\infty) - \alpha(-\infty)}
 = \frac{\beta(\eta) - \beta(-\infty)}{\beta(\infty) - \beta(-\infty)}
 \label{SigmaCS}
\end{eqnarray}
For $Pr =1$, the most natural choice for a shear layer is
\begin{eqnarray}
\Sigma = \frac{u(\eta) - u_{\infty}}{u_{\infty}- u_{-\infty}}
\end{eqnarray}
For a general value of $Pr$, $\Sigma$ as a functional of $u$ can still be given.
These results apply for the single diffusion flame as well as for a multiple-flame configuration. For the diffusion flame case with oxygen and fuel only coming in opposing flows, the mixture fraction $Z$ commonly used is simply the normalized $\alpha$ conserved scalar. Thereby, $Z = \Sigma$ in that case; however, in a broader set of problems, they are not always the same. $Z$ will not vary from zero to unity if the bounding compositions are not pure fuel and pure oxidizer.

In similar fashion to \cite{Peters},
the independent variable $y$ can be replaced by  $\Sigma(y)$ in Equation (\ref{YF}). The result is
\begin{eqnarray}
 \chi
\frac{d^2 Y_F}{d \Sigma^2}+ Pr\frac{\omega_F x}{u_{\infty} }&=& 0
\nonumber \\
 \chi \frac{d^2 \tilde{h}}{d  \Sigma^2}- Pr\frac{Q}{h_{\infty}}\frac{\omega_F x}{u_{\infty} }&=& 0
\nonumber \\
\chi \equiv \frac{1}{2} \Big(\frac{d \Sigma}{d \eta}\Big)^2 &= &
\frac{1}{2 \rho^2}  \Big(\frac{d \Sigma}{d y}\Big)^2 =
 \frac{1}{2} \frac{e^{-2I(\eta)}}{ J^2(\infty) }
\label{Norbert}
\end{eqnarray}
where $\chi$ is commonly named the scalar dissipation rate. However, for laminar mixing-layer and boundary-layer flows at $Pr =1$, it is better described as one half of the square of the strain rate. In Equation (\ref{Norbert}), $\eta(\Sigma) = J^{Inv} (\Sigma J(\infty))$ must be substituted where $J^{Inv}$ is the inverse function of $J$ which must be determined  numerically or approximated.

Equation (\ref{Blasius}) also presents $f'$ as a "conserved scalar"; thus, it will also be linear in $\Sigma$. In fact, the velocity field can be use to determine $\Sigma$.
\begin{eqnarray}
\Sigma \equiv \frac{J(\eta)}{J(\infty)}  =  \frac{\int^{\eta}_{-\infty} \big(f''(\zeta)\big)^{Pr} d\zeta}{ \int^{\infty}_{-\infty} \big(f''(\eta)\big)^{Pr} d\eta  }    \nonumber  \\
\chi =   \frac{1}{2}  \frac{\Big(d u/d \eta\Big)^{2 Pr}}{ \Big(\int^{\infty}_{-\infty} \Big(d u/d \eta\Big)^{Pr} d\eta \Big)^2    }
\end{eqnarray}
The ODEs for $Y_F$ and $\tilde{h}$ given in (\ref{Norbert}) still apply. The only parameter affecting the solutions for $Y_F(\Sigma)$ and $\tilde{h}(\Sigma)$ is the product $PrDa= 2.693$x$10^6 PrK$. For most of the $\Sigma$ space, the reaction rate is negligible and a linear relation without dependence on any parameter appears. A dependence on $PrK$ can only occur in the narrow reaction zone between two larger linear domains.

$\tilde{h}$ and $Y_F$ will have significant variation within a narrow region around each of the $\Sigma$ values where a reaction zone exists. On both sides of that narrow region, $\tilde{h}$ and $Y_m$ will be linear in $\Sigma$. Equations for the conserved scalars,  $\alpha$ and $\beta$, can be created, producing homogeneous equations with linear solutions in $\Sigma$.  There is little reason though to solve the ODEs in $\Sigma$ space since they couple back to $f(\eta)$ and $G(\eta)$. It is more sensible to solve the Equations (\ref{Blasius}) through (\ref{betatilde}) or (\ref{Blasius}) through (\ref{YF}) and (\ref{CS}) with use of (\ref{SigmaCS}) than to integrate the ODEs in $\Sigma$ space. Equation (\ref{Norbert}) is interesting but really is only useful for producing solutions with constant-density counterflow where $f$ becomes linear in $y$ and $E$ is not present.

\subsection{Numerical Method}  \label{numerical}

The system of ordinary differential equations is solved numerically  using a relaxation method and central differences. Typically, solution over the range $-3 \leq \eta \leq 3$ provides adequate fittings to the asymptotic behaviors.  Most calculations have $K = 1, Pr = 1,    u_{-\infty}/ u_{\infty} =0.25 $, and $G_{\infty}=1$ with emphasis on the effect of variation on composition of the free streams. However, the effects of $K, Pr, u_{-\infty}/ u_{\infty} $, $G_{\infty}$  and $G_{-\infty}$ variations are shown as well. The ambient temperatures in the two incoming streams are generally taken to be identical here at a value of 300 K with  calorically-perfect-gas relations yielding density and enthalpy.  Typically, except where noted, the temperatures, densities, and enthalpies of the two ambient incoming flows are identical to each other.  \cite{Sirignano2019a} has shown that the difference in ambient temperatures can have some importance on the flow.  Several qualitatively different cases with regard to the compositions of the two free streams are examined.

\section{Results}\label{results}

The results for a non-reacting case will be discussed first in the next subsection. In the following subsections, the results for a configuration with a mixing layer containing a diffusion flame will be considered followed by an examination of results for  mixing layers with two and three flames. Consequences of the imposed normal strain, differences in the two free-stream velocities,  the difference between  kinematic viscosity and thermal diffusivity will be considered through the variations of $G, u_{-\infty}/u_{\infty},$ and $Pr$, respectively. The importance of compressibility and viscous dissipation will be considered in the non-reacting case through the variation of $M$ and the consideration of kinetic energy and dissipated energy.  In the configurations with flames, the impact of chemical reaction rates will be examined through variation of $Da$.

\newpage
\subsection{Non-reacting Mixing Layer} \label{nonreact}

For the non-reacting case, the effects of Mach number $M$ are considered. Thereby, kinetic energy and viscous dissipation are not neglected although they are for reacting flows.  The impact of imposed normal strain via counterflow and the thermal diffusivity are addressed through variation of $G$ and $Pr$. Both the velocity  and scalar fields are examined. The consequences of the generalized Crocco relation is presented.

\begin{figure}[thbp]
		\centering
		\subfigure[$h/h_{\infty}$]{
			\includegraphics[height =4.5cm, width=0.48\linewidth]{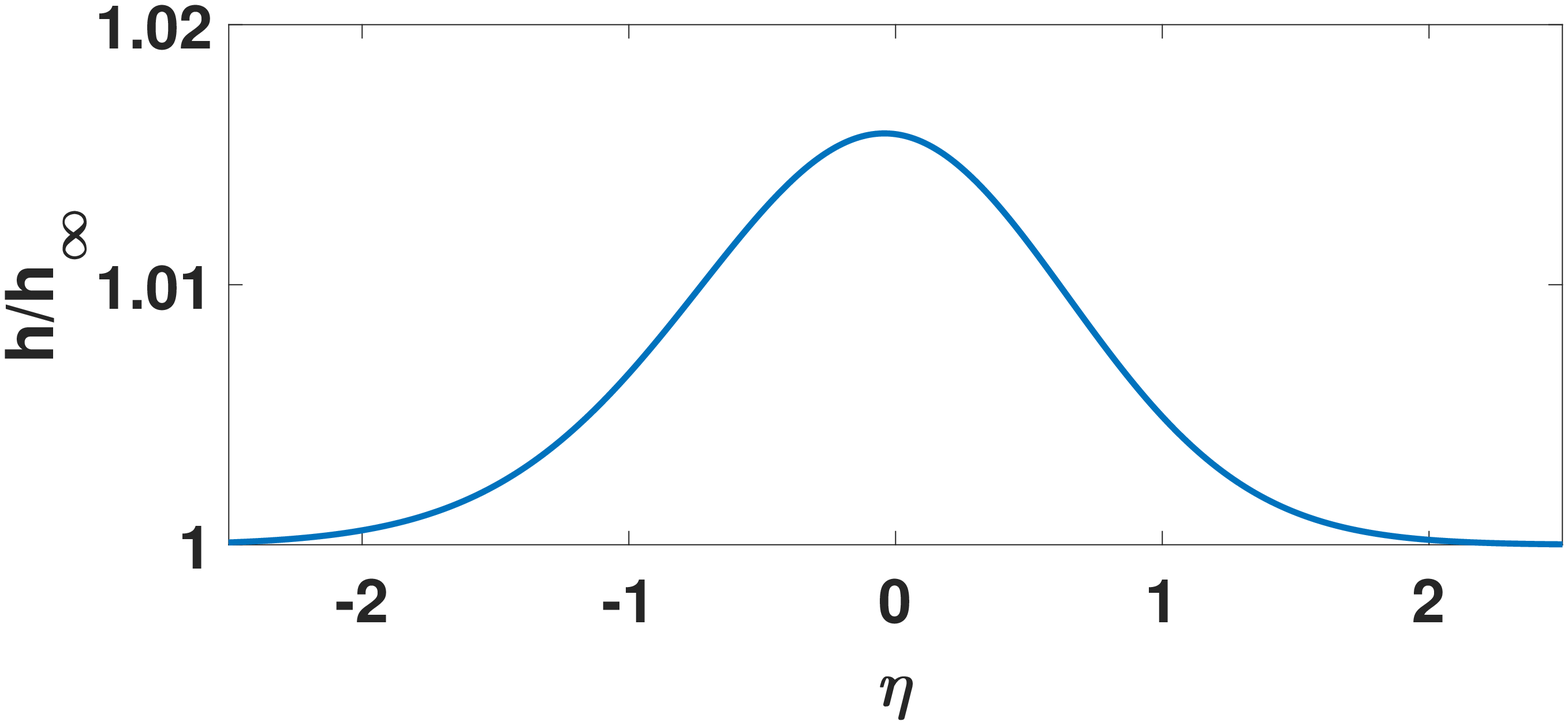}}
		\subfigure[$h_o(u)/h_{\infty}$]{
			\includegraphics[height =4.5cm, width=0.48\linewidth]{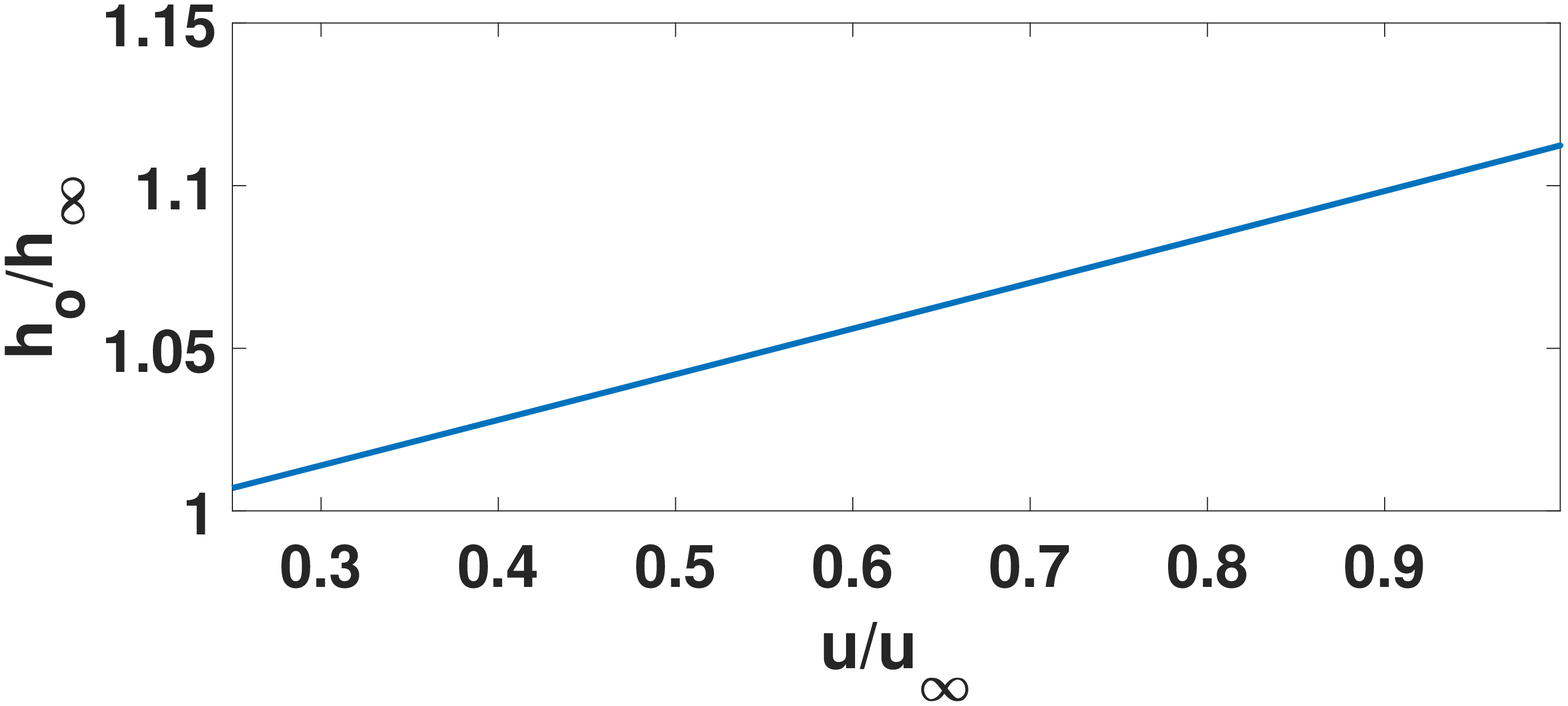}}
		\vspace{-0.1cm}
		\caption{Non-reacting mixing layer with imposed normal strain, effect of viscous dissipation and Crocco Integral solution:   $h(\eta)/h_{\infty}$ and  \\
 $h_o/h_{\infty}$ vs  $u/u_{\infty}$. $Pr =1.0  ;  G_{\infty} = 1.0; u_{-\infty} / u_{\infty}= 0.25; M =0.5 $.}
		\label{Nonreact}
	\end{figure}

\begin{figure}[thbp]
		\centering
		\subfigure[$h/h_{\infty}$]{
			\includegraphics[height =4.7cm, width=0.48\linewidth]{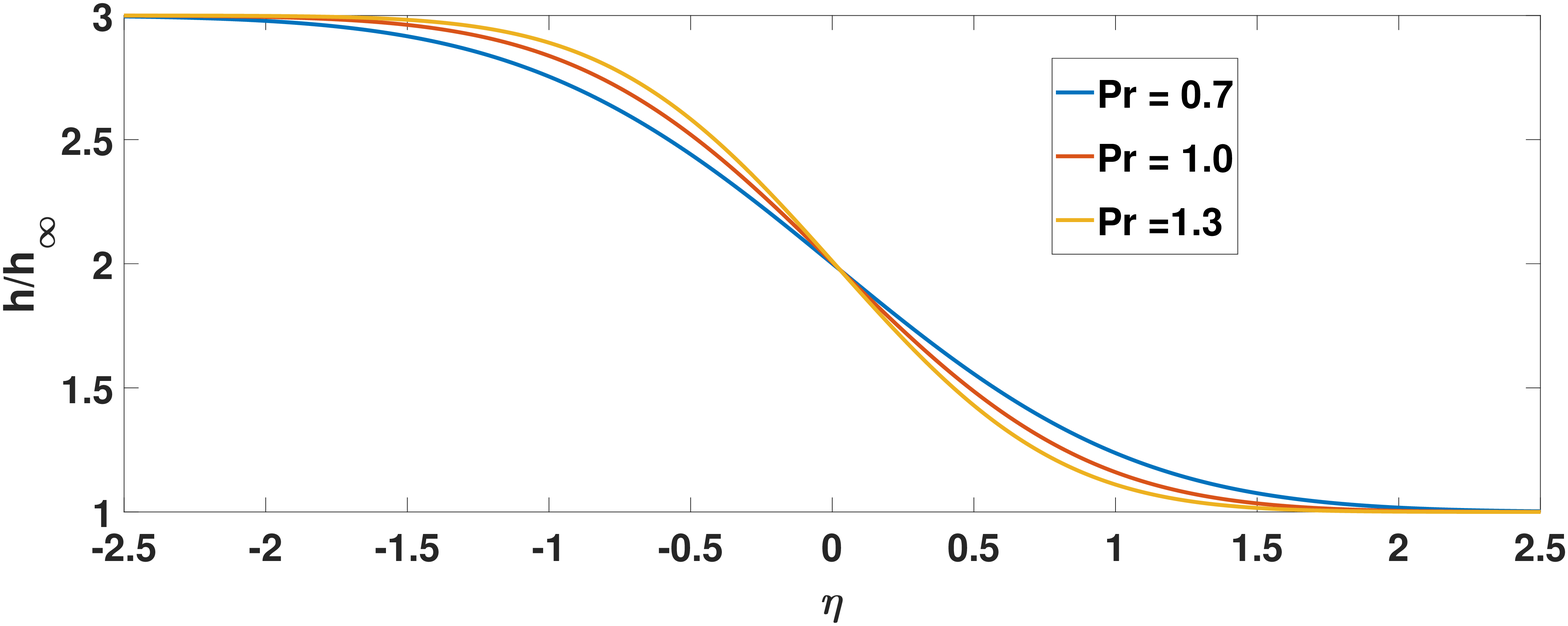}}
		\subfigure[$h_o/h_{\infty}$]{
			\includegraphics[height =4.7cm, width=0.48\linewidth]{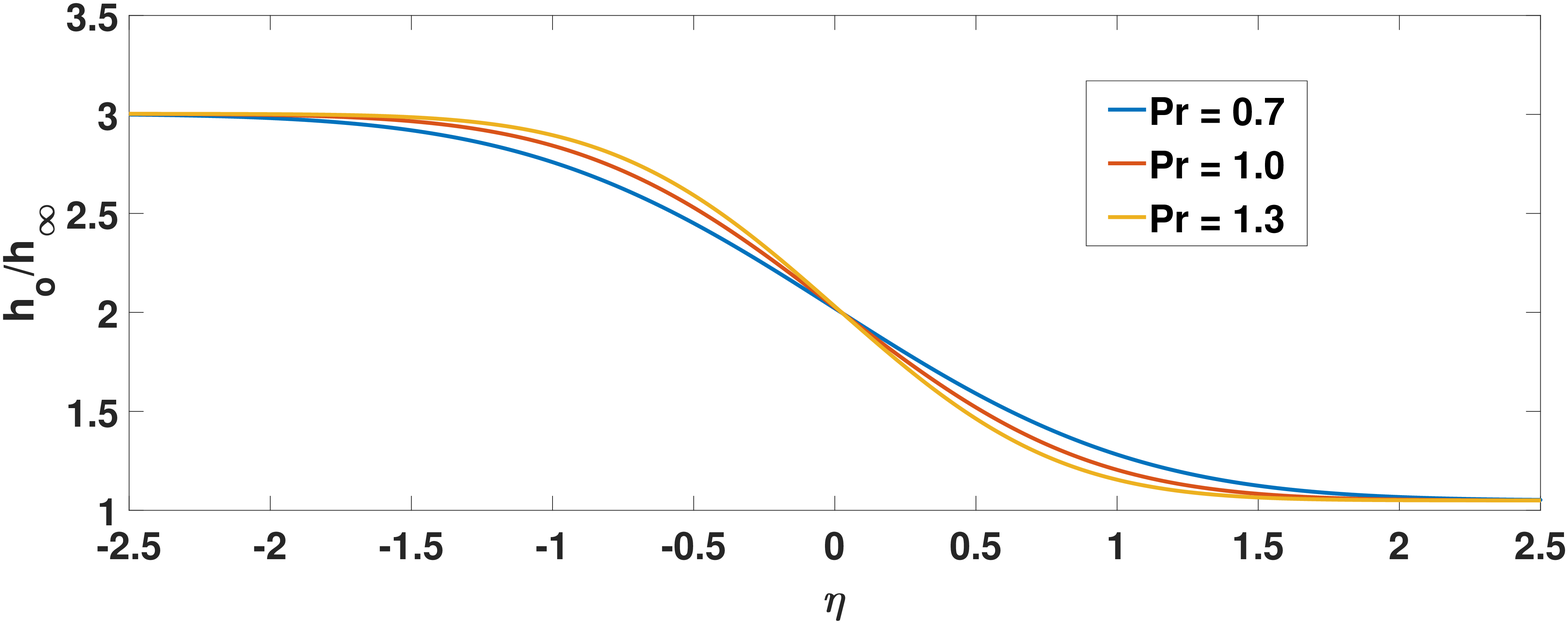}}
		\vspace{-0.1cm}
		\subfigure[$h_o(u)/h_{\infty}$]{
			\includegraphics[height =4.7cm, width=0.48\linewidth]{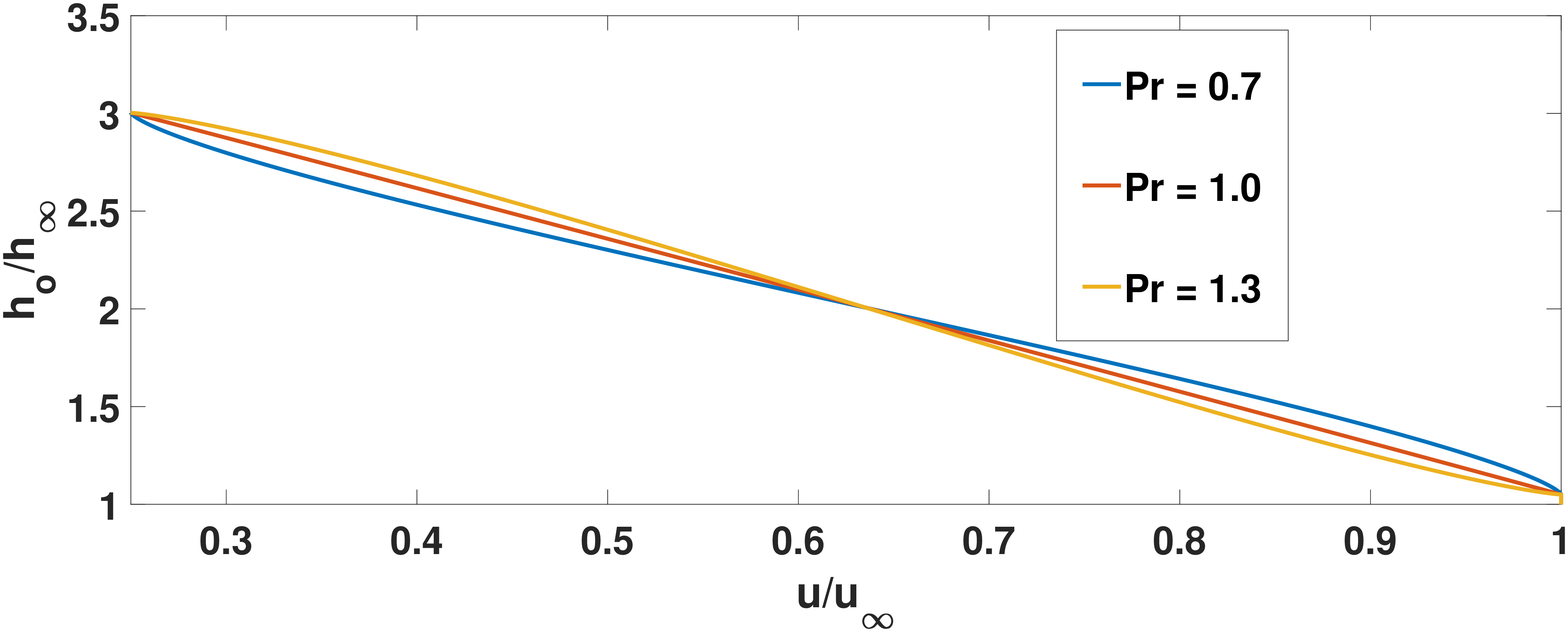}}
		\vspace{-0.1cm}
		\caption{Non-reacting mixing layer with imposed normal strain, effect of Prandtl number:   $h(\eta)/h_{\infty}, h_o(\eta)/h_{\infty}$ and $h_o/h_{\infty}$ vs  $u/u_{\infty}.  \;  \\ Pr =0.7, 1.0, 1.3  ;  G_{\infty} = 1.0; u_{-\infty} / u_{\infty}= 0.25; M =0.5 $.}
		\label{Nonreact2}
	\end{figure}

\begin{figure}[thbp]
		\centering
		\subfigure[$f$]{
			\includegraphics[height =4.7cm, width=0.48\linewidth]{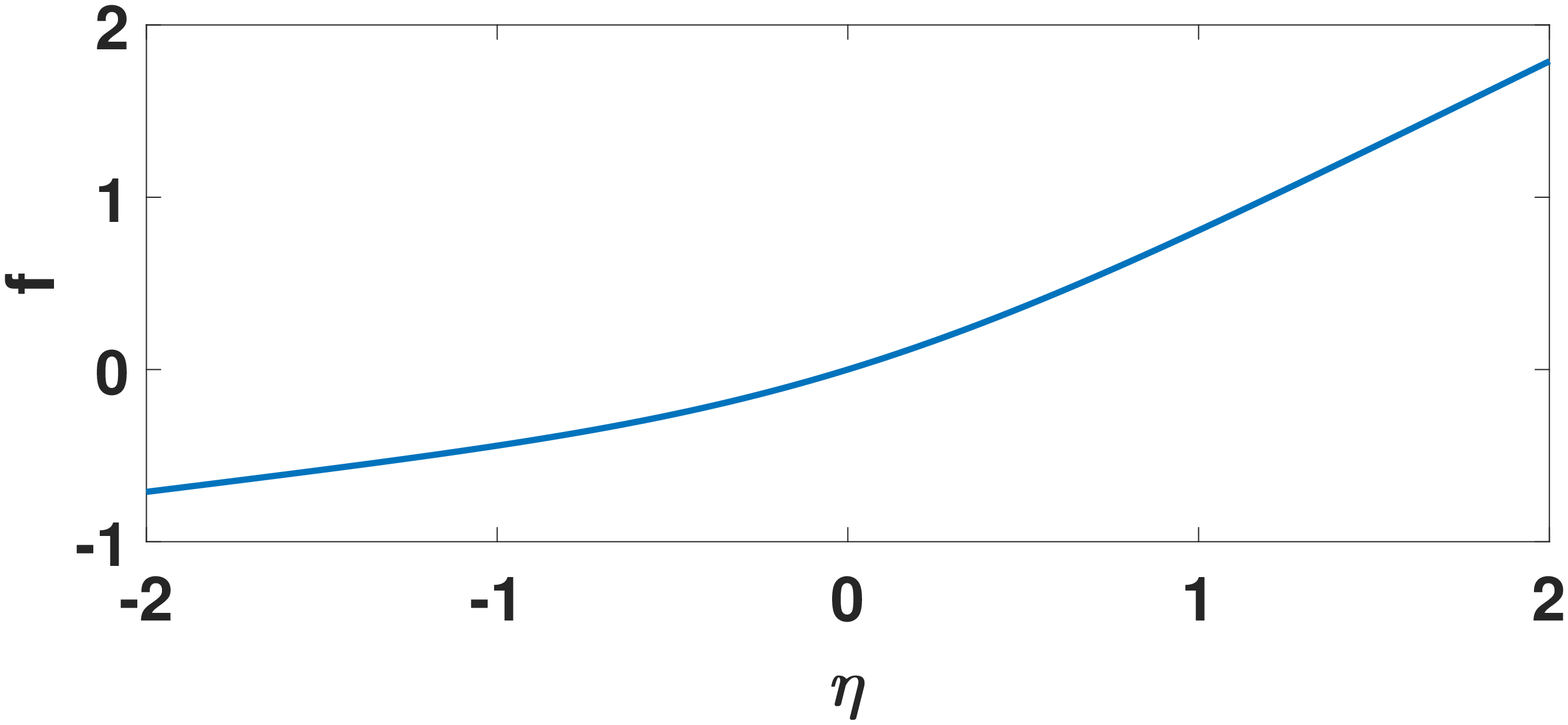}}
		\subfigure[$E$]{
			\includegraphics[height =4.7cm, width=0.48\linewidth]{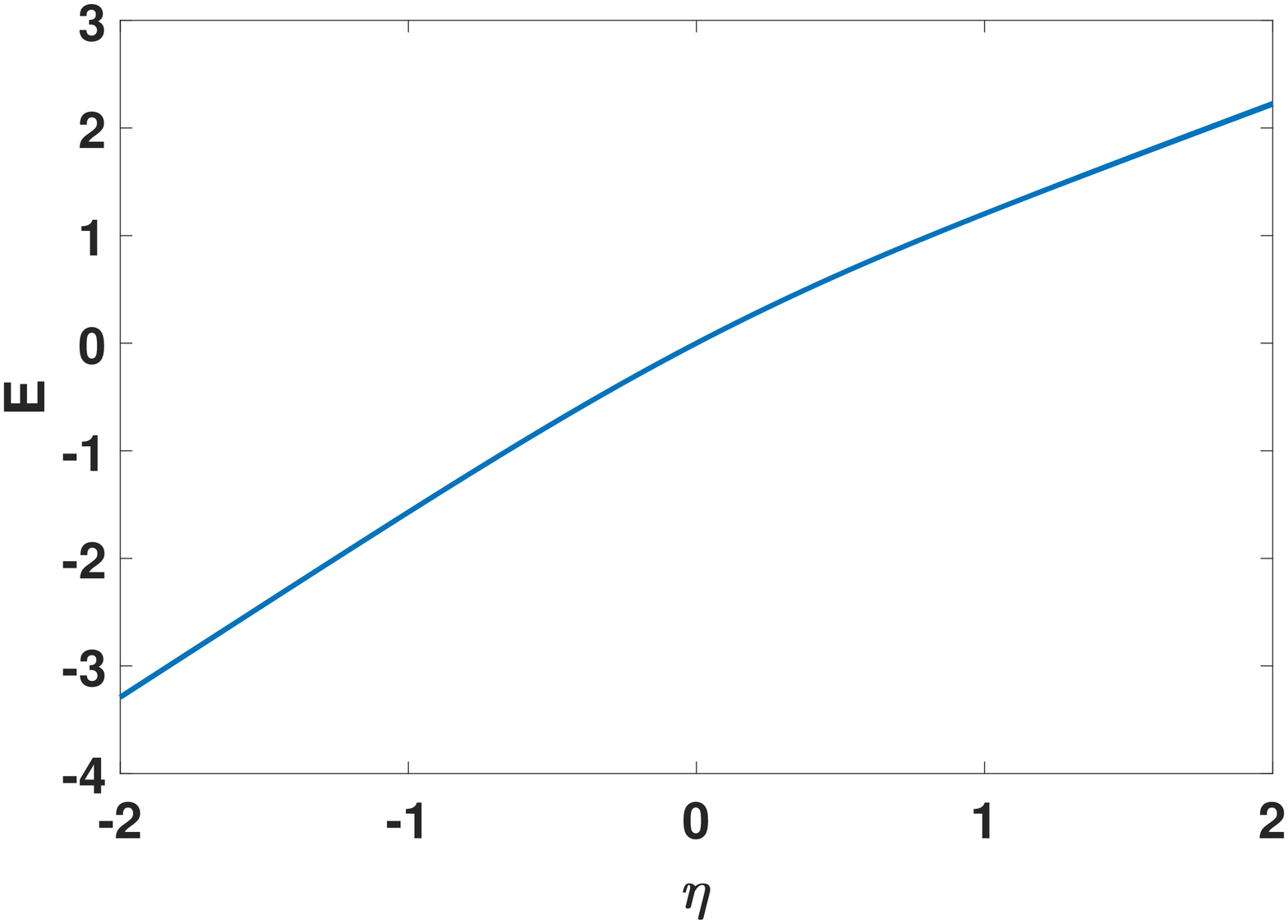}}
		\vspace{-0.1cm}
		\subfigure[$u/u_{\infty}$]{
			\includegraphics[height =4.7cm, width=0.48\linewidth]{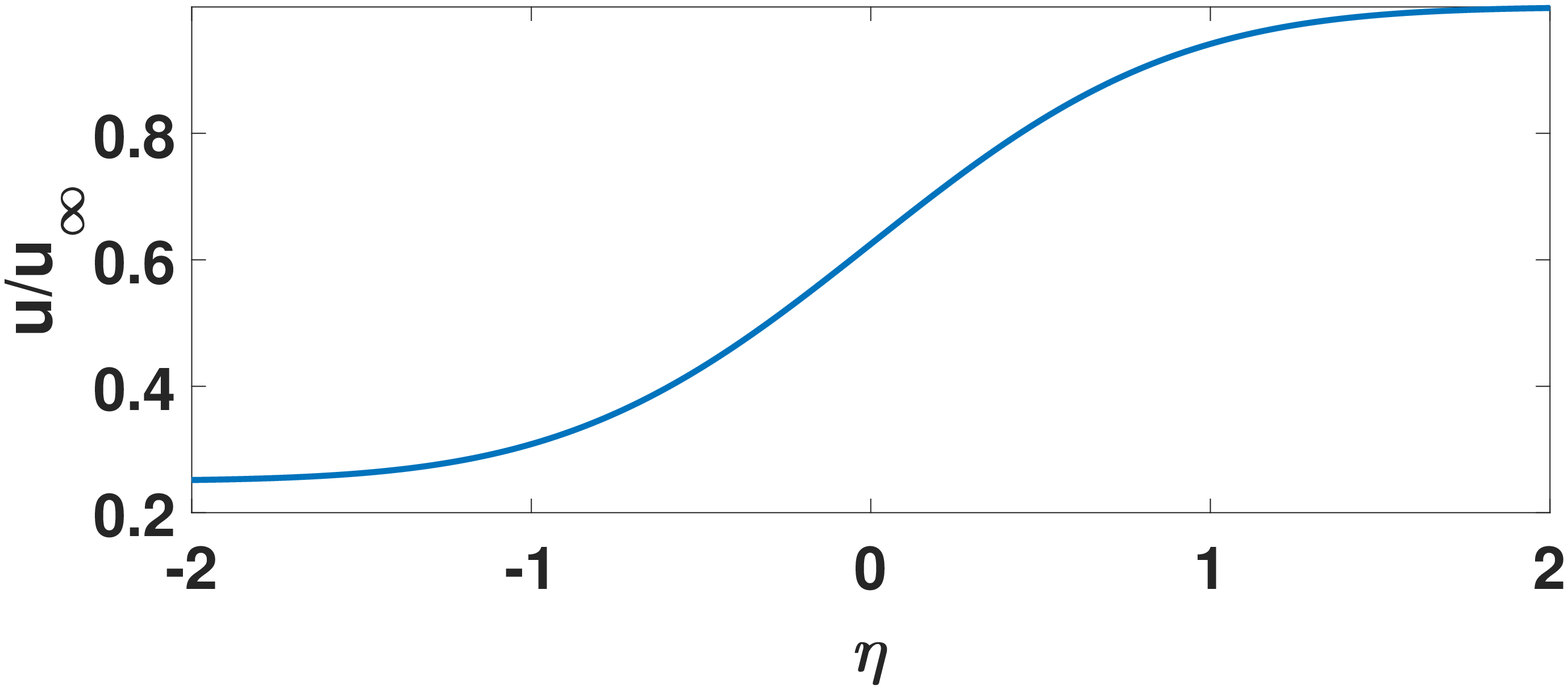}}
		\subfigure[$vg/\mu_{\infty}$]{
			\includegraphics[height =4.7cm, width=0.48\linewidth]{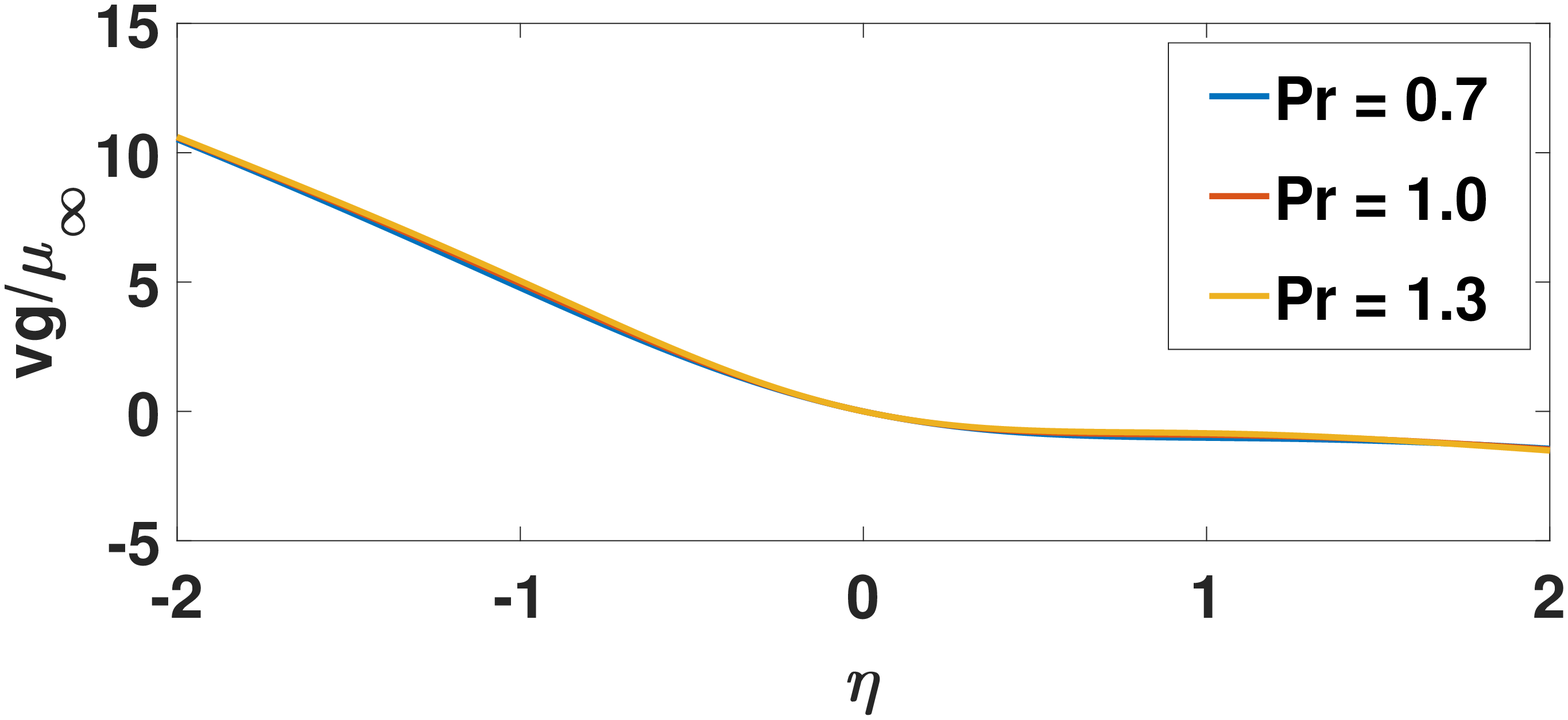}}
		\vspace{-0.1cm}
		\subfigure[$G$]{
			\includegraphics[height =4.7cm, width=0.48\linewidth]{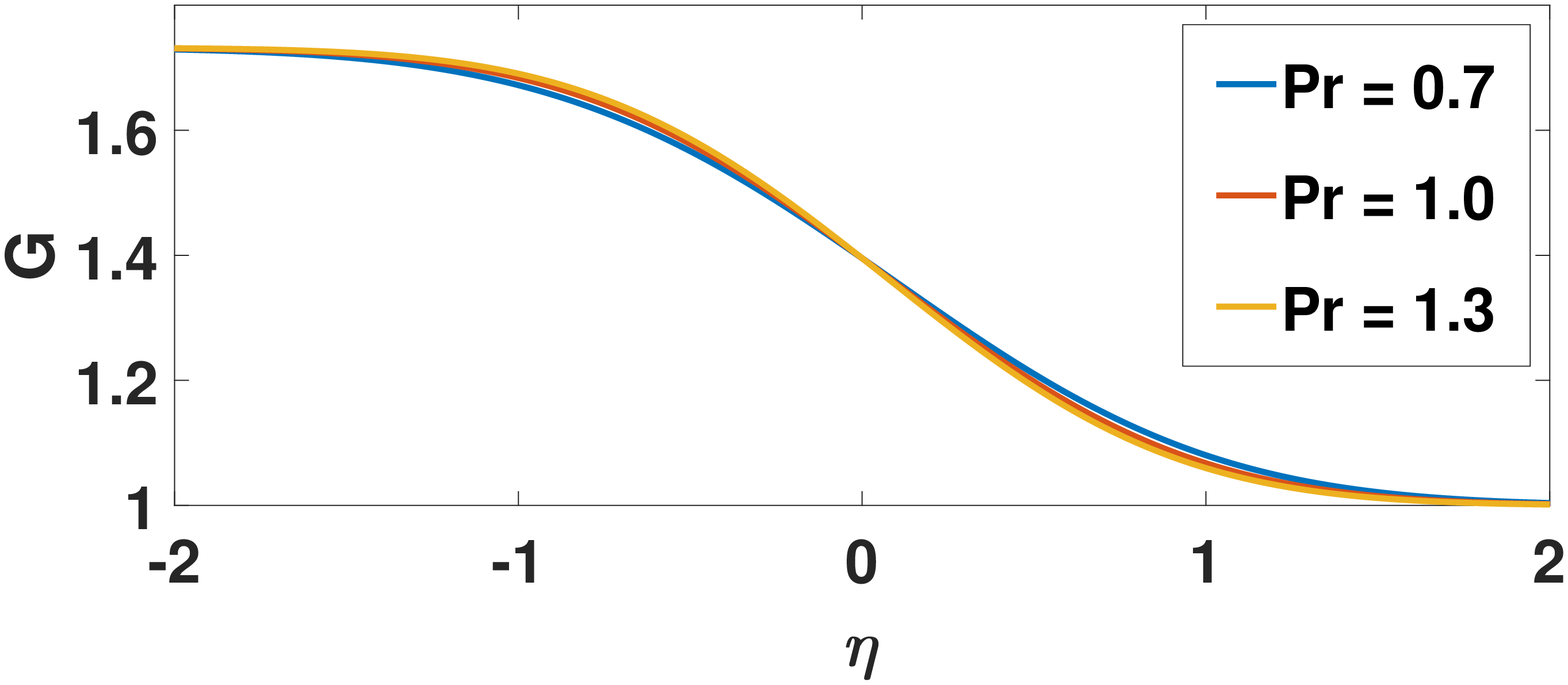}}
		\vspace{-0.1cm}
		\caption{Non-reacting mixing layer with imposed normal strain, effect of Prandtl number:   $f(\eta), E(\eta), u(\eta)/u_{\infty}, v(\eta)g/\mu_{\infty}, $ and $G(\eta)$. \\ $Pr =0.7, 1.0, 1.3  ;  G_{\infty} = 1.0; u_{-\infty} / u_{\infty}= 0.25; M =0.5 $.}
		\label{Nonreact3}
	\end{figure}

\begin{figure}[thbp]
		\centering
		\subfigure[$h/h_{\infty}$]{
			\includegraphics[height =5.3cm, width=0.48\linewidth]{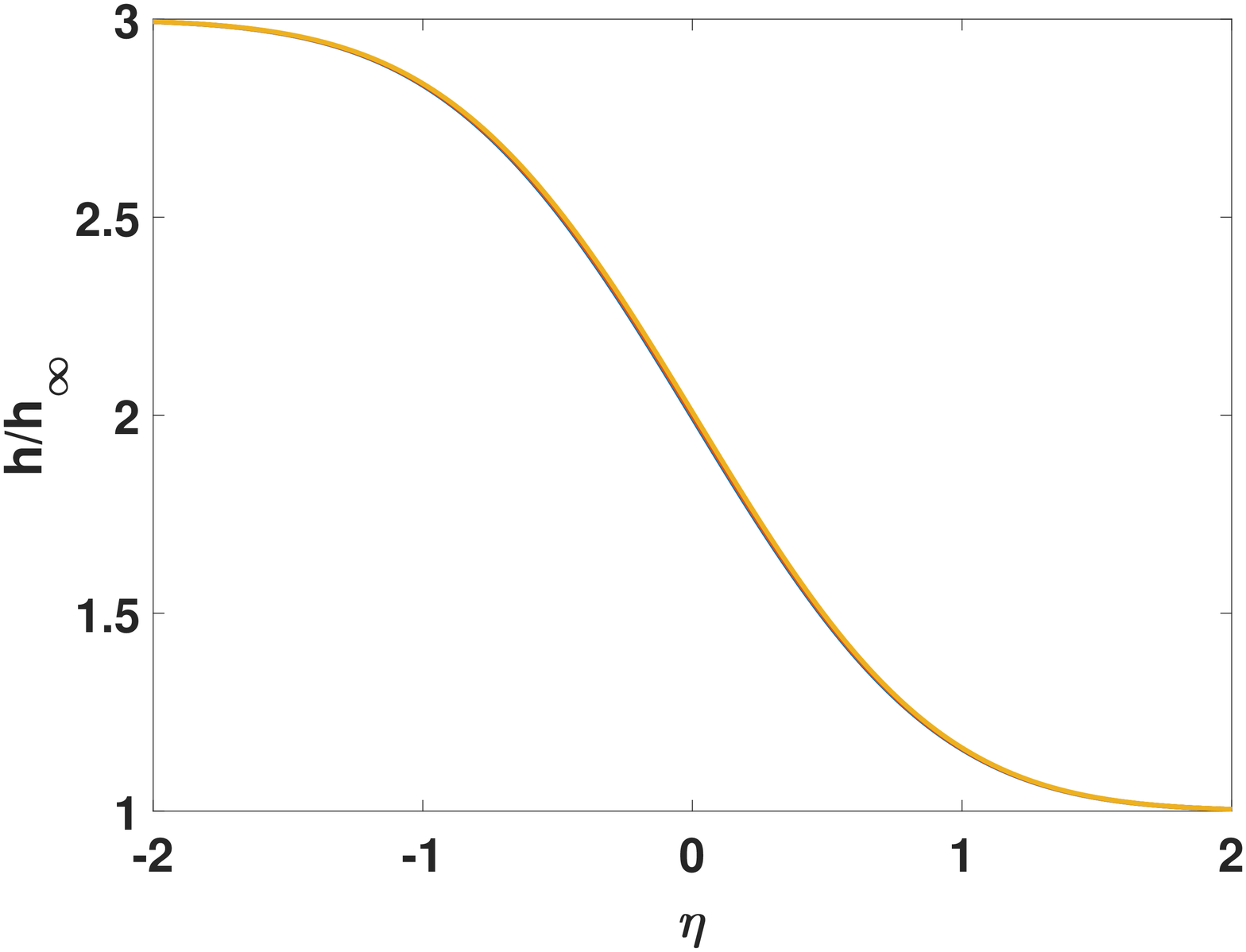}}
		\subfigure[$h_o/h_{\infty}$]{
			\includegraphics[height =5.3cm, width=0.48\linewidth]{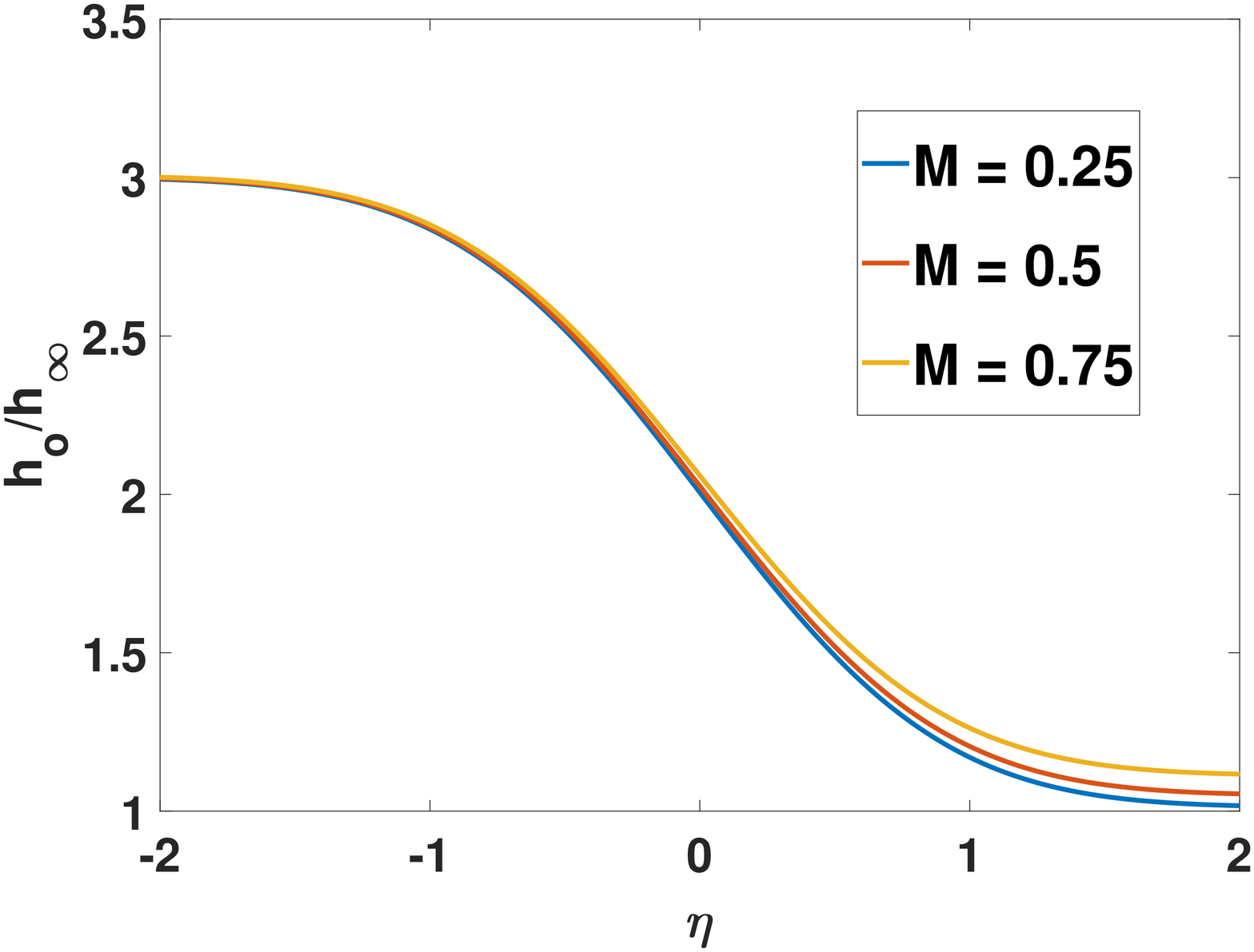}}
		\vspace{-0.1cm}
		\subfigure[$h_o(u)/h_{\infty}$]{
			\includegraphics[height =5.3cm, width=0.48\linewidth]{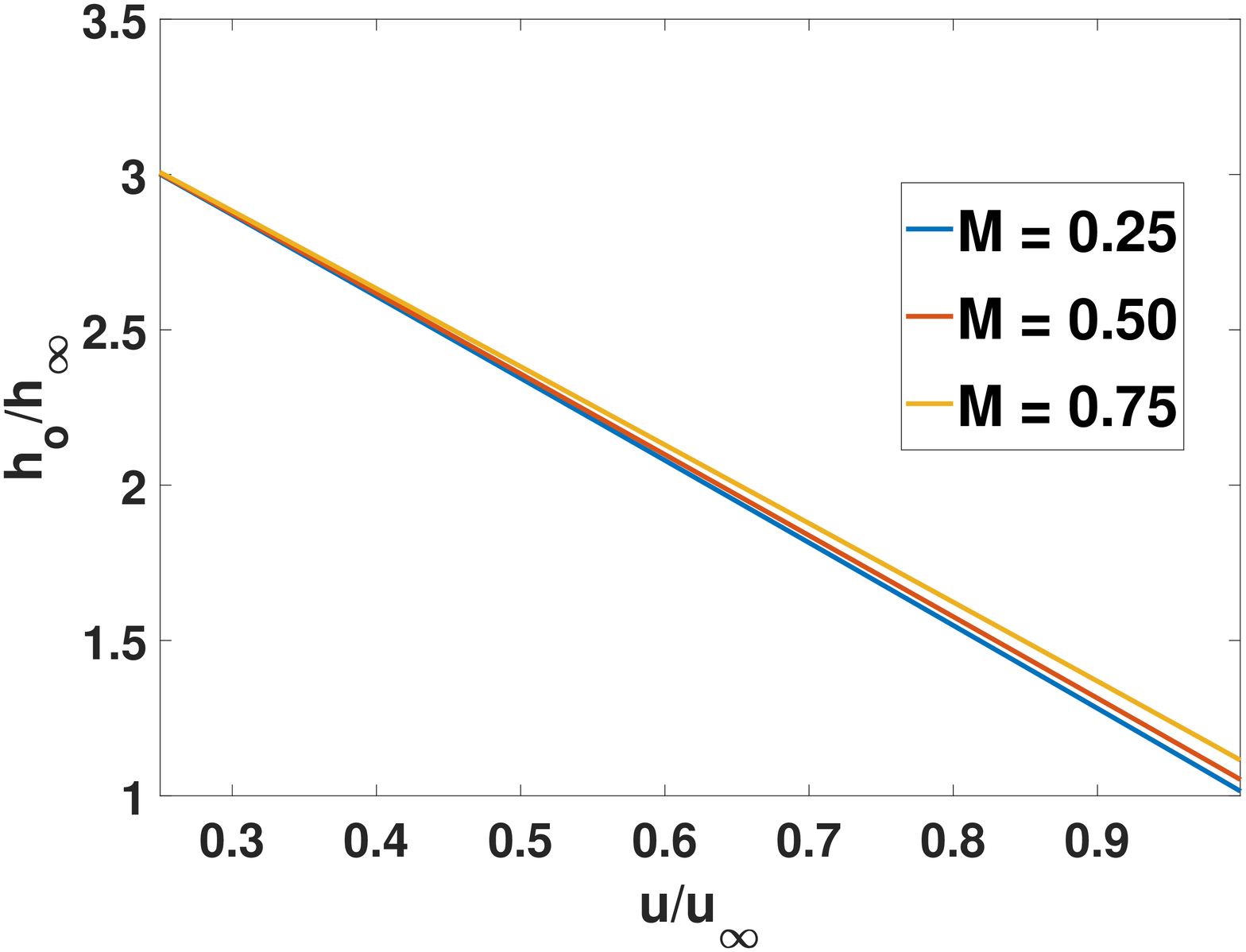}}
		\vspace{-0.1cm}
		\caption{Non-reacting mixing layer with imposed normal strain, effect of Mach number:   $h(\eta)/h_{\infty}, h_o(\eta)/h_{\infty}$ and $h_o/h_{\infty}$ vs  $u/u_{\infty}. \; Pr =1.0 ;  G_{\infty} = 1.0; u_{-\infty} / u_{\infty}= 0.25; M =0.25, 0.50, 0.75 $.}
		\label{Nonreact4}
	\end{figure}

\begin{figure}[thbp]
		\centering
		\subfigure[$h/h_{\infty}$]{
			\includegraphics[height =5.3cm, width=0.48\linewidth]{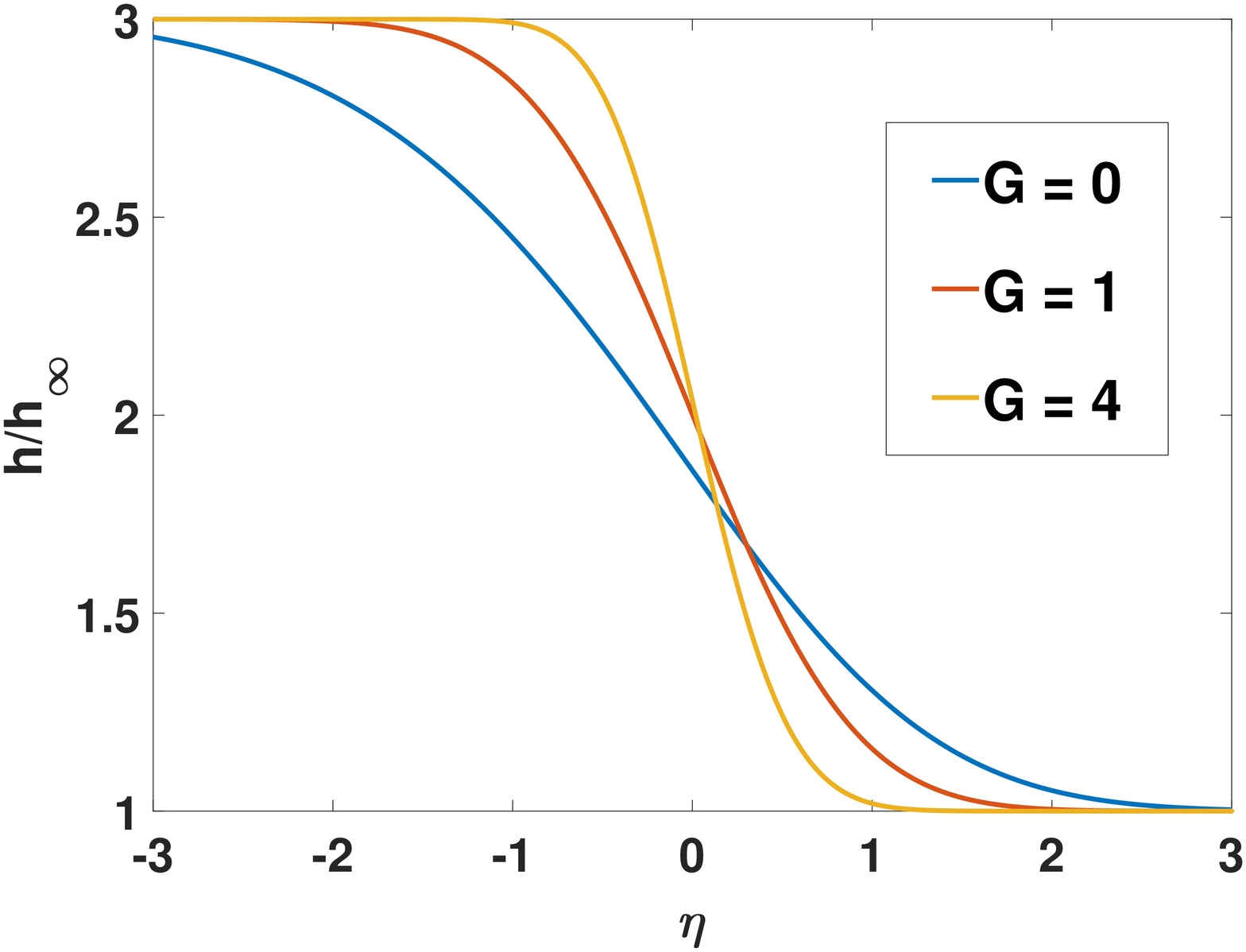}}
		\subfigure[$h_o/h_{\infty}$]{
			\includegraphics[height =5.3cm, width=0.48\linewidth]{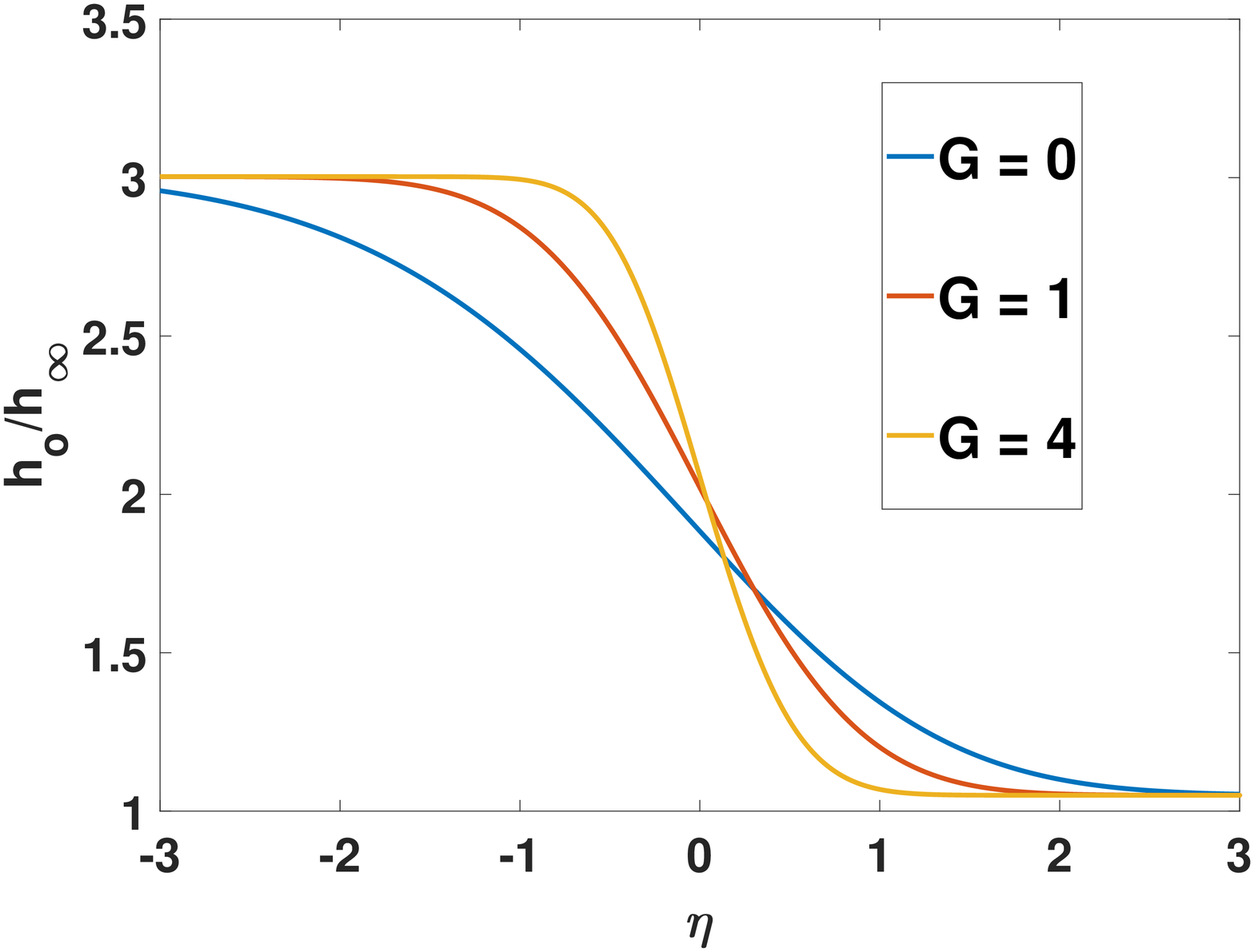}}
		\vspace{-0.1cm}
		\subfigure[$h_o(u)/h_{\infty}$]{
			\includegraphics[height =5.3cm, width=0.48\linewidth]{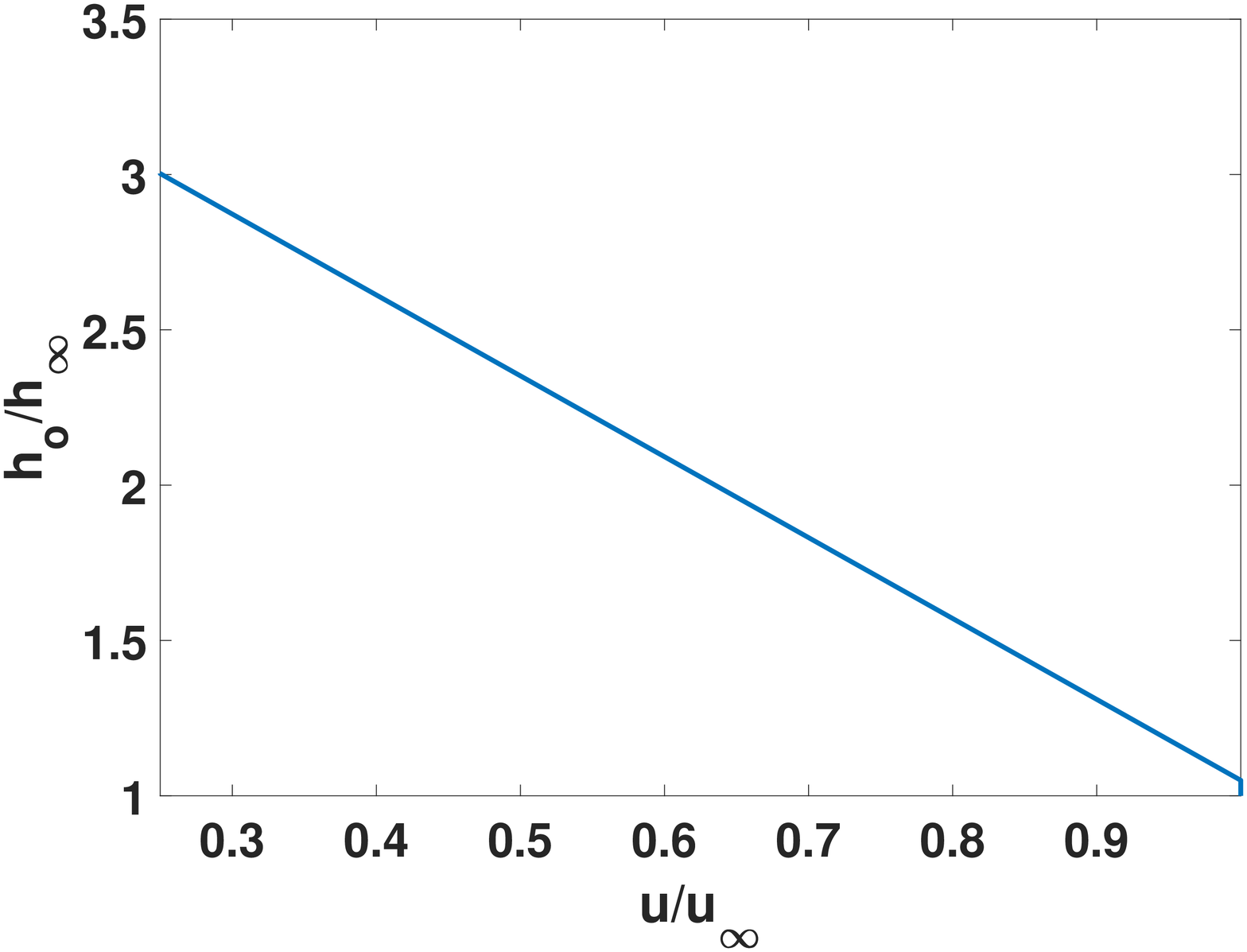}}
		\vspace{-0.1cm}
		\caption{Effect of normal strain rate on non-reacting mixing layer:   $h(\eta)/h_{\infty}, h_o(\eta)/h_{\infty}$ and $h_o/h_{\infty}$ vs  $u/u_{\infty}. \; Pr =1.0 ;  G_{\infty} = 0, 1.0,4.0 ; u_{-\infty} / u_{\infty}= 0.25; M =0.5 $.}
		\label{Nonreact5}
	\end{figure}

\begin{figure}[thbp]
		\centering
		\subfigure[$f$]{
			\includegraphics[height =5.3cm, width=0.48\linewidth]{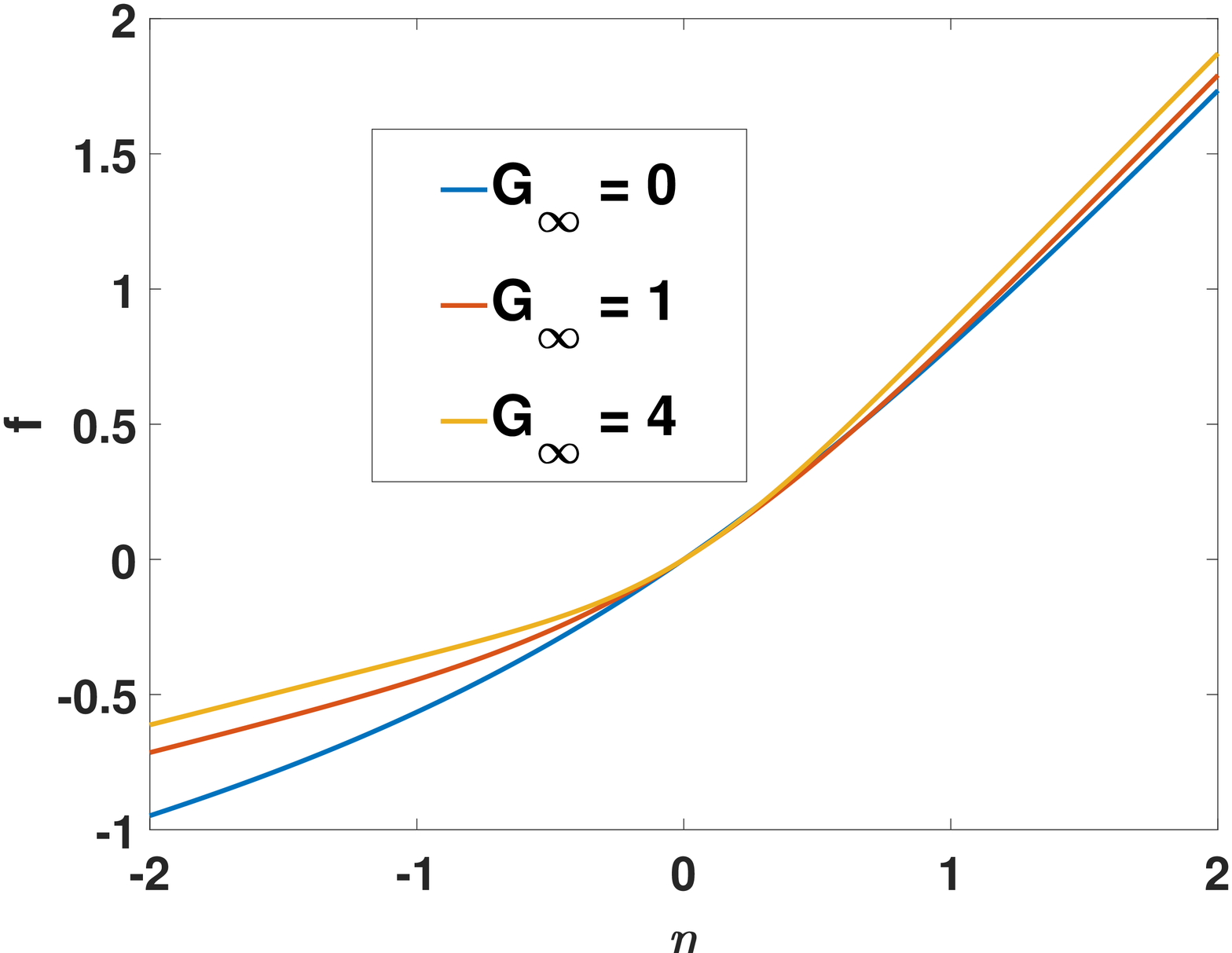}}
		\subfigure[$E$]{
			\includegraphics[height =5.3cm, width=0.48\linewidth]{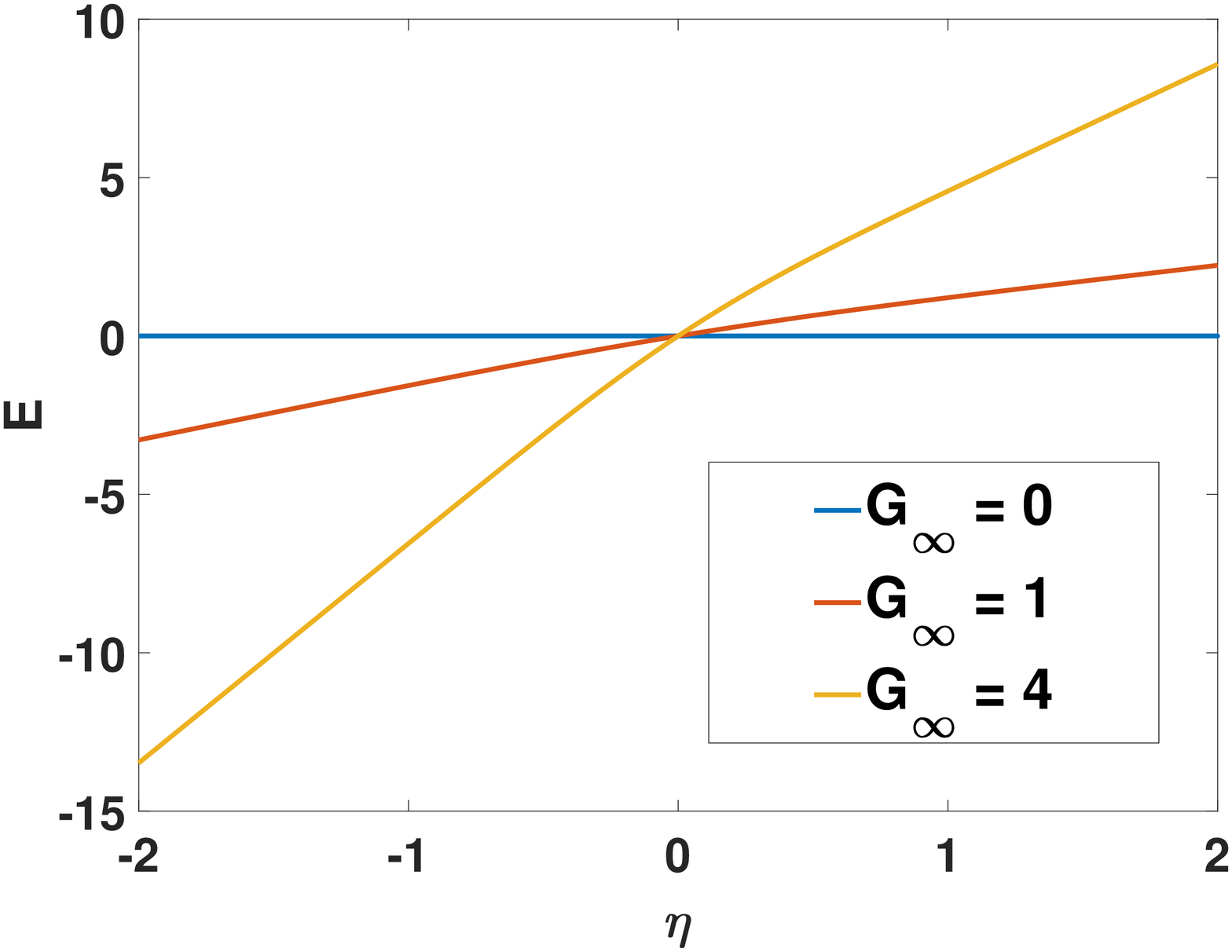}}
		\vspace{-0.1cm}
		\subfigure[$u/u_{\infty}$]{
			\includegraphics[height =5.3cm, width=0.48\linewidth]{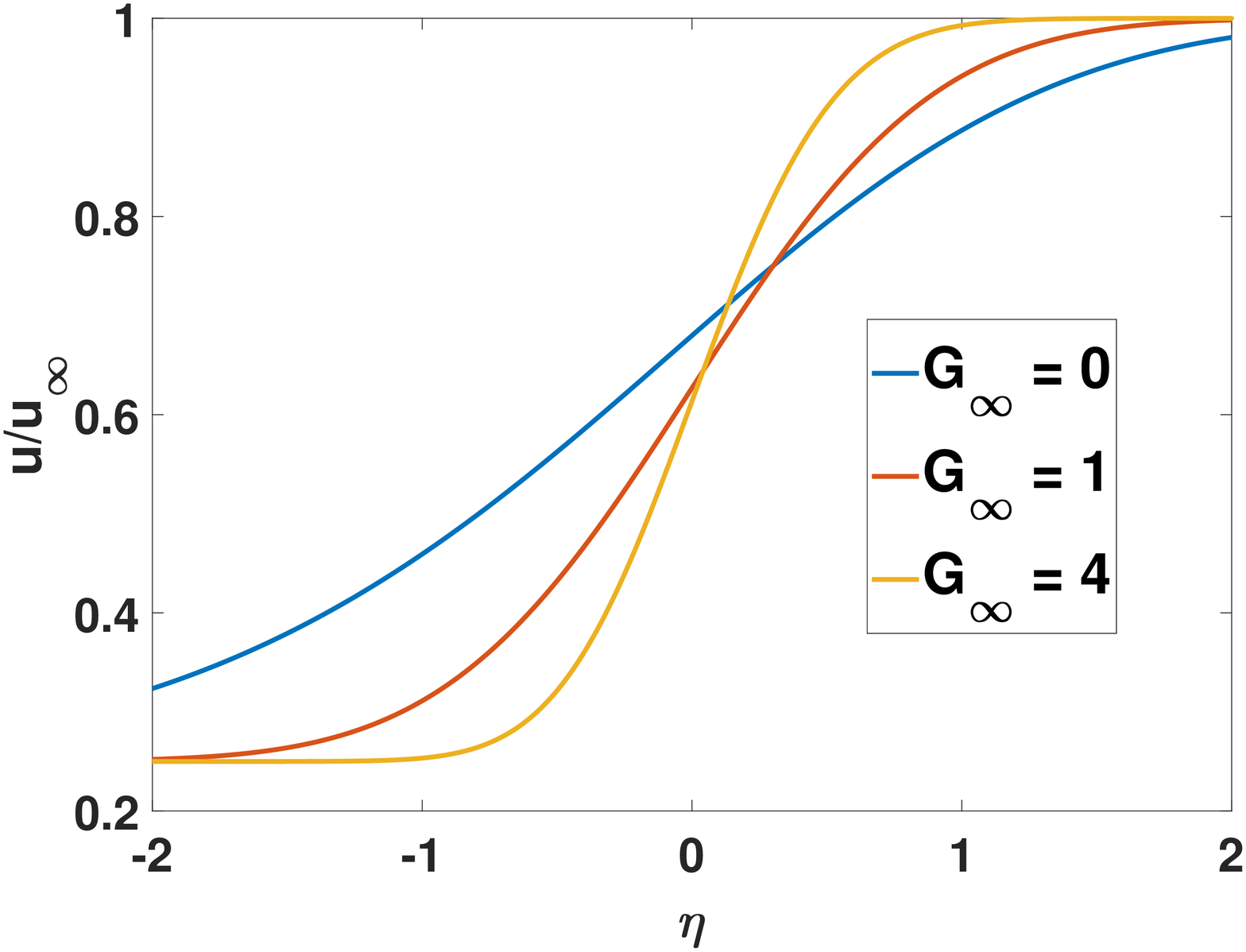}}
		\subfigure[$vg/\mu_{\infty}$]{
			\includegraphics[height =5.3cm, width=0.48\linewidth]{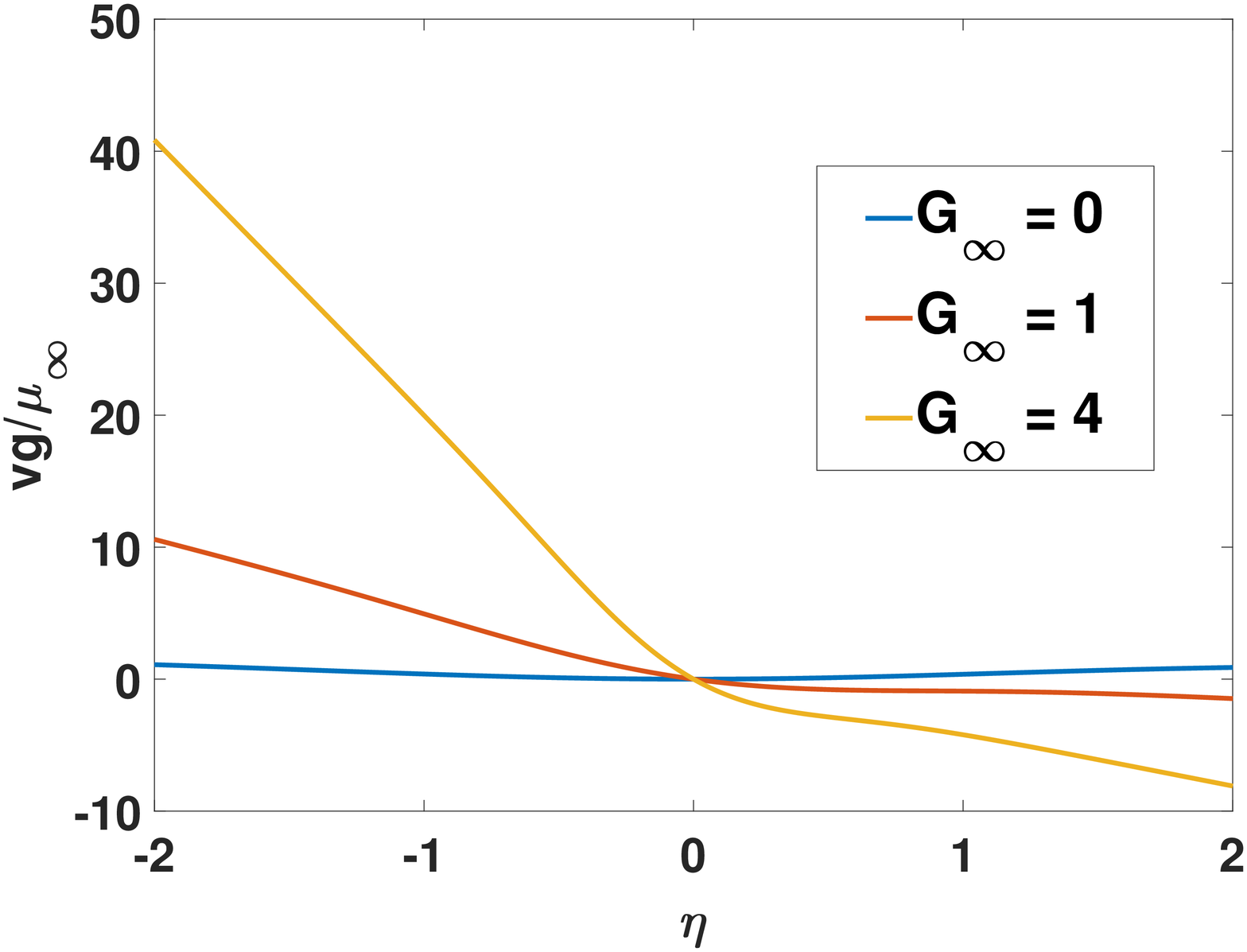}}
		\vspace{-0.1cm}
		\subfigure[$G$]{
			\includegraphics[height =5.3cm, width=0.48\linewidth]{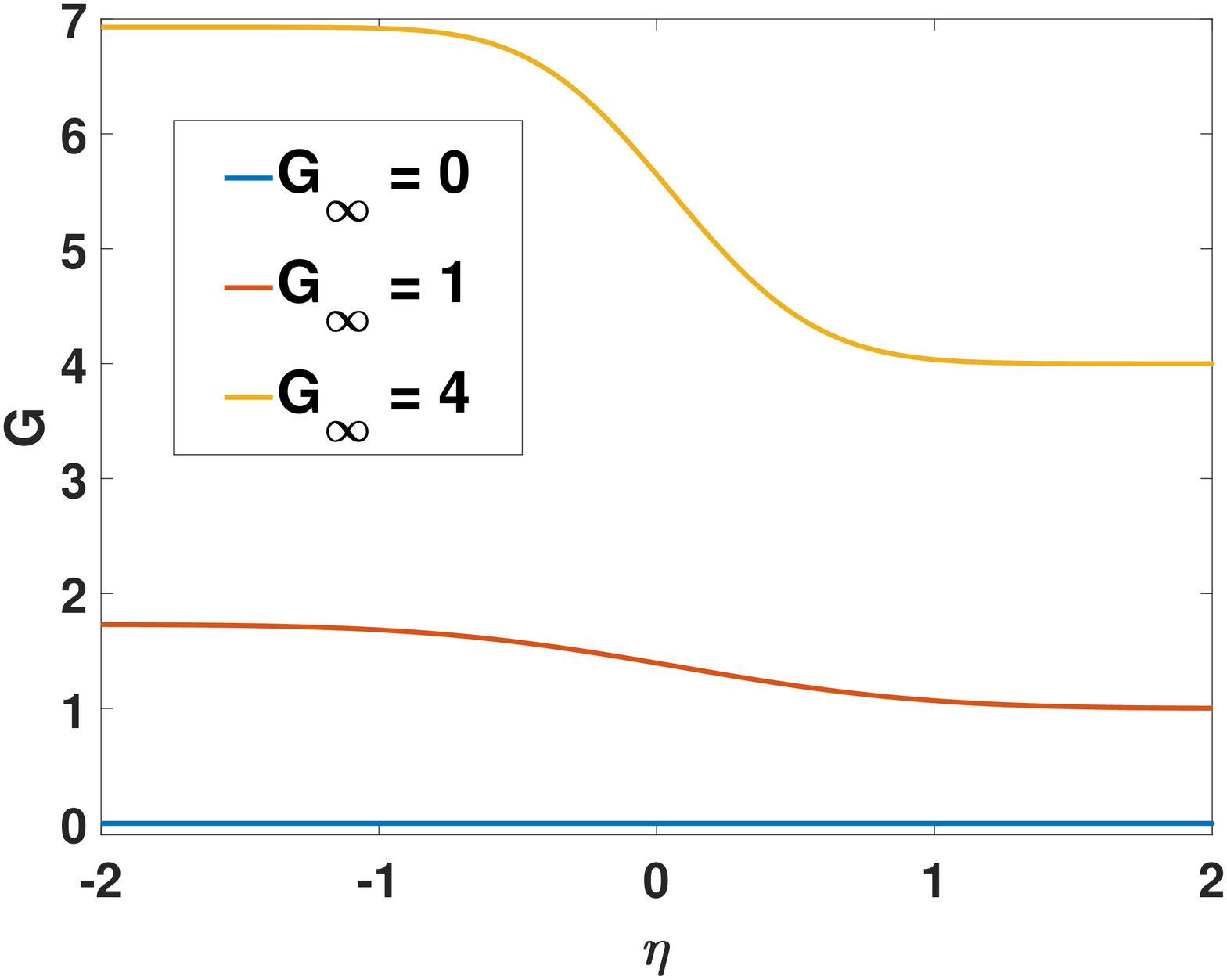}}
		\vspace{-0.1cm}
		\caption{Effect of normal strain rate on non-reacting mixing layer:   $f(\eta), E(\eta), u(\eta)/u_{\infty}, v(\eta)g/\mu_{\infty}, $ and $G(\eta)$. $Pr = 1.0  ;  G_{\infty} = 0, 1.0, 4.0;   u_{-\infty} / u_{\infty}= 0.25; M =0.5 $.}
		\label{Nonreact6}
	\end{figure}

Figure \ref{Nonreact}  applies to a case where the two free streams have the same thermodynamic conditions. The small increase of temperature and enthalpy within the mixing layer in subfigure \ref{Nonreact}a here is due only to viscous dissipation. The amount of energy dissipated is not large compared to the thermal energy in the free stream, even for the significant value of $M = 0.5$. In \ref{Nonreact}b, it is shown that, here with $Pr =1$, the classical Crocco relation holds. Figures \ref{Nonreact2} through \ref{Nonreact6} have results for cases where the two ambient temperatures are distinct and the impact of viscous dissipation is not so readily seen in the plots where enthalpy is monotonic across the layer. The value of $M$ is determined by using the higher free-stream velocity at $y =\infty$ and the lower free-stream speed of sound at $y= \infty$; Therefore, it is the highest Mach number in the flow field.  Figures \ref{Nonreact2} and \ref{Nonreact3} show variations of $h, h_o, f, E, u/u_{\infty}, vg/\mu_{\infty},$ and $G$ across the mixing layer for $Pr = 0.7, 1.0,$ and $1.3$. The effect of $Pr$ on $h$ and therefore on $h_o$ is significant; its effect on the transverse velocities $v$ and $w$ is slight, as shown via $vg/\mu_{\infty}$ and $G$; and the effect on $f, E,$ and $ u/u_{\infty}$ is negligible.
The classical linear relation between $h_o$ and $u$ is not followed in subfigure \ref{Nonreact2}c unless $Pr =1$. The differences in values for $h_o$ appear to be $O(Pr -1)$ as expected from Equation (\ref{Crocco2}). The  viscous dissipation levels remain low throughout this $Pr$ range.

Now, the effect of $M$ is shown in Figure \ref{Nonreact4} for $Pr =1$. There is negligible effect on static enthalpy but the influence on kinetic energy and therefore on $h_o$ is clear. The  slope of $h_o(u)$ increases with $M$.

The rate of normal strain is varied in Figures \ref{Nonreact5} and \ref{Nonreact6}. As the nondimensional rate of strain $G$ increases with increase in $G_{\infty}$, the mixing layer gets thinner and the slopes of both scalars and velocity increase. The linear relation between $h_o$ and $u$ does not change as $G_{\infty}$ increases for $Pr = 1$.  For $Pr\neq 1$, $h_o$ is no longer a conserved scalar; that is, viscous dissipation would appear in an equation describing it. Thereby, the function $h_o(u)$ becomes influenced by $Pr$ and linearity can be lost. The gradients of $h$ and $h_o$ increase as $Pr$ increases.
As $G_{\infty} $ increases, the values of $G, E,$ and $v$ tend to increase substantially affecting layer thickness, gradients, and transport rates. In the case where $G=0$ and $E = 0$, the value of $v$ is very small; however, it is not identically zero everywhere. Without imposed normal strain, a normal strain rate and a non-zero $v$ occur because the accelerating flow for $\eta <0$ has converging streamlines while the decelerating flow for $\eta>0$ has diverging streamlines, each producing a positive $v$. Realize that we have arbitrarily taken $v=0$ at $y=0$ (i.e., $\eta =0$) to represent the dividing streamline between the two streams. The exact (small) $y(x)$ value for the dividing streamline can easily be found by a well-known correction that sets the $y$-momentum flux of the two streams in balance. The correction does not modify strain rates or mixing rates; thus, we bypass it here.
\newpage

\subsection{Diffusion Flame}  \label{diff}

\begin{figure}[thbp]
		\centering
		\subfigure[$h/h_{\infty}$]{
			\includegraphics[height =5.5cm, width=0.48\linewidth]{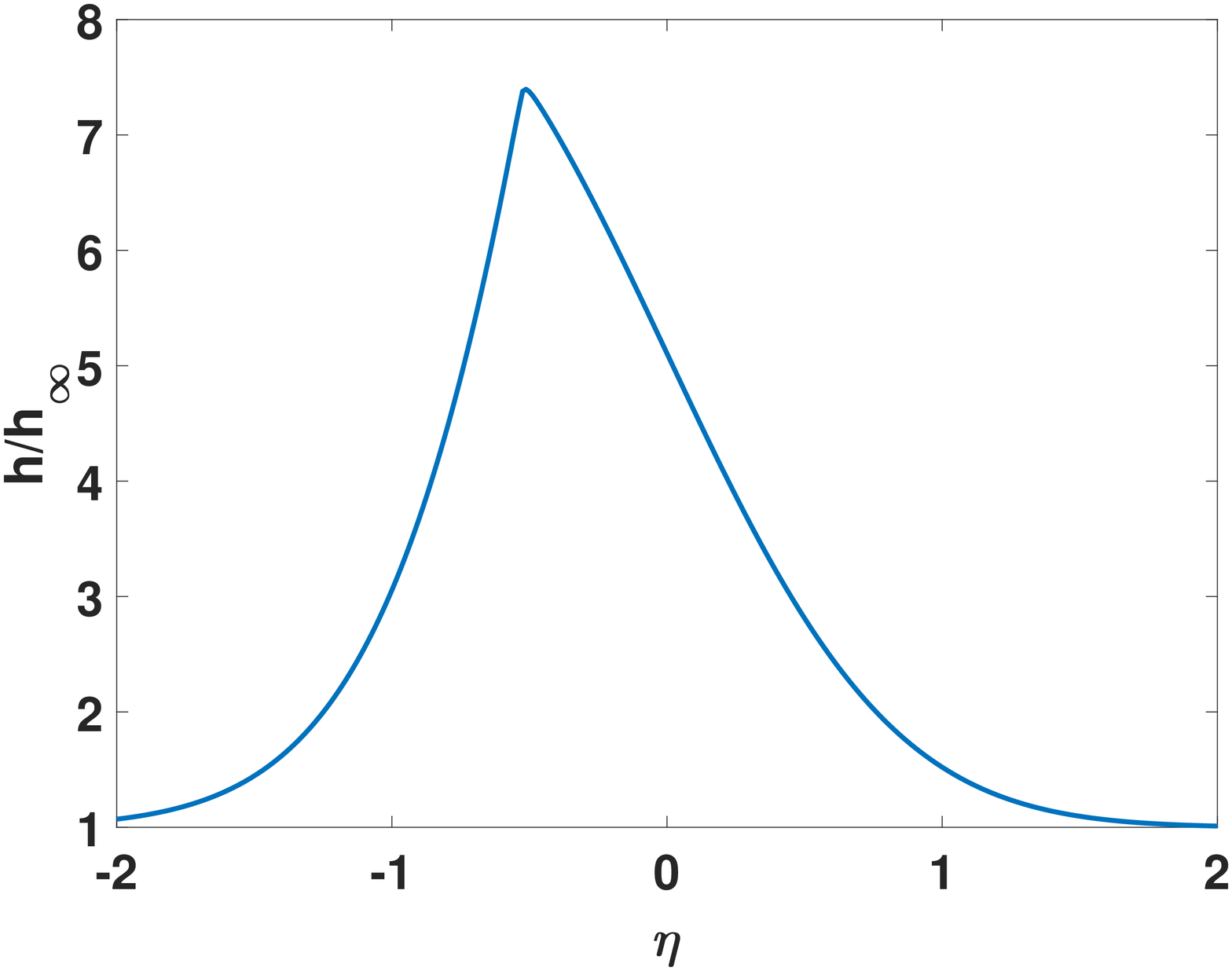}}
		\subfigure[$\omega_F x / u_{\infty}$]{
			\includegraphics[height =5.5cm, width=0.48\linewidth]{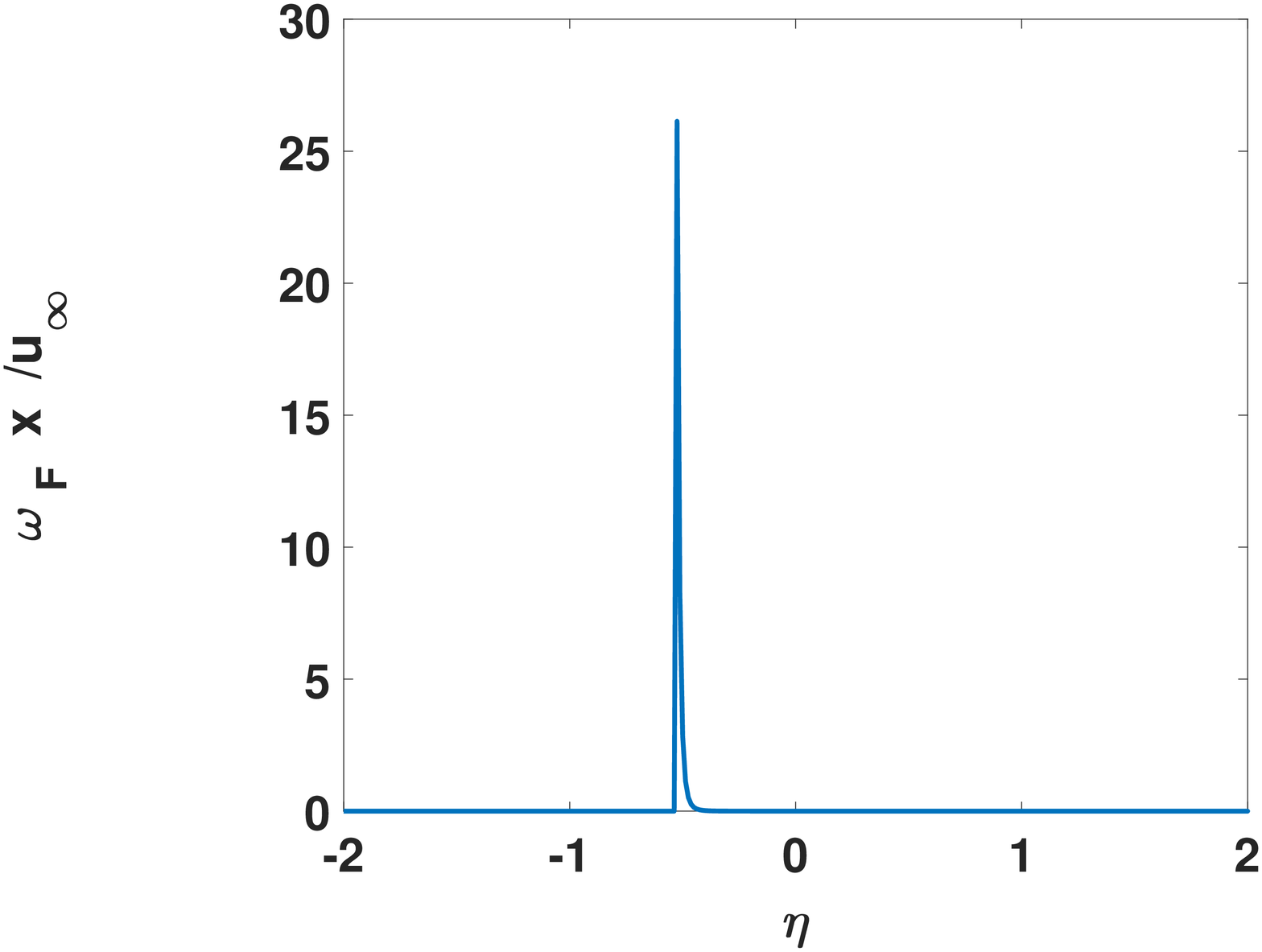}}
		\vspace{-0.1cm}
		\subfigure[$Y_F$]{
			\includegraphics[height =5.5cm, width=0.48\linewidth]{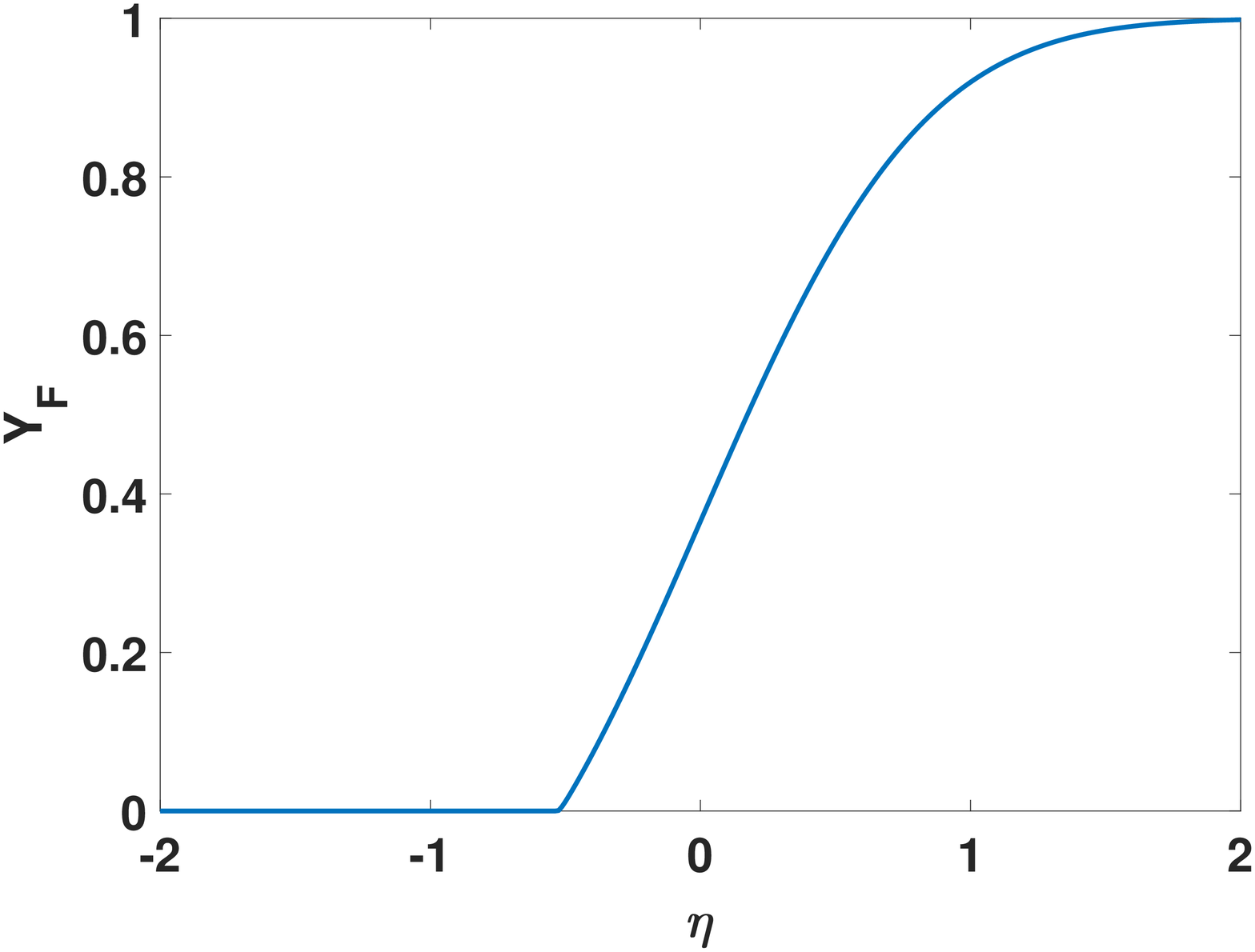}}
		\subfigure[$\nu Y_O$]{
			\includegraphics[height =5.5cm, width=0.48\linewidth]{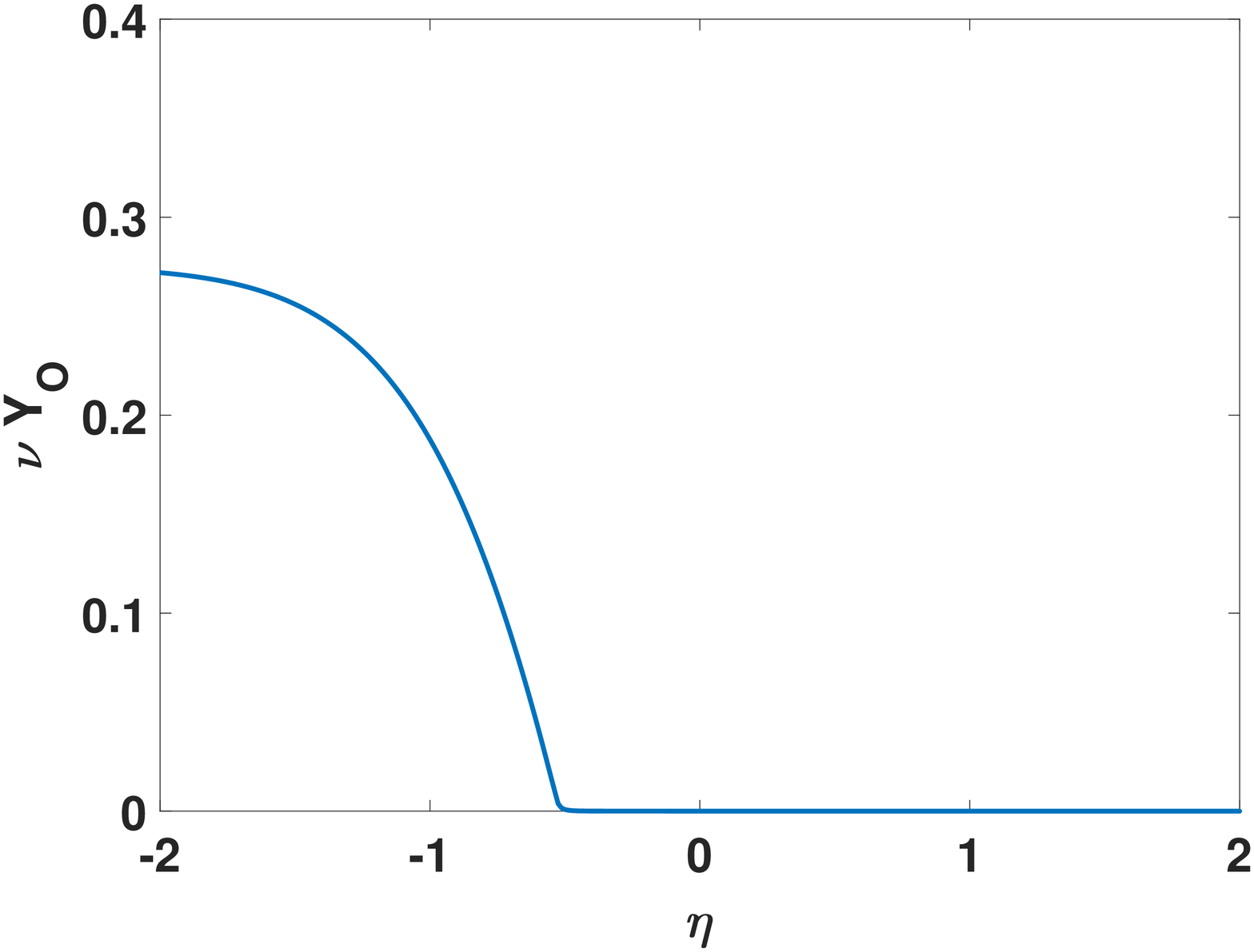}}
		\vspace{-0.1cm}
		\subfigure[$\alpha$]{
			\includegraphics[height =5cm, width=0.48\linewidth]{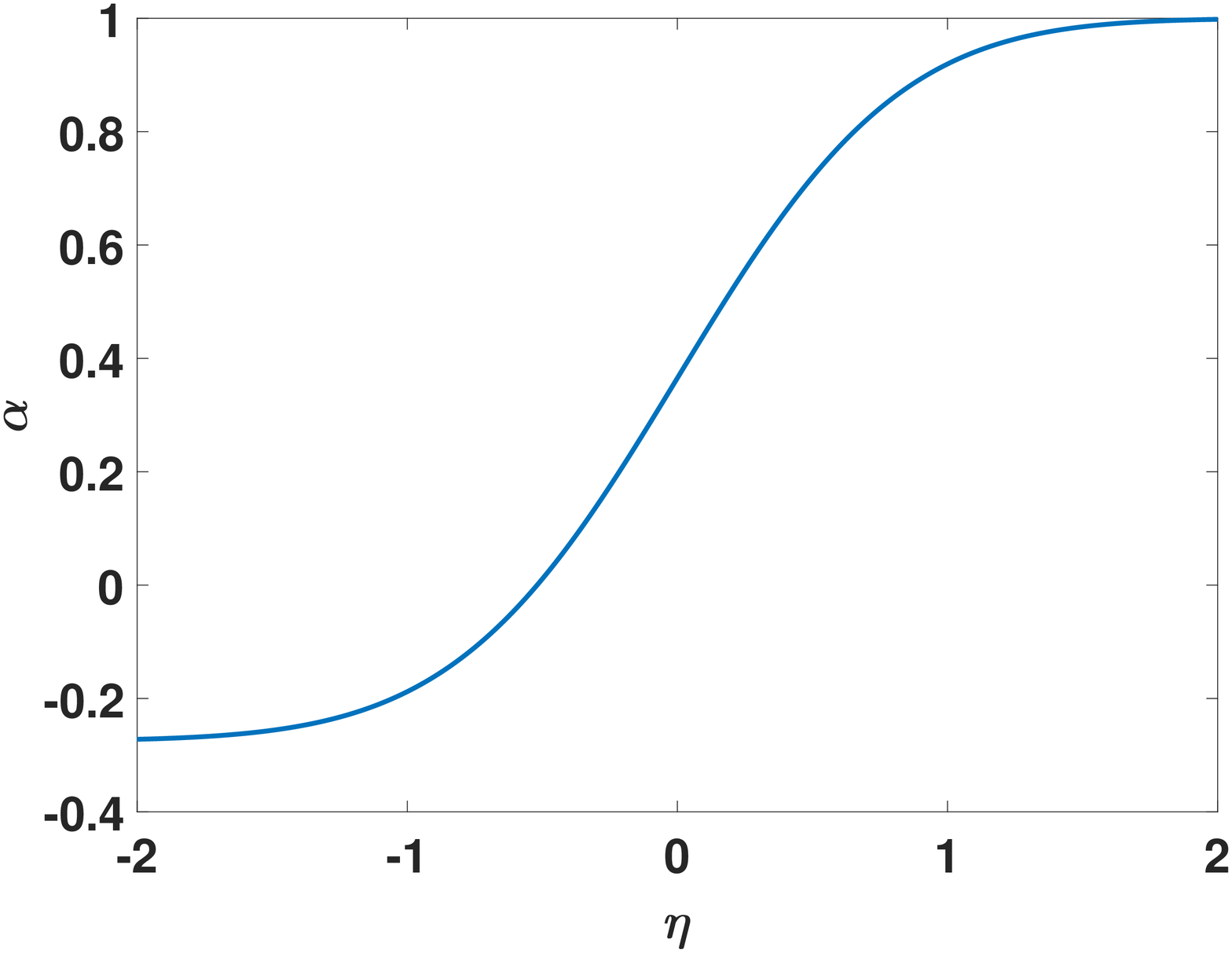}}
		\vspace{-0.1cm}
         \subfigure[$\beta$]{
			\includegraphics[height =5cm, width=0.48\linewidth]{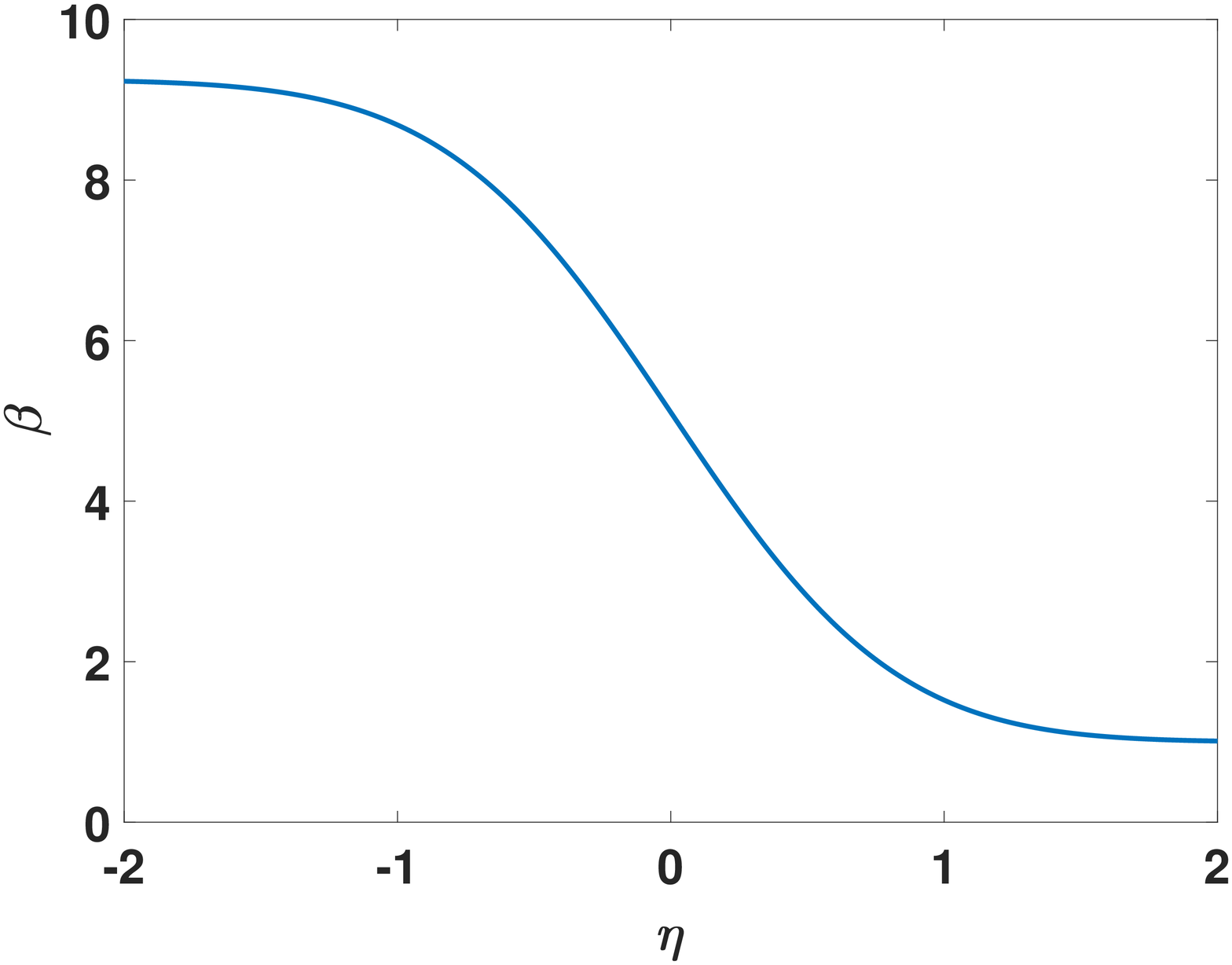}}
		\vspace{-0.1cm}
		\caption{Single diffusion flame solutions:  enthalpy $h/h_{\infty}$; reaction rate $\omega_F x/u_{\infty}$; mass fractions $\nu Y_O, Y_F; and $ scalars $\alpha, \beta.$    $K =1.0; Pr =1.0;  G_{\infty} = 1.0; u_{-\infty} / u_{\infty}= 0.25 $.}
		\label{DiffusionFlame}
	\end{figure}

\begin{figure}[thbp]
		\centering
		\subfigure[$f$]{
			\includegraphics[height =5.3cm, width=0.48\linewidth]{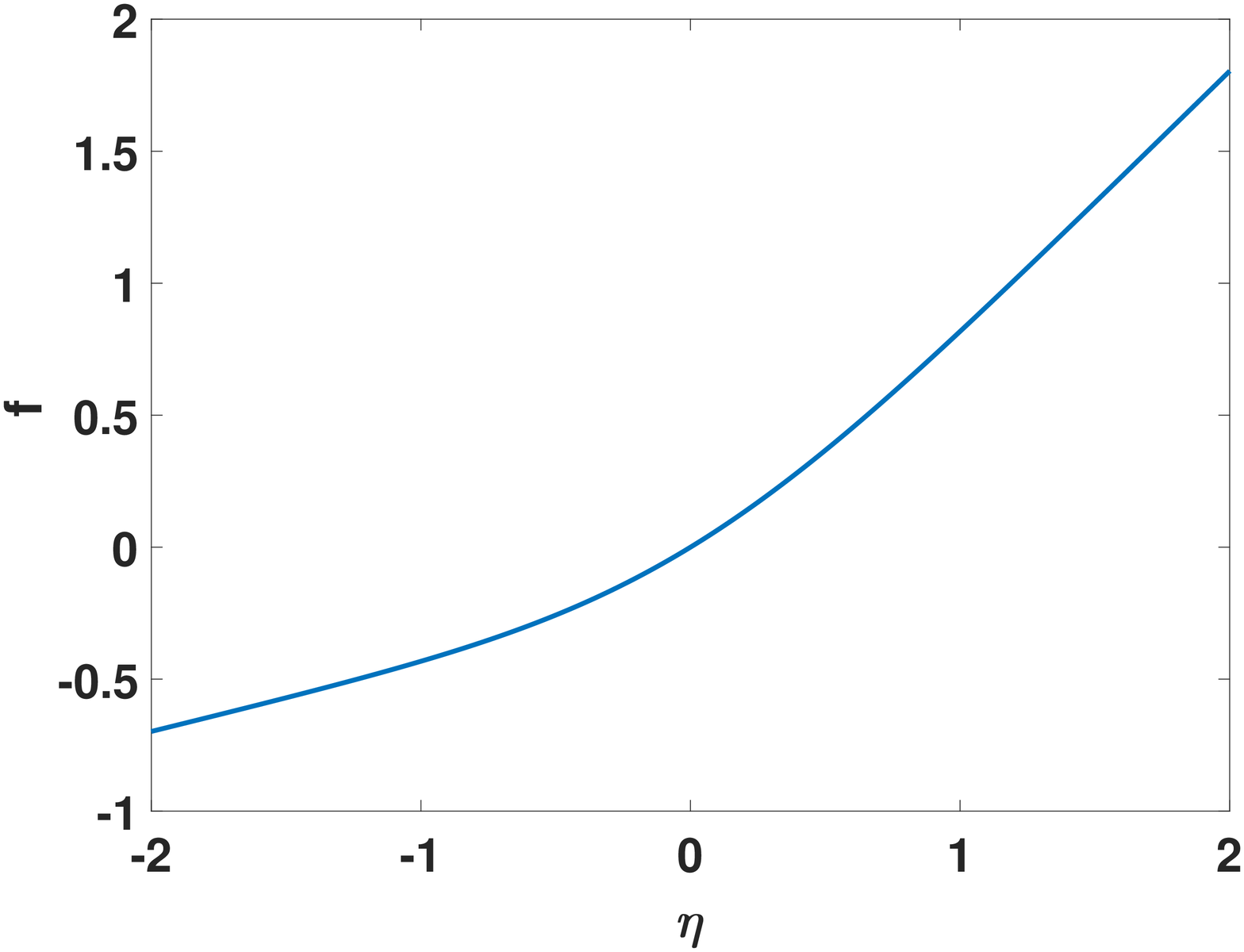}}
		\subfigure[$E$]{
			\includegraphics[height =5.3cm, width=0.48\linewidth]{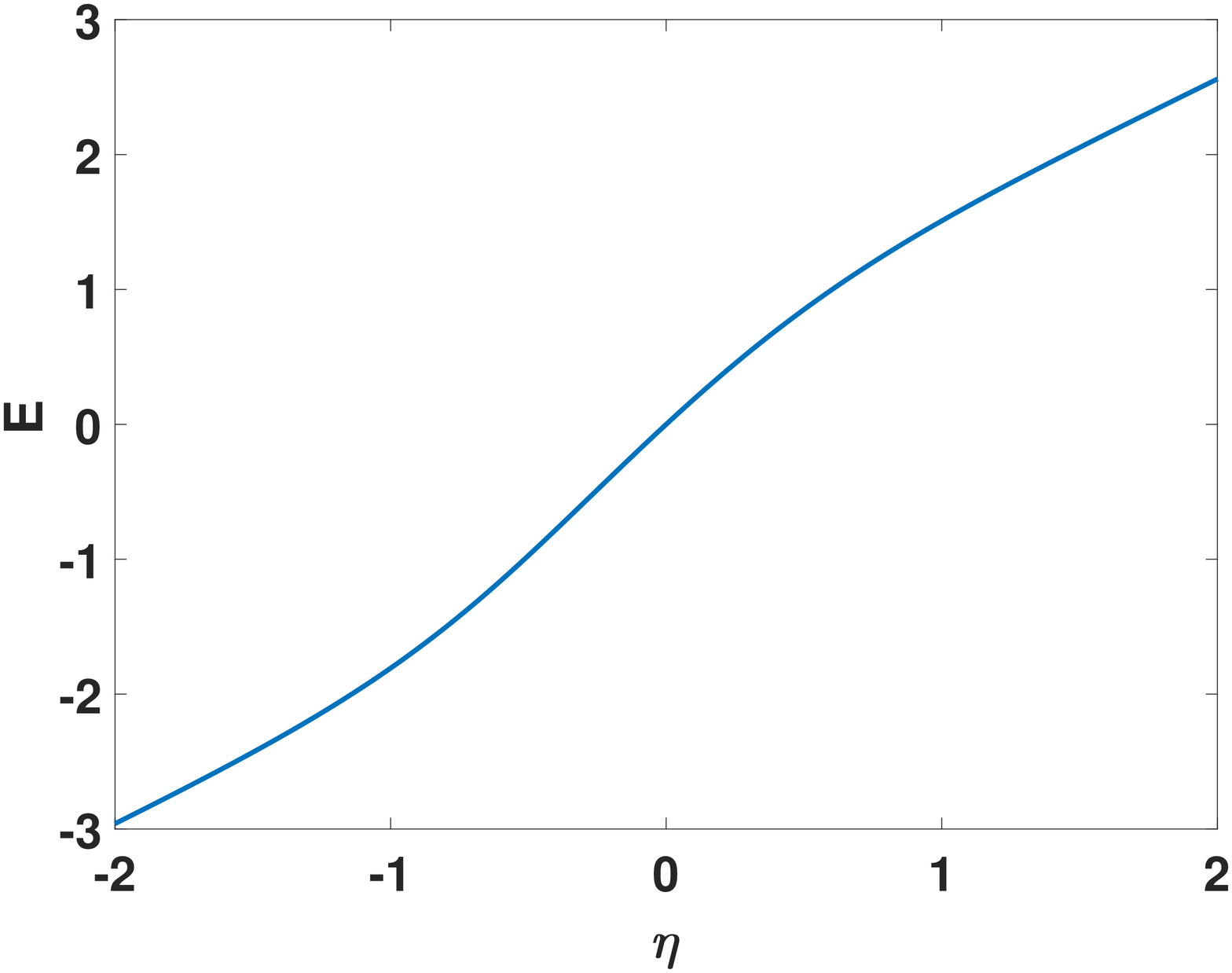}}
		\vspace{-0.1cm}
         \subfigure[$u/u_{\infty}$]{
			\includegraphics[height =5.3cm, width=0.48\linewidth]{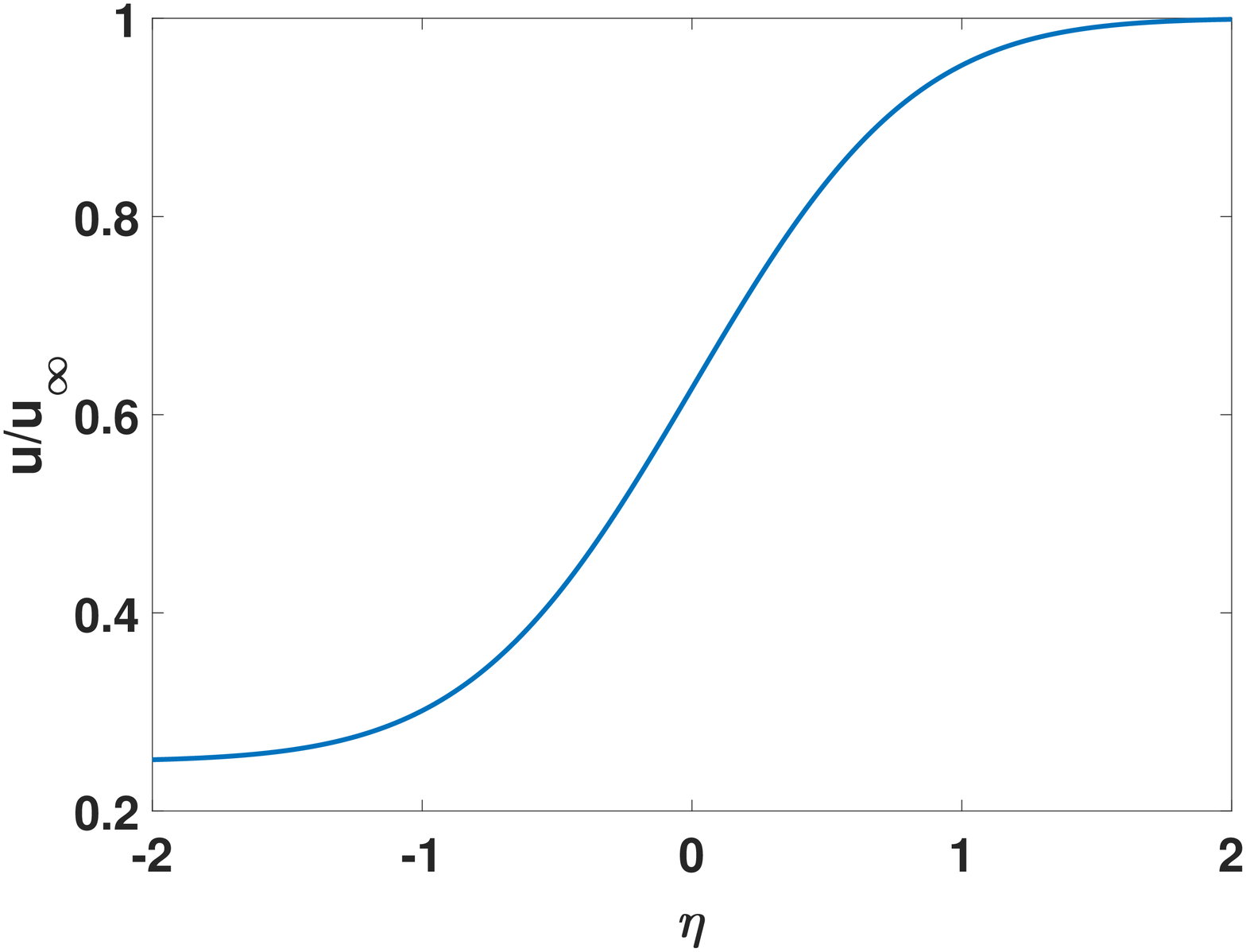}}
        \subfigure[$vg/\mu_{\infty}$]{
			\includegraphics[height =5.3cm, width=0.48\linewidth]{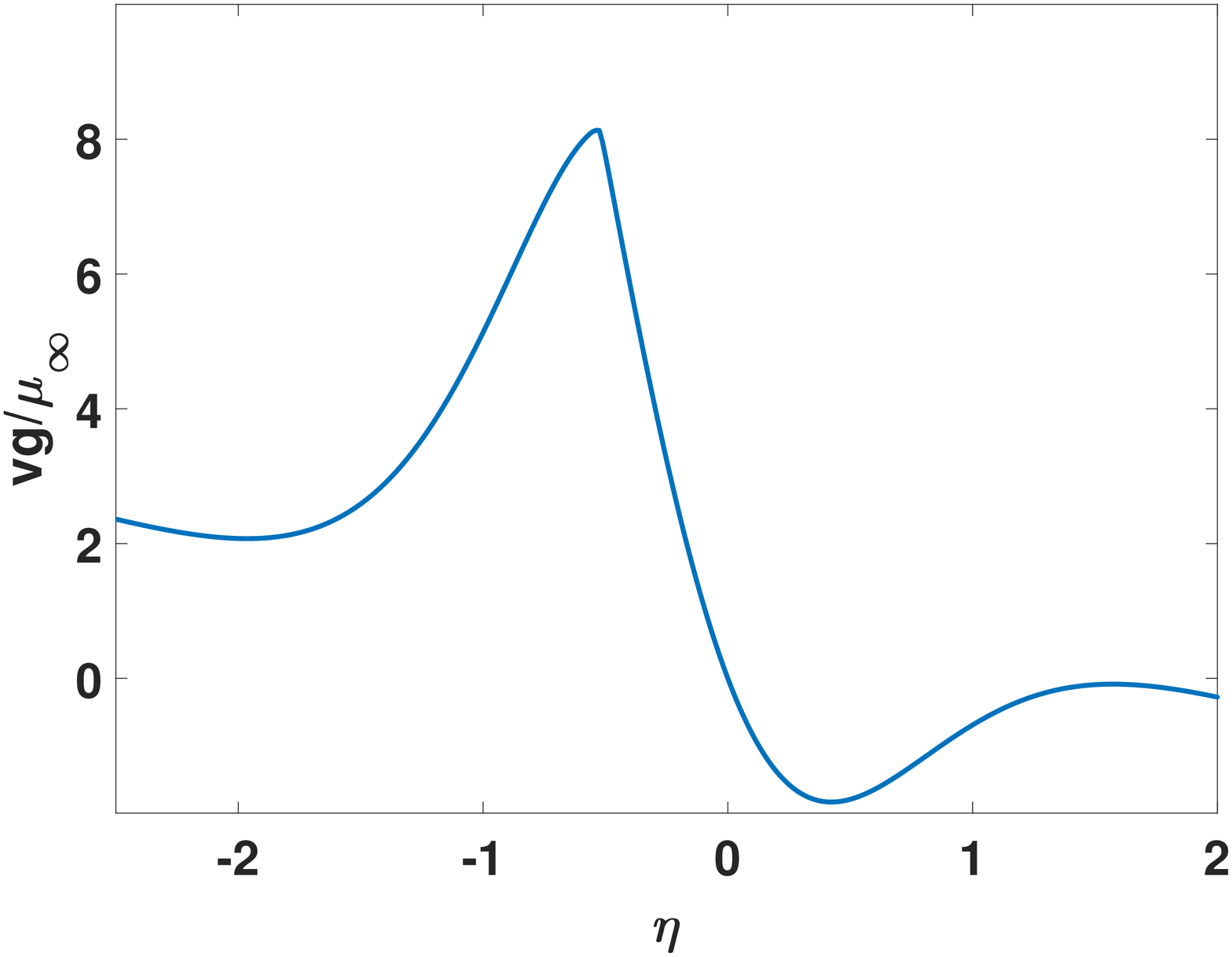}}
        \vspace{-0.1cm}
        \subfigure[$G$]{
			\includegraphics[height =5.0cm, width=0.48\linewidth]{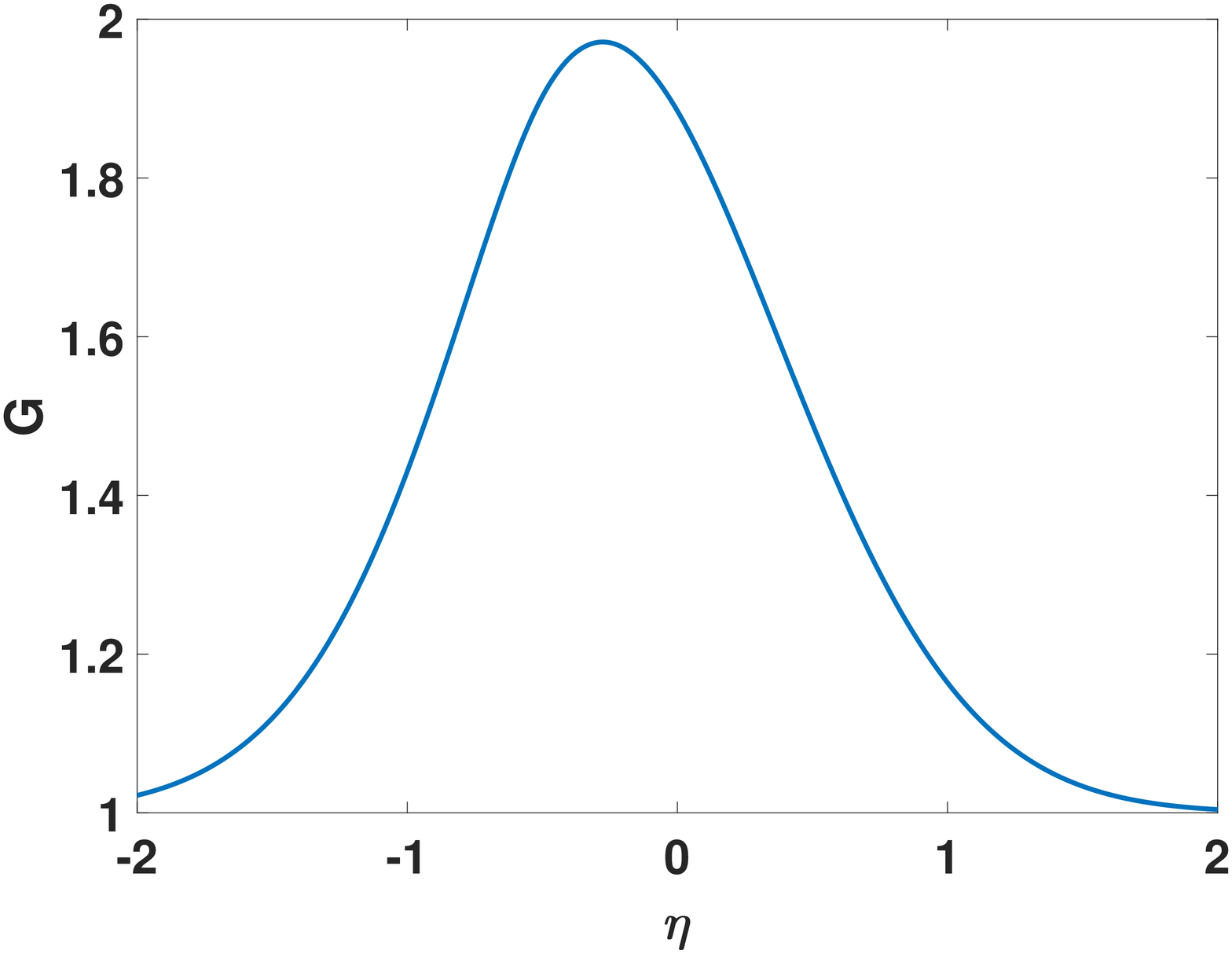}}
        \vspace{-0.1cm}
		\caption{Single diffusion flame solutions for dynamic field variables: $f, E, u/u_{\infty}, vg/\mu_{\infty},  G$.
			$ K =1.0 ;  Pr =1.0  ;  G_{\infty} = 1.0; u_{-\infty} / u_{\infty}= 0.25 $.}
		\label{DiffusionFlame2}
	\end{figure}

\begin{figure}[thbp]
		\centering
		\subfigure[$h/h_{\infty}$]{
			\includegraphics[height =5.5cm, width=0.48\linewidth]{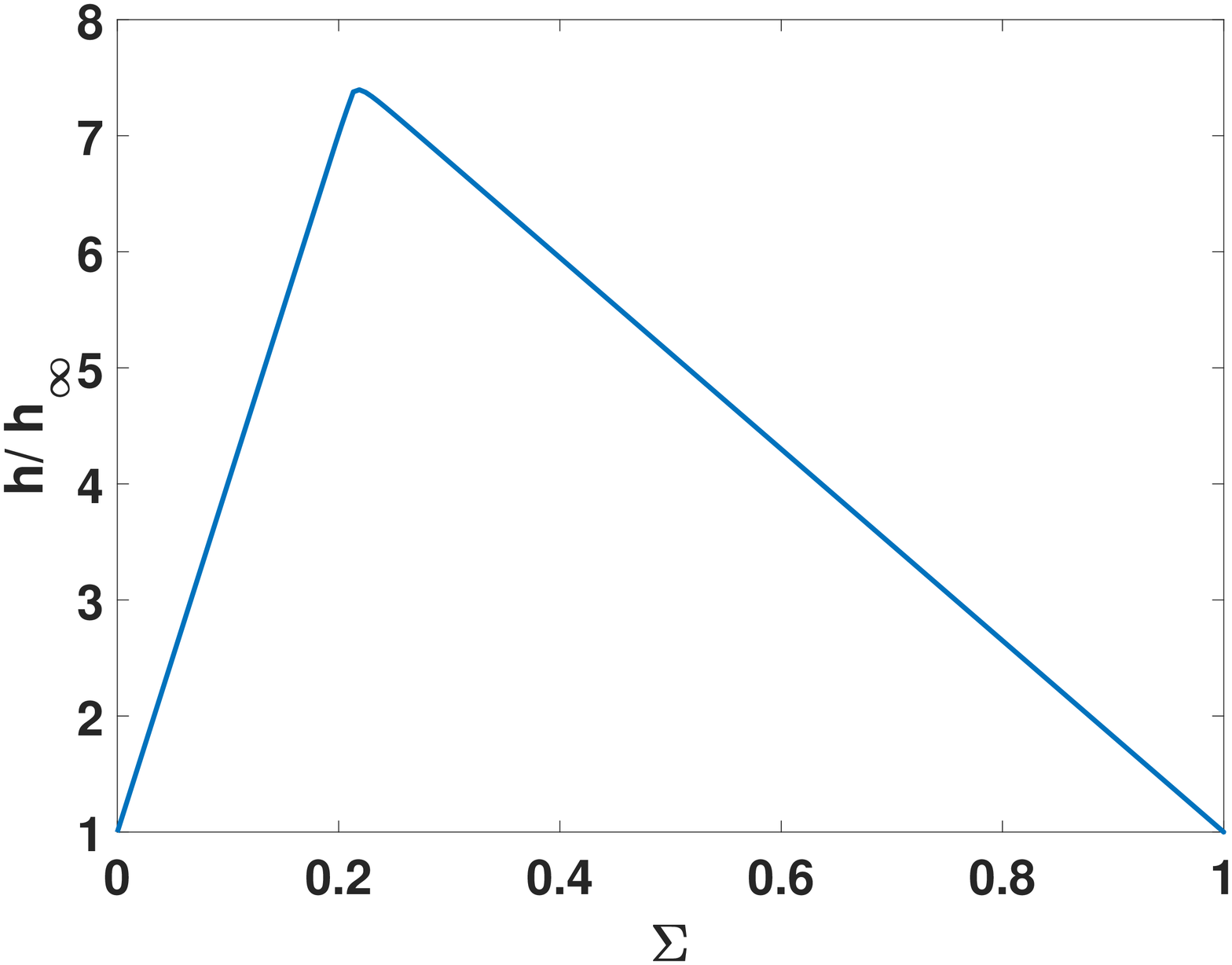}}
		\subfigure[$Y_F$]{
			\includegraphics[height =5.5cm, width=0.48\linewidth]{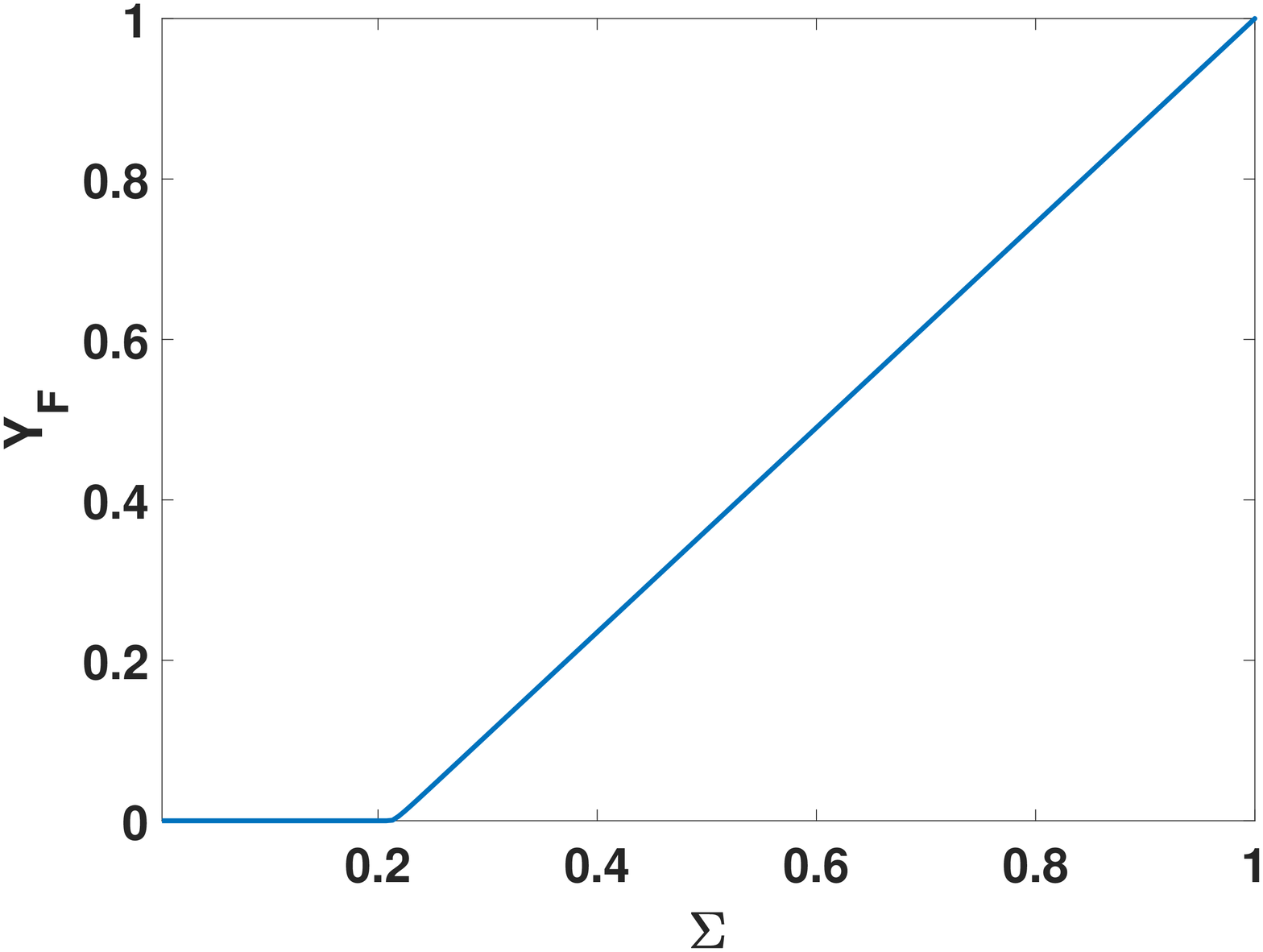}}
        \vspace{-0.1cm}
		\subfigure[$\nu Y_O$]{
			\includegraphics[height =5.5cm, width=0.48\linewidth]{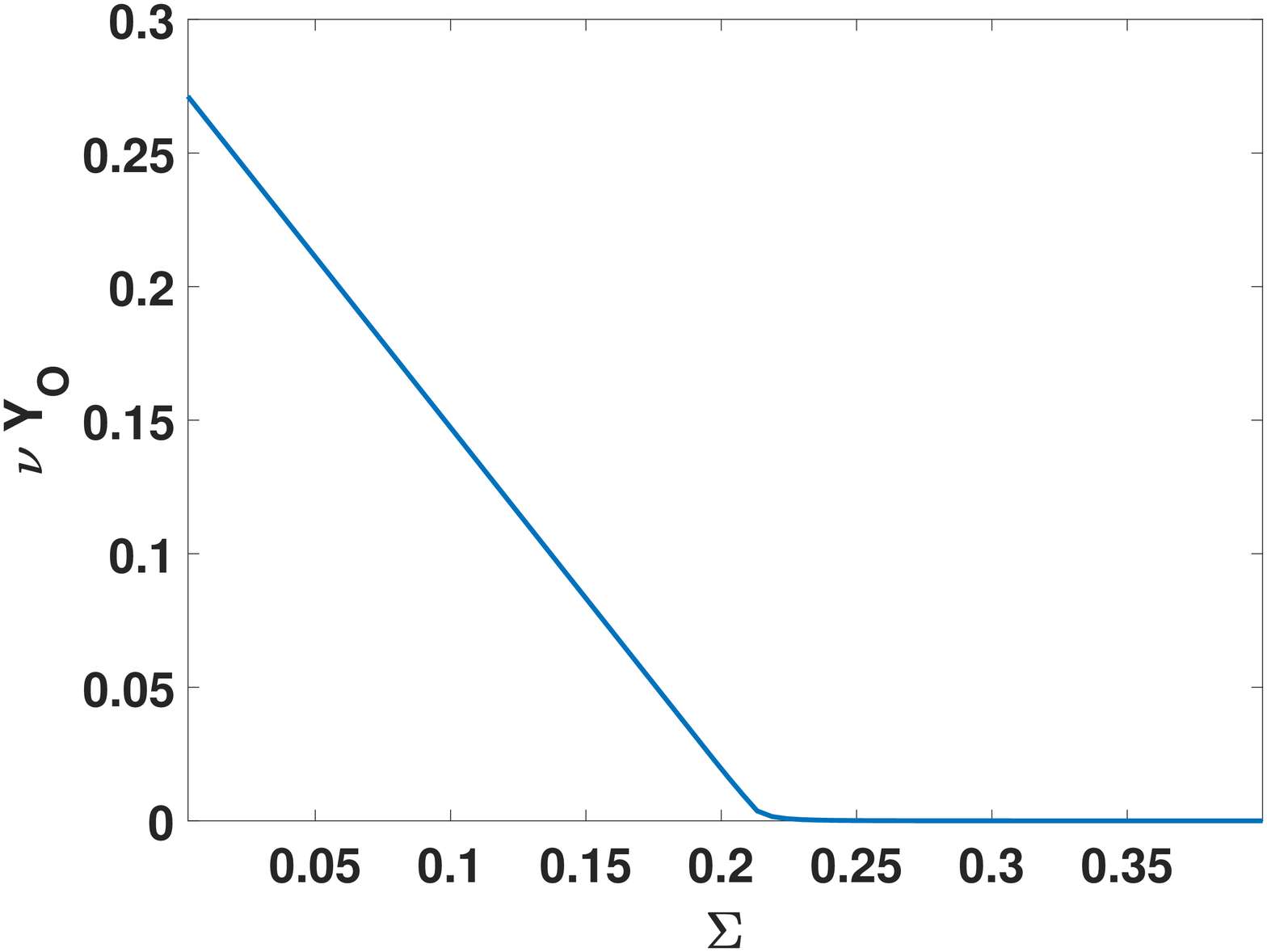}}
		\vspace{-0.1cm}
		\caption{Single diffusion flame solutions in $\Sigma$ space:  enthalpy $h/h_{\infty}$; and mass fractions $\nu Y_O, Y_F. $    $ K =1.0; Pr =1.0  ;  G_{\infty} = 1.0; u_{-\infty} / u_{\infty}= 0.25 $.}
		\label{DiffusionFlame3}
\end{figure}

Results for the basic parameter case with a single diffusion flame are shown in Figures \ref{DiffusionFlame}, \ref{DiffusionFlame2}, and \ref{DiffusionFlame3}. The higher velocity stream at $y = \infty$ is composed purely of  propane (i.e., $Y_{F,\infty} =1$) while the slower stream at $y = -\infty$ contains pure oxygen (i.e., $Y_{O,-\infty} =1$).
In the basic case, the parameters are $ K =1.0 , Pr =1.0  , G_{\infty} = 1.0,$ and
$ u_{-\infty} / u_{\infty}= 0.25 $.  The reference temperature is the $300 K$ value taken as boundary conditions at $\eta = \infty$ and $\eta = - \infty$. This temperature determines the ambient values for enthalpy, density, and dynamic viscosity.

The first figure shows from the thermochemical variables that the reaction zone stationed around $\eta = -0.5$ is quite narrow and orders of magnitude thinner than the mixing layer. So, clearly, diffusion is rate controlling here, supporting the use of a similar solution.

Figure \ref{DiffusionFlame2} shows that the velocity $u$ is monotonic through the shear layer, as expected in the absence of an imposed pressure gradient in flow direction. The Blasius function $f$ is also monotonic. However, the transverse velocities do not vary monotonically  with $\eta$ as shown in the sub-figures for $vg/\mu_{\infty}$ and $G$; the variation of density due to heat release cause overshoots of this velocity component similar to findings for the counterflow \citep{Sirignano2019b, Sirignano2019a}. In particular, the  strain rates near the reaction zone are substantially augmented in magnitude compared to those imposed by the same free-stream flows in a non-reacting case.  The integral $E$ is monotonic by nature. Its magnitude  compares with $f$ in its effect on the solution. In fact, in this case, the imposed normal strain represented through $E$ has  somewhat more influence than the shear strain represented through $f$.

The behavior of enthalpy and mass fractions of the reactants are shown in Figure \ref{DiffusionFlame3} to be linear in $\Sigma$ space except within the narrow reaction zones. For our perfect gas, temperature will have the qualitatively identical behavior.
Thus, in analogy to the counterflow case, the $\Sigma$ space provides interesting information but it can only be calculated with coupling to the system of ODEs.
Here, $\Sigma$ could be a conserved scalar or could be formed from the velocity field via $J(\eta)$.

\begin{figure}[thbp]
		\centering
		\subfigure[$h/h_{\infty}$]{
			\includegraphics[height =5.5cm, width=0.48\linewidth]{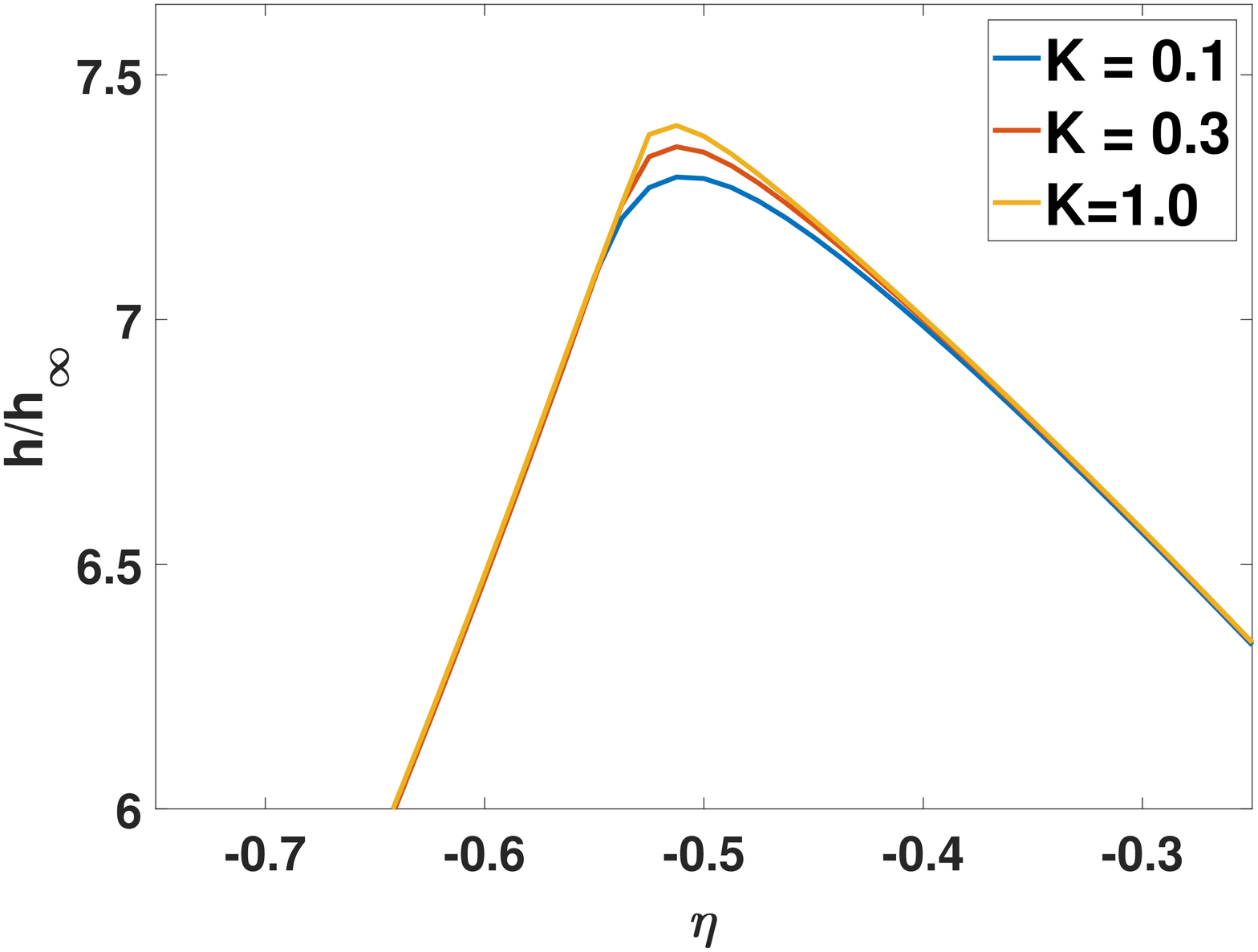}}
		\subfigure[$Y_F$]{
			\includegraphics[height =5.5cm, width=0.48\linewidth]{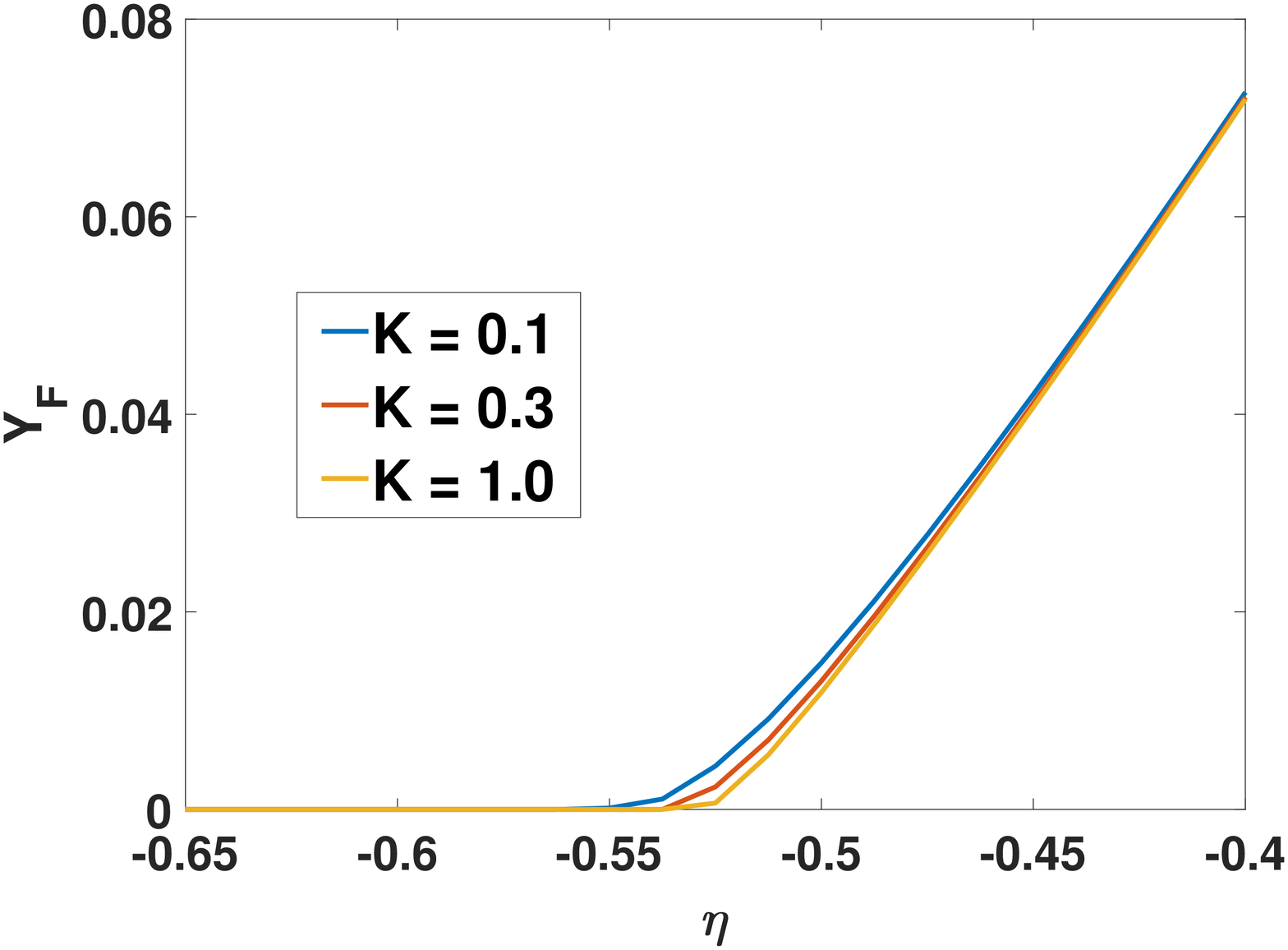}}
        \vspace{-0.1cm}
		\subfigure[$\nu Y_O$]{
			\includegraphics[height =5.5cm, width=0.48\linewidth]{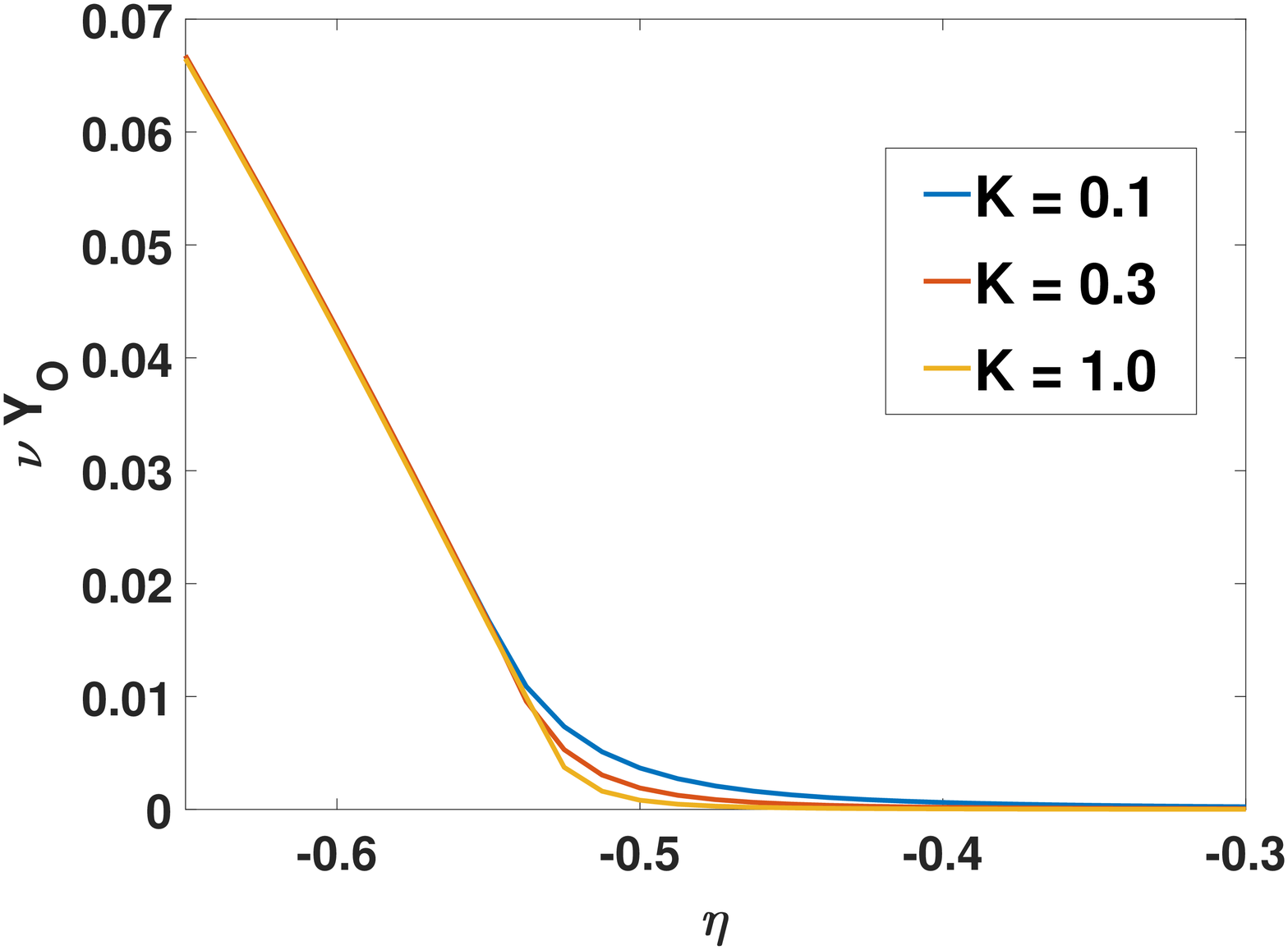}}
		\vspace{-0.1cm}
		\caption{Effect of Damk\"{o}hler number on Single diffusion flame solutions:  enthalpy $h/h_{\infty}$; and mass fractions $\nu Y_O, Y_F. $    $ K =0.1, 0.3, 1.0 ;  Pr =1.0  ;  G_{\infty} = 1.0; u_{-\infty} / u_{\infty}= 0.25 $.}
		\label{DiffusionFlame4}
\end{figure}

In order to assure that the similarity approximation is reasonable here, we reduce the $Da$ severalfold up to an order of magnitude. In Figure \ref{DiffusionFlame4}, the results for the $h/h_{\infty}, Y_f, $ and $\nu Y_O$ are displayed in a fashion that zooms with an expansion on values around the reaction zone. The reaction zone increases somewhat in size and the peak value of enthalpy decreases with decreasing $Da$ but the zone still remains very narrow compared to mixing-layer dimensions. Furthermore, the asymptotic behaviors at the edge of the reaction zone remain independent of $Da$, indicating that the integral of the reaction rate is unchanged with changing $Da$. Thereby, we know that the similar solution is quite satisfactory for this diffusion-flame configuration.  With $Da$ varying by an order of magnitude, the fields for $u/u_{\infty},  G, f,$ and $E$ show no differences. The plot of scalars in $\Sigma$-space shows no significant difference either.

\begin{figure}[thbp]
		\centering
		\subfigure[$h/h_{\infty}$]{
			\includegraphics[height =5.5cm, width=0.48\linewidth]{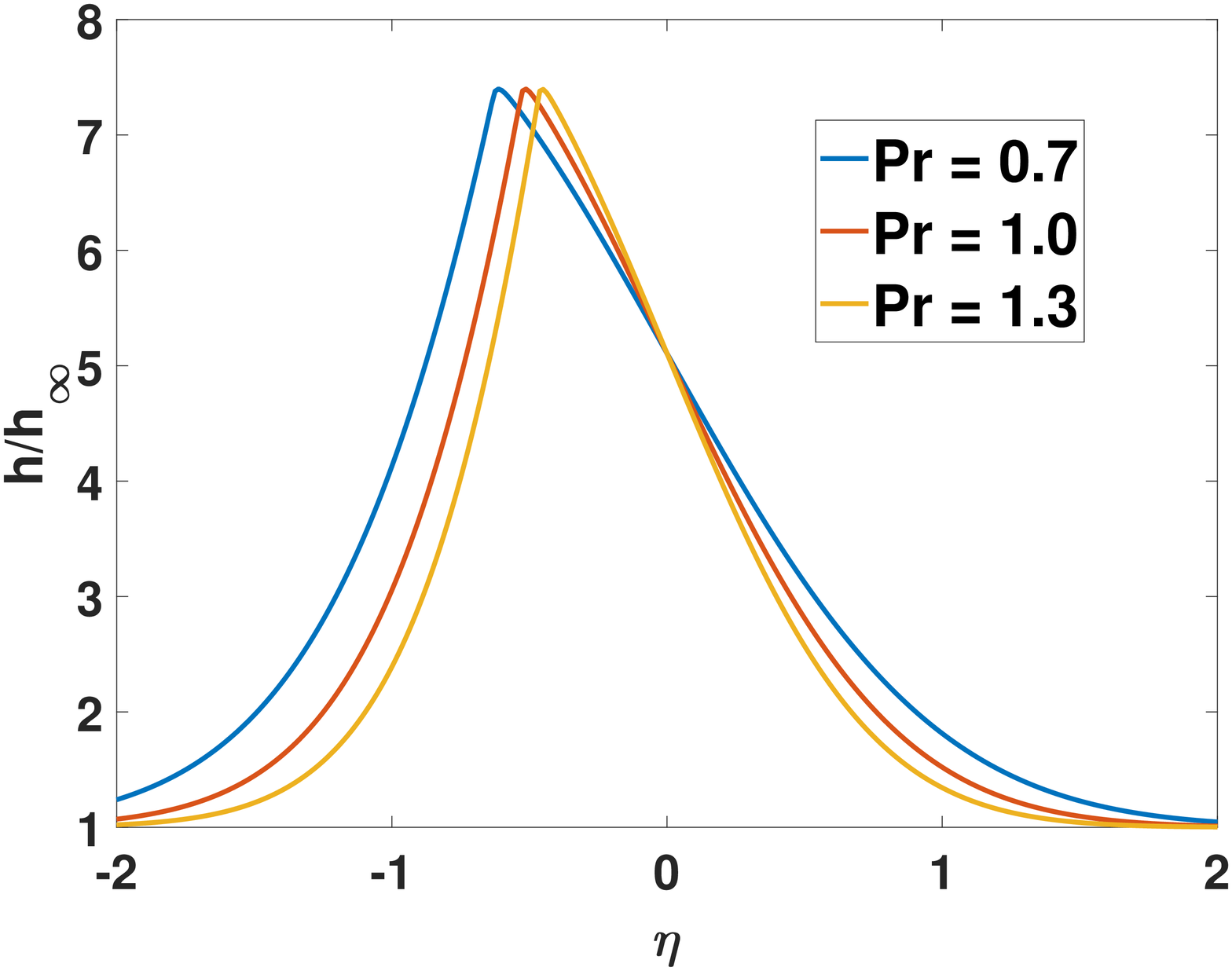}}
		\subfigure[$\omega_F x / u_{\infty}$]{
			\includegraphics[height =5.5cm, width=0.48\linewidth]{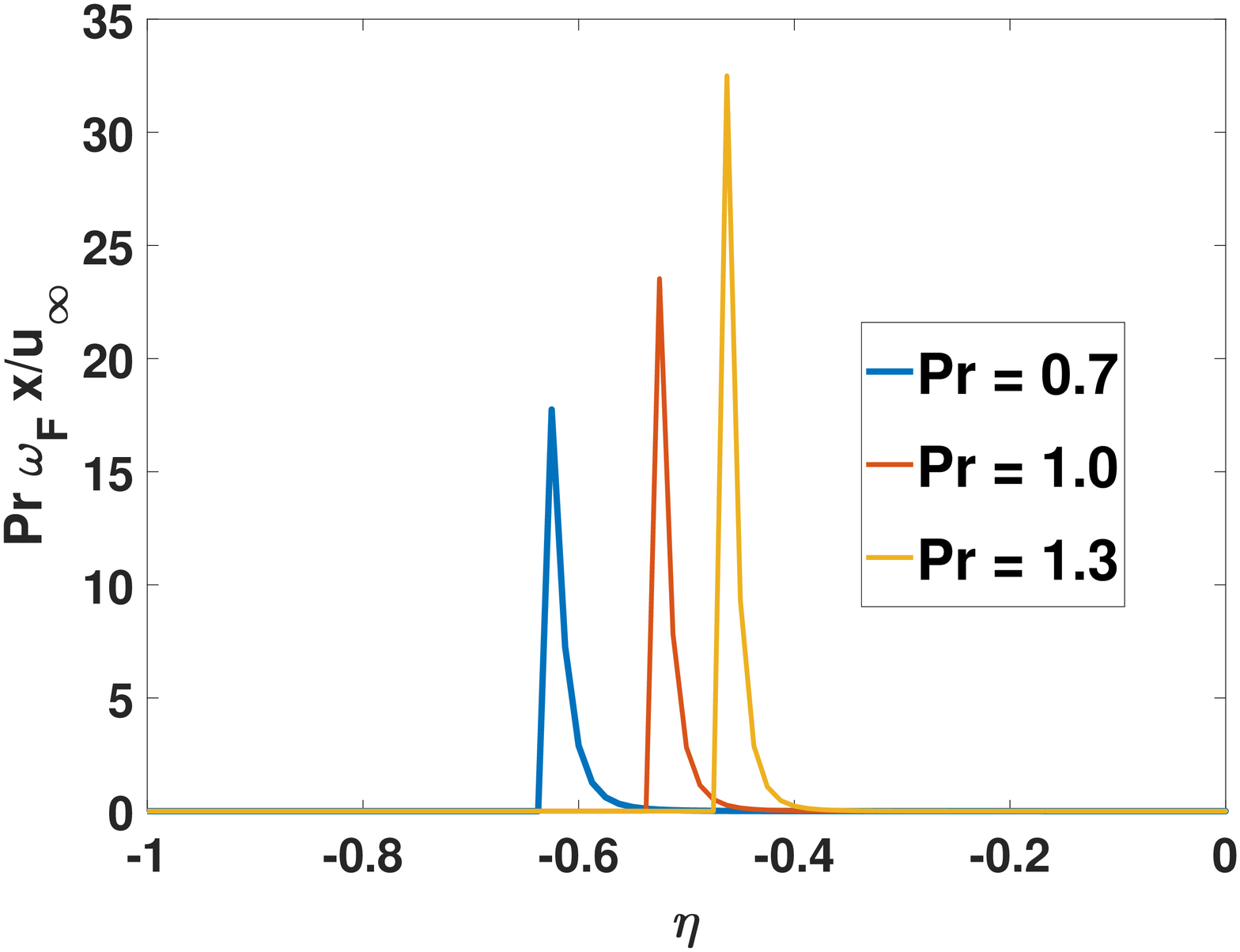}}
		\vspace{-0.1cm}
		\subfigure[$Y_F$]{
			\includegraphics[height =5.5cm, width=0.48\linewidth]{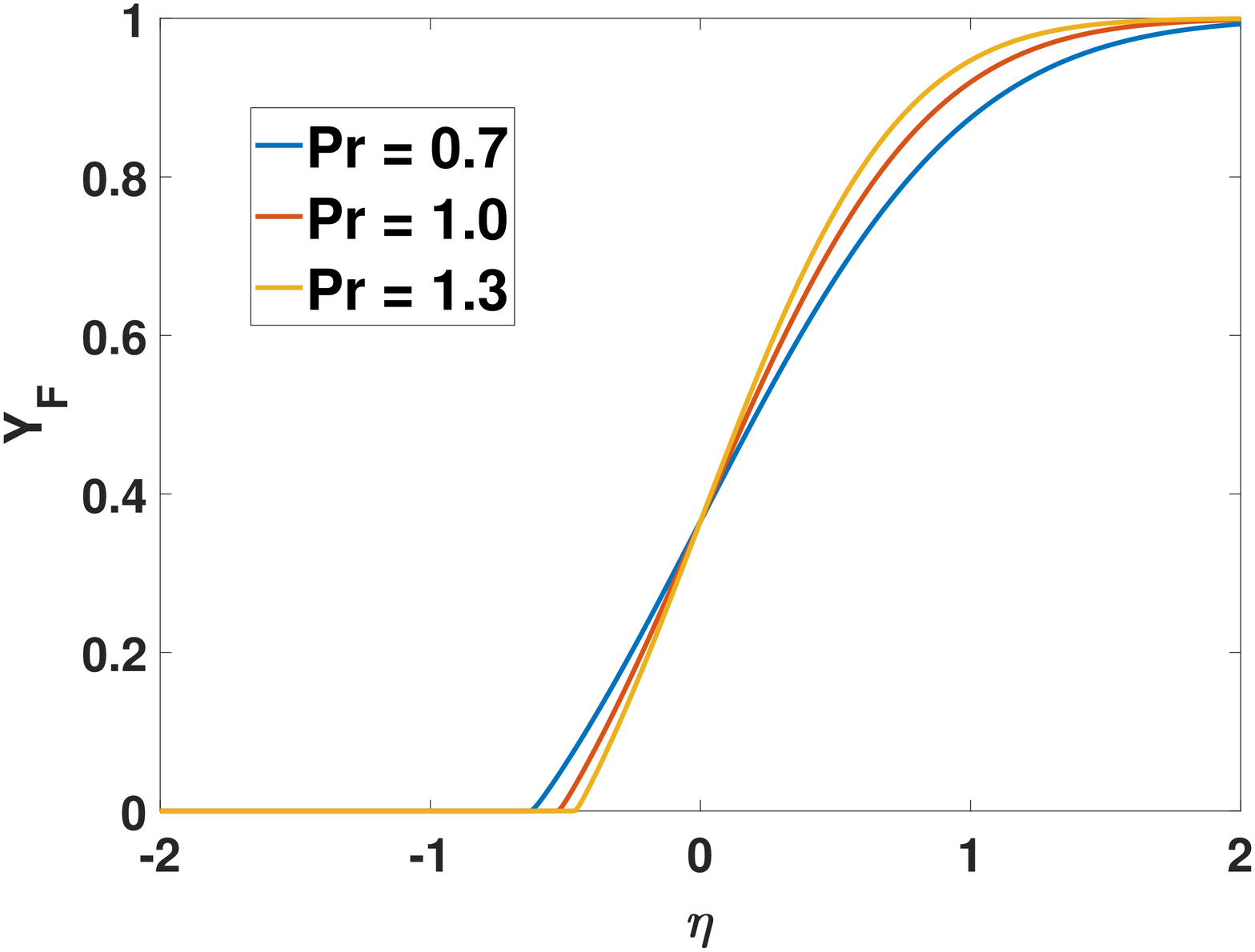}}
		\subfigure[$\nu Y_O$]{
			\includegraphics[height =5.5cm, width=0.48\linewidth]{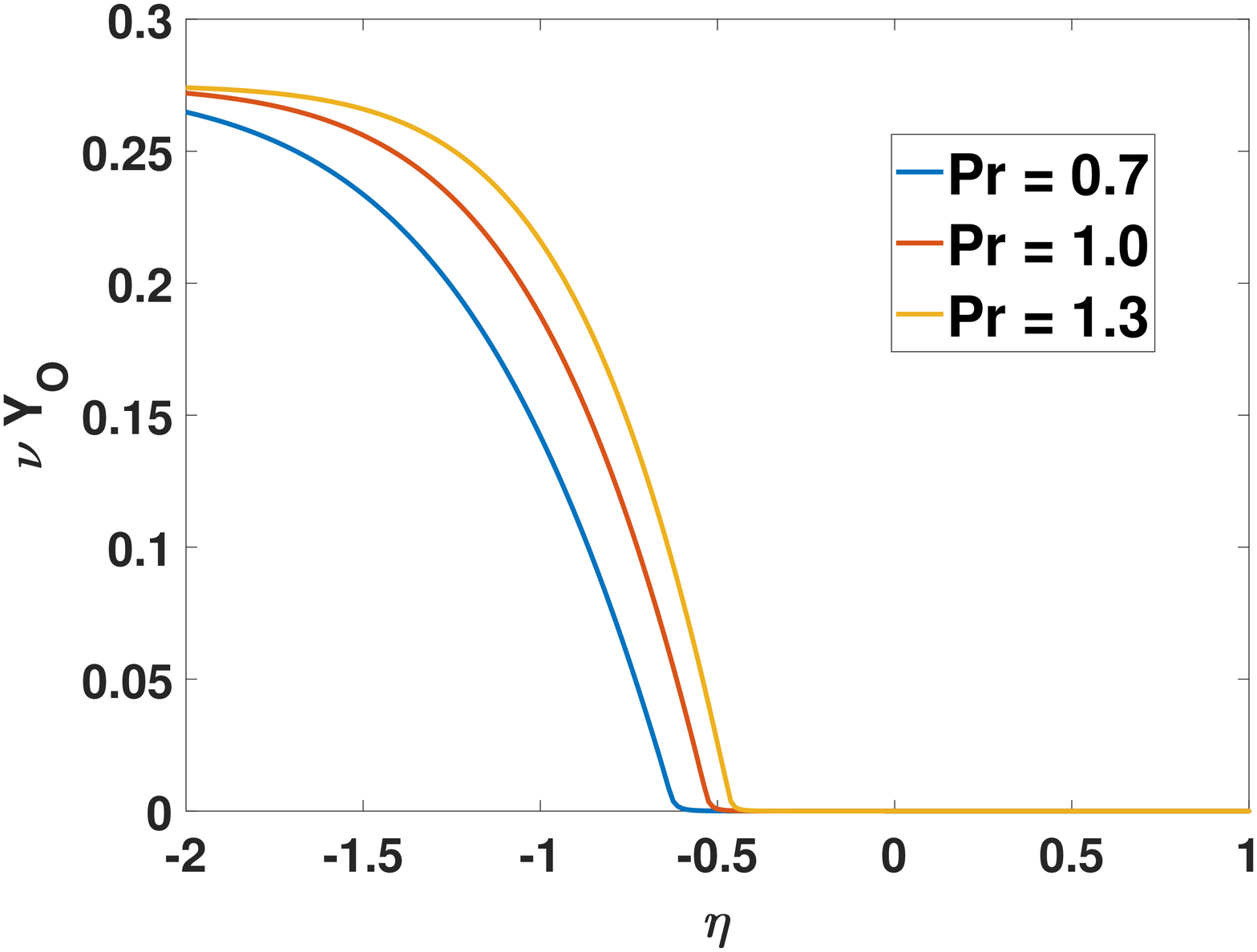}}
		\vspace{-0.1cm}
		\caption{Effect of Prandtl number on single diffusion flame solutions:  enthalpy $h/h_{\infty}$ ; $\omega_F x/u_{\infty}$; mass fractions $\nu Y_O, Y_F$.   $ K =1.0 ; Pr= 0.7, 1.0, 1.3; G_{\infty} = 1.0; u_{-\infty} / u_{\infty}= 0.25 $.}
		\label{DiffusionFlame5}
	\end{figure}

\begin{figure}[thbp]
		\centering
		\subfigure[$E$]{
			\includegraphics[height =5.3cm, width=0.48\linewidth]{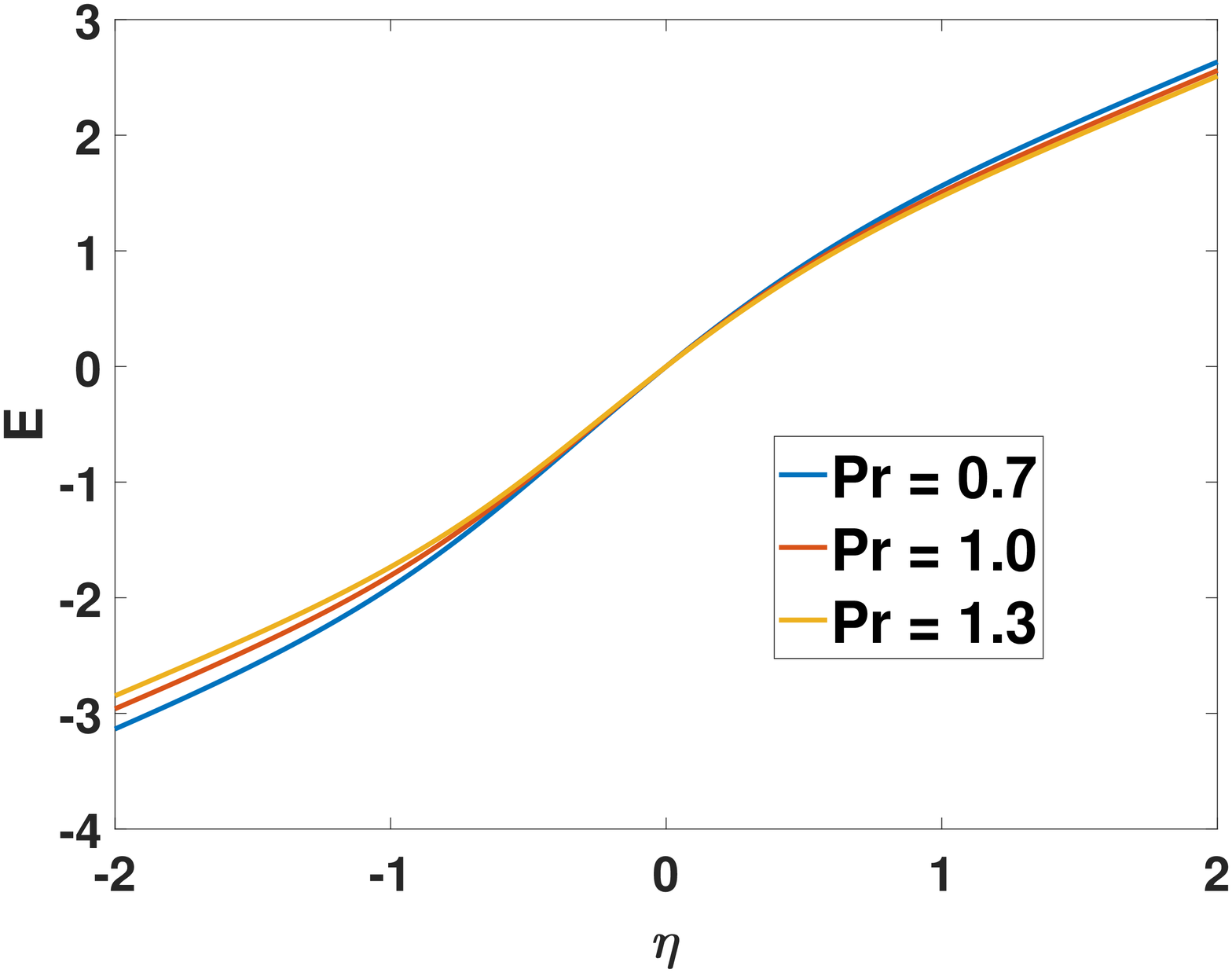}}
		\subfigure[$G$]{
			\includegraphics[height =5.3cm, width=0.48\linewidth]{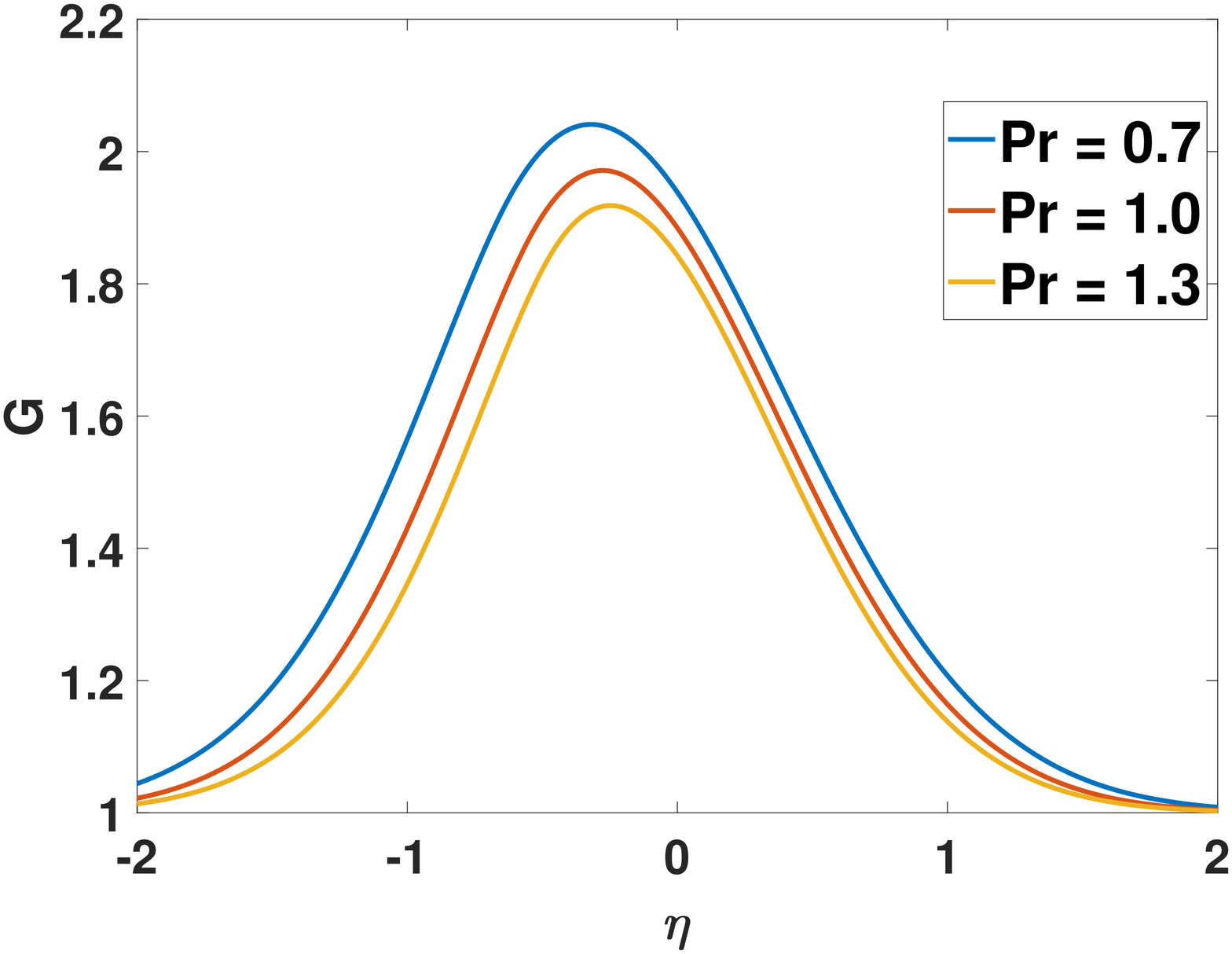}}
		\vspace{-0.1cm}
		\caption{Effect of Prandtl number on single diffusion flame solutions for dynamic field variables:  $E,  G$.    $ K =1.0 ; Pr= 0.7, 1.0, 1.3; G_{\infty} = 1.0; u_{-\infty} / u_{\infty}= 0.25 $.}
		\label{DiffusionFlame6}
	\end{figure}

The Prandtl number $Pr$ is a parameter of importance. In practical situations, its magnitude will not vary that widely; however, it has consequences and its understanding is valuable. Increases in $Pr$ make scalar gradients increase as shown in Figure \ref{DiffusionFlame5}. If $Pr > 1(<1) $, the normalized scalar gradients become greater than (smaller than) the normalized velocity gradients. Accordingly, as $Pr$ increases from 0.7 to 1.3, the  peak value of the source in the energy equation (\ref{h}) becomes  larger as shown in Sub-figure \ref{DiffusionFlame5}b since it contains $Pr$ multiplied by the reaction rate. The steeper gradients need not imply increasing transport rates since the diffusivities for mass and energy decrease as $Pr$ increases.  There is a shift of the reaction zone towards the fuel side with increased $Pr$. The $u$-component of velocity has insignificant change due to variation of $Pr$ but more change is seen in the transverse components indicated by the behavior of $G$  indicated in  Figure \ref{DiffusionFlame6}.

\begin{figure}[thbp]
		\centering
		\subfigure[$h(\eta)/h_{\infty}$]{
			\includegraphics[height =5.5cm, width=0.48\linewidth]{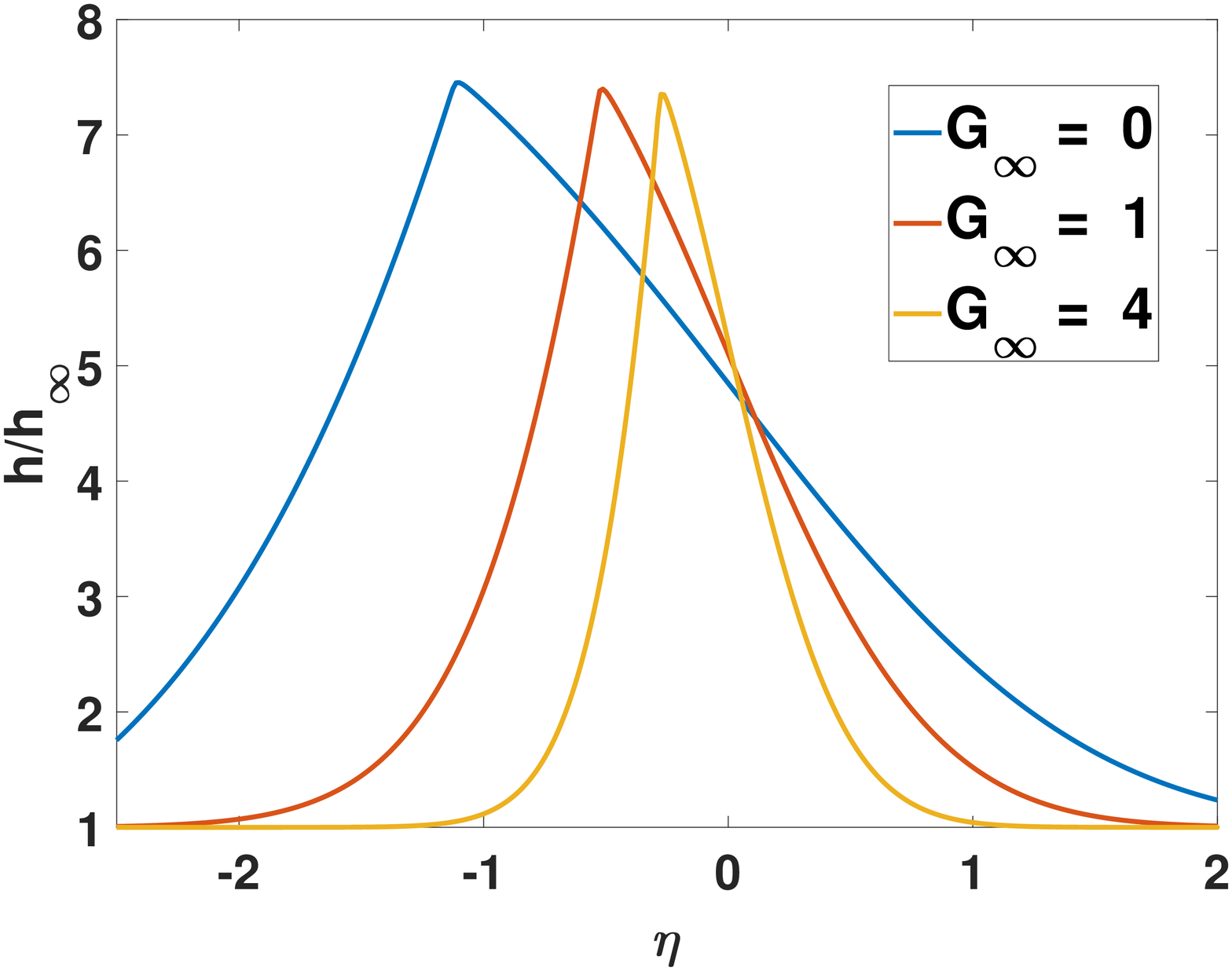}}
		\subfigure[$h(\Sigma)/h_{\infty}$ ]{
			\includegraphics[height =5.5cm, width=0.48\linewidth]{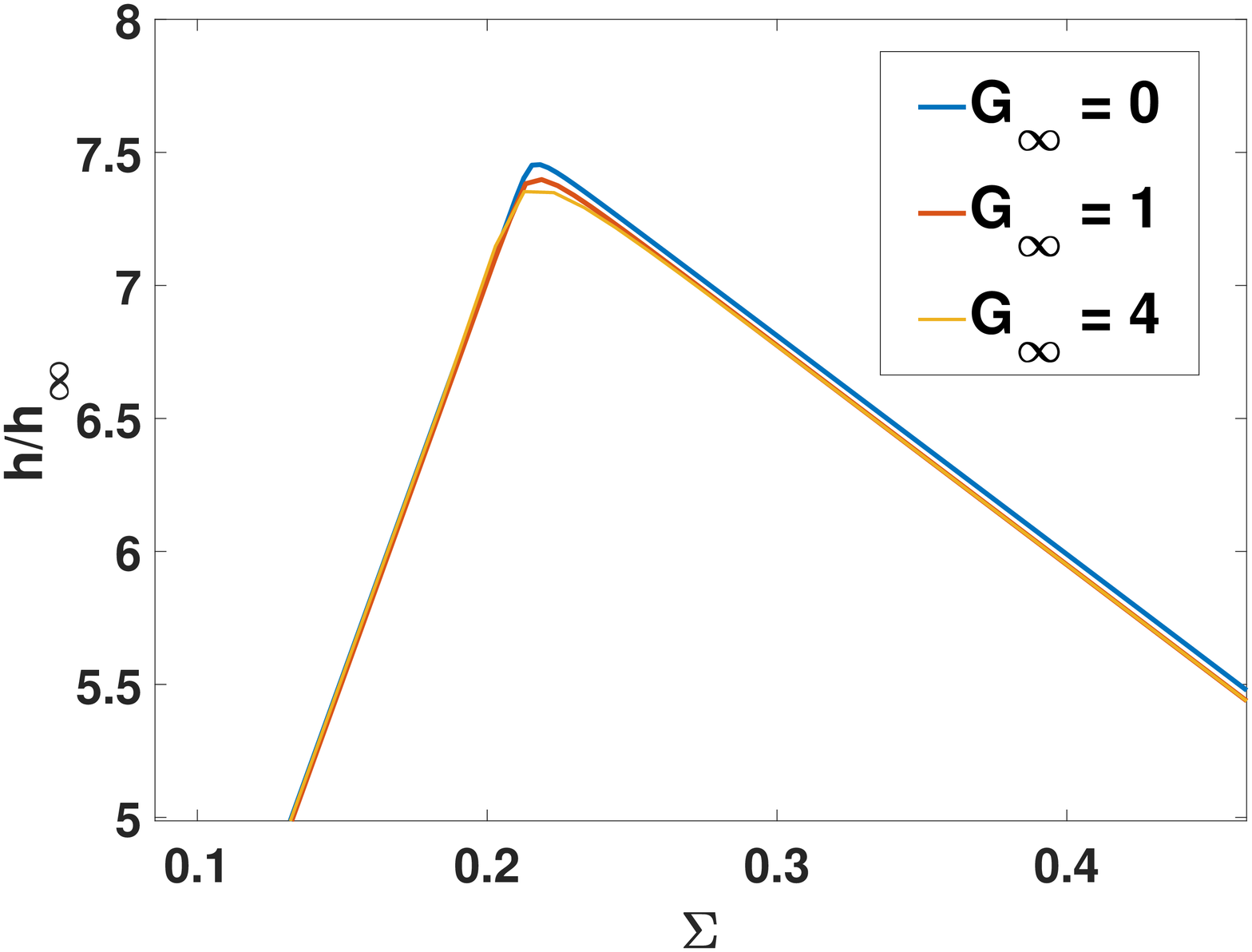}}
		\vspace{-0.1cm}
		\subfigure[$Y_F$]{
			\includegraphics[height =5.5cm, width=0.48\linewidth]{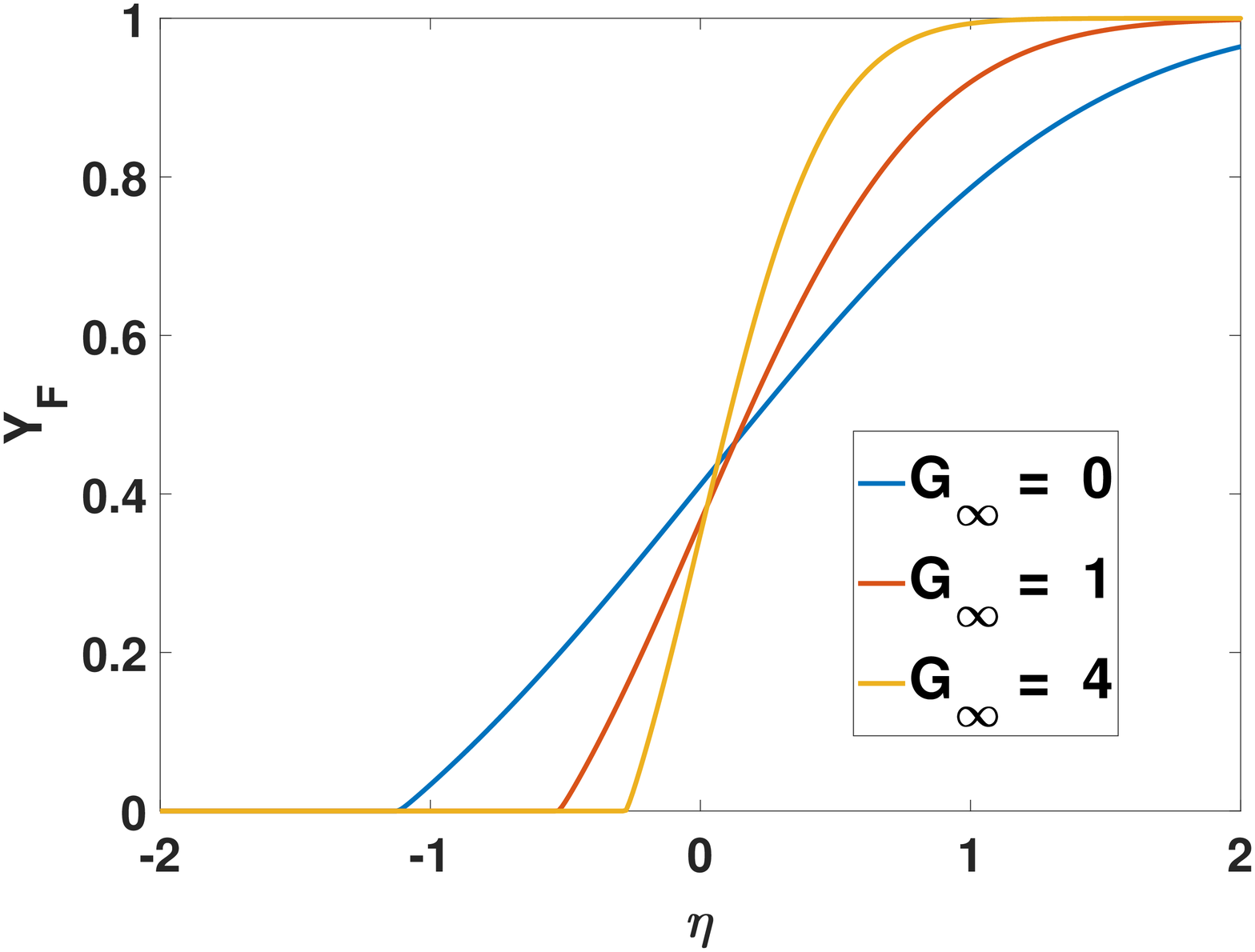}}
		\subfigure[$\nu Y_O$]{
			\includegraphics[height =5.5cm, width=0.48\linewidth]{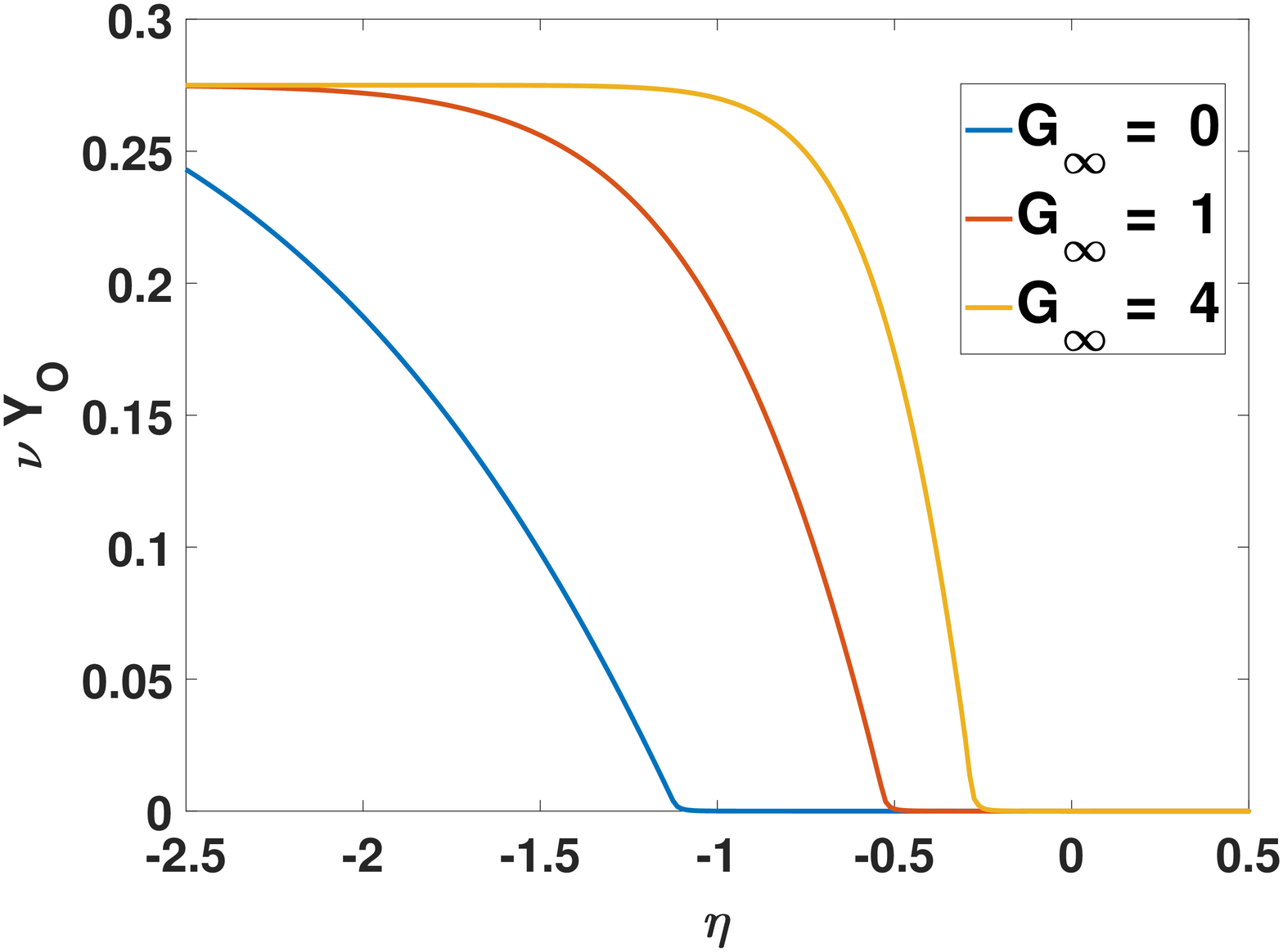}}
		\vspace{-0.1cm}
		\caption{Effect of normal strain rate on single diffusion flame solutions:  enthalpy $h/h_{\infty}$ in both $\eta$ and $\Sigma$ spaces; mass fractions $\nu Y_O, Y_F$.
       $ K =1.0 ;   Pr= 1.0;   G_{\infty} = 0, 1.0, 4.0; u_{-\infty} / u_{\infty}= 0.25 $.}
		\label{DiffusionFlame7}
	\end{figure}

\begin{figure}[thbp]
		\centering
		\subfigure[$f$]{
			\includegraphics[height =5.4cm, width=0.48\linewidth]{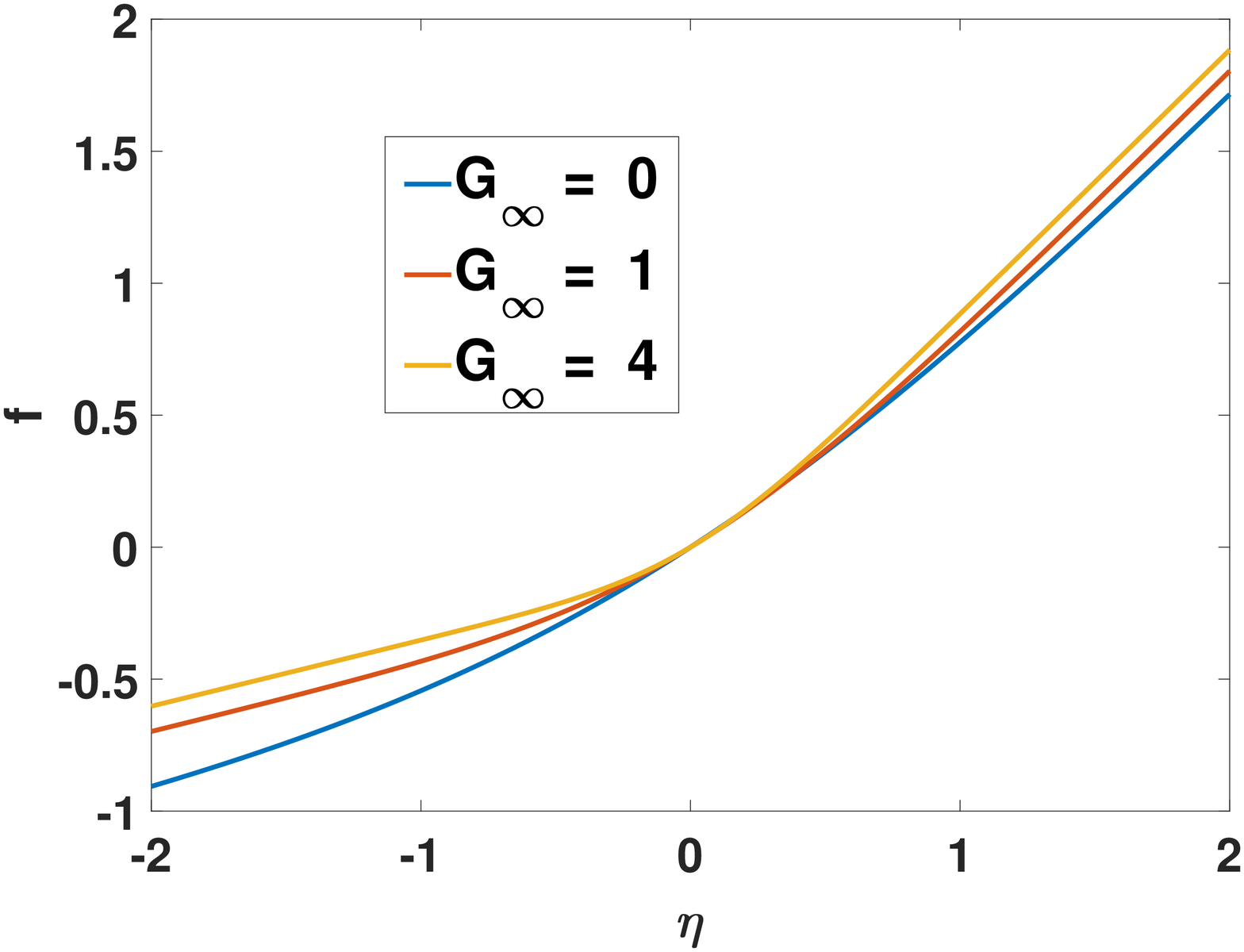}}
		\subfigure[$E$]{
			\includegraphics[height =5.4cm, width=0.48\linewidth]{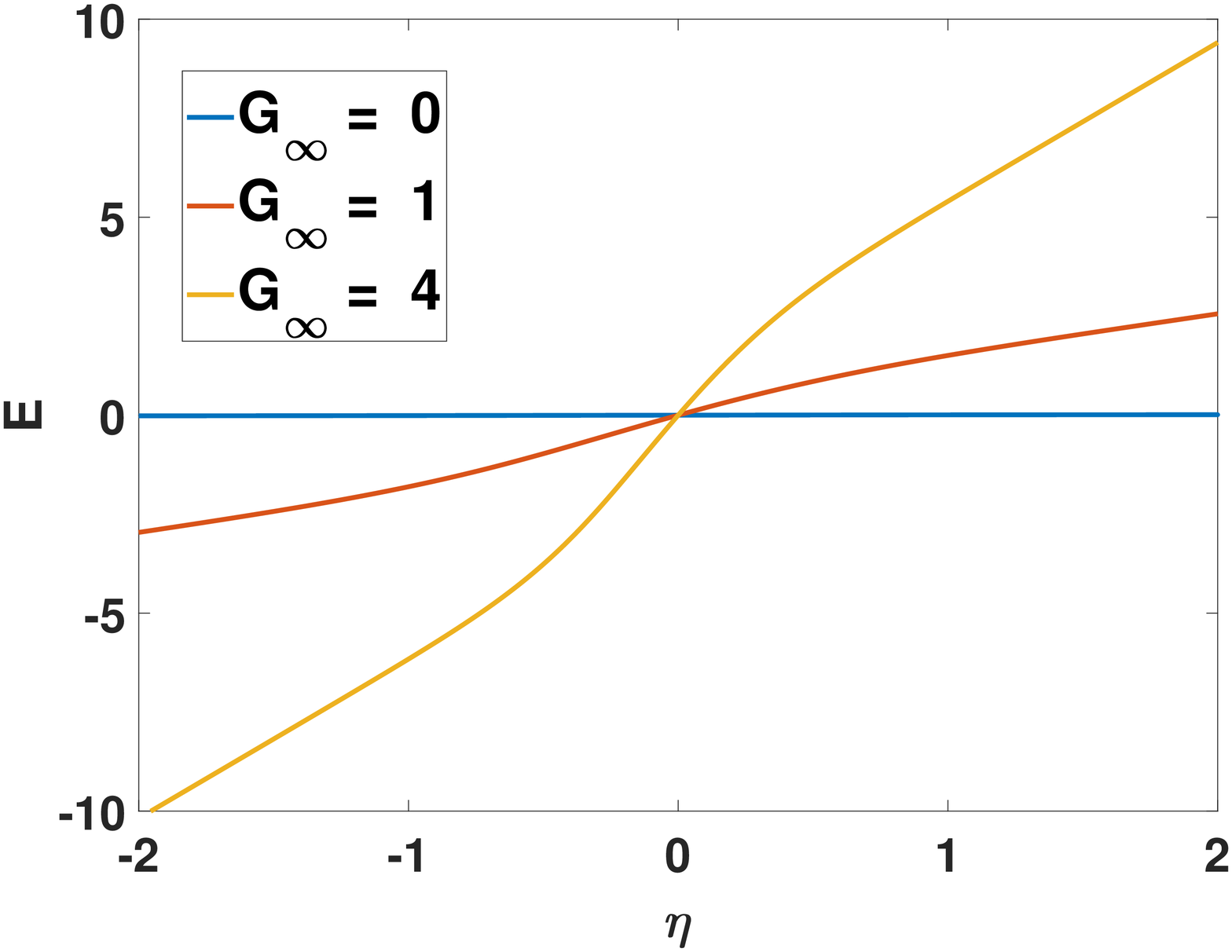}}
		\vspace{-0.1cm}
         \subfigure[$u/u_{\infty}$]{
			\includegraphics[height =5.4cm, width=0.48\linewidth]{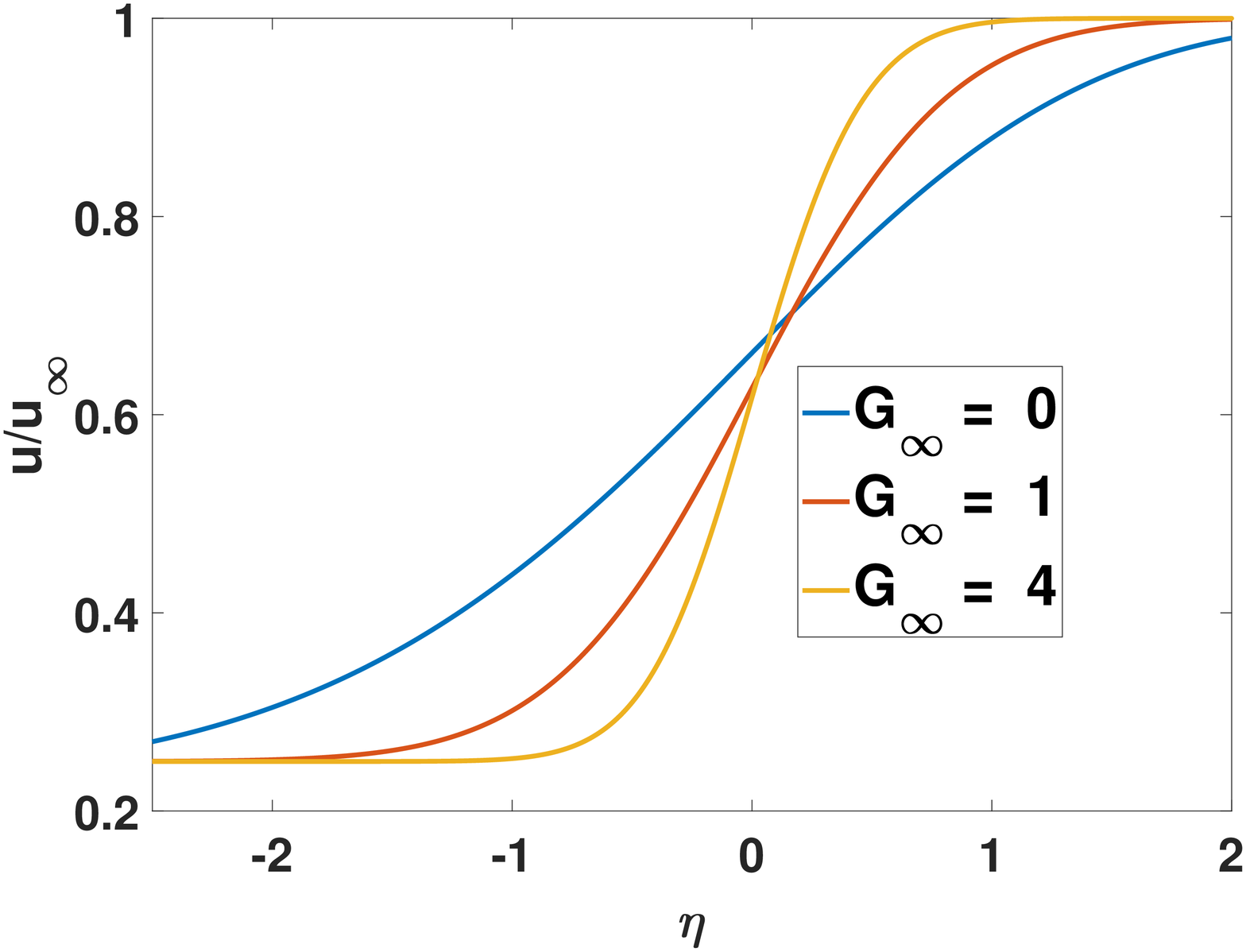}}
        \subfigure[$vg/\mu_{\infty}$]{
			\includegraphics[height =5.4cm, width=0.48\linewidth]{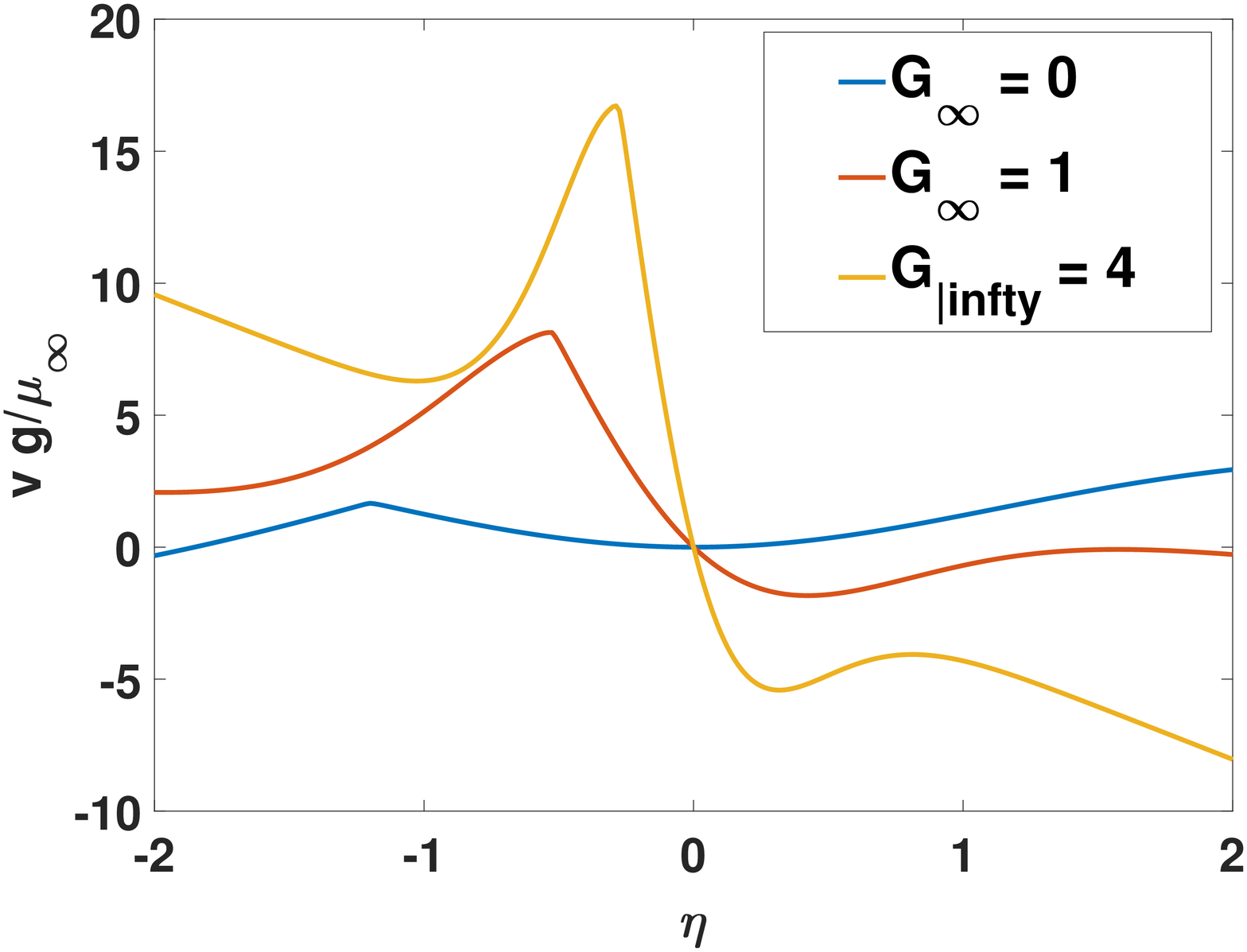}}
        \vspace{-0.1cm}
        \subfigure[$G$]{
			\includegraphics[height =5.0cm, width=0.48\linewidth]{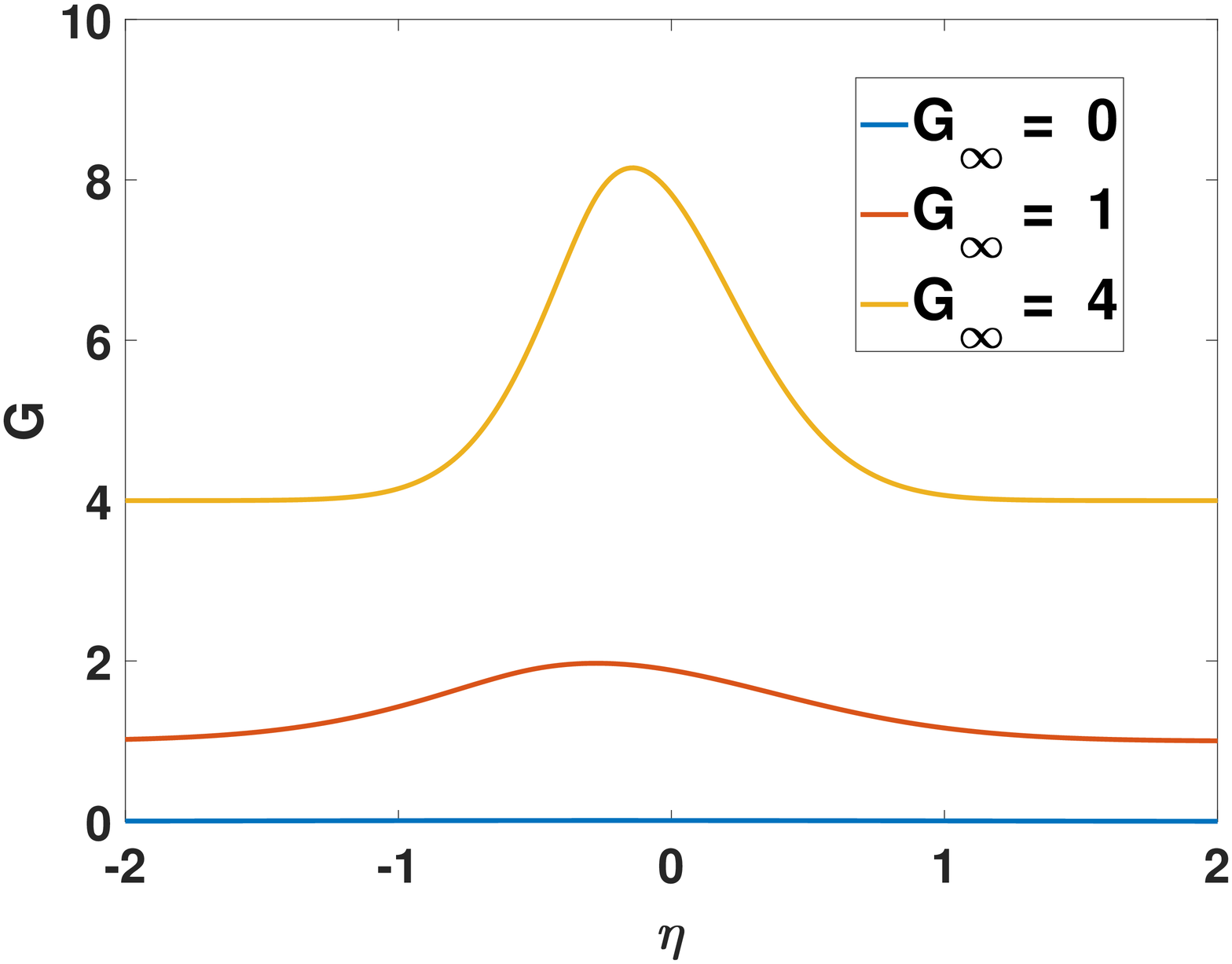}}
          \vspace{-0.1cm}
		\caption{Effect of normal strain rate on single diffusion flame solutions  for  dynamic field  variables $f, E, u/u_{\infty}, G, vg/\mu_{\infty}$. $ K =1.0 ; Pr= 1.0; G_{\infty} = 0, 1.0, 4.0;  u_{-\infty} / u_{\infty}= 0.25  $.}
		\label{DiffusionFlame8}
	\end{figure}

\newpage
The transverse normal strain rate imposed through the counterflow has major effects as shown in Figures \ref{DiffusionFlame7} and \ref{DiffusionFlame8}. Scalar variables and their transport rates are affected in major ways. With increasing normal strain, i.e., increasing value of $G_{\infty}$, the mixing layer gets thinner, increasing transport rates. Heat production rate and fuel consumption rate in the diffusion flame accommodate to the increased transport rate. Note that flame location moves towards the transverse-velocity stagnation plane at $\eta = 0$ but the velocity at the peak reaction rate actually increases with increasing strain rate. Without imposed normal strain, $G=0$ and $E=0$ throughout the flow. Thus, $w=0$ everywhere in that case.  However, the transverse velocity component $v$ has non-zero values due to converging and diverging streamlines and gas expansion caused by heat release. As before, we arbitrarily set $v(x,0) =0$ and bypass the opportunity to correct it based on transverse momentum balance.

\begin{figure}[thbp]
		\centering
		\subfigure[$h/h_{\infty}$]{
			\includegraphics[height =5.3cm, width=0.48\linewidth]{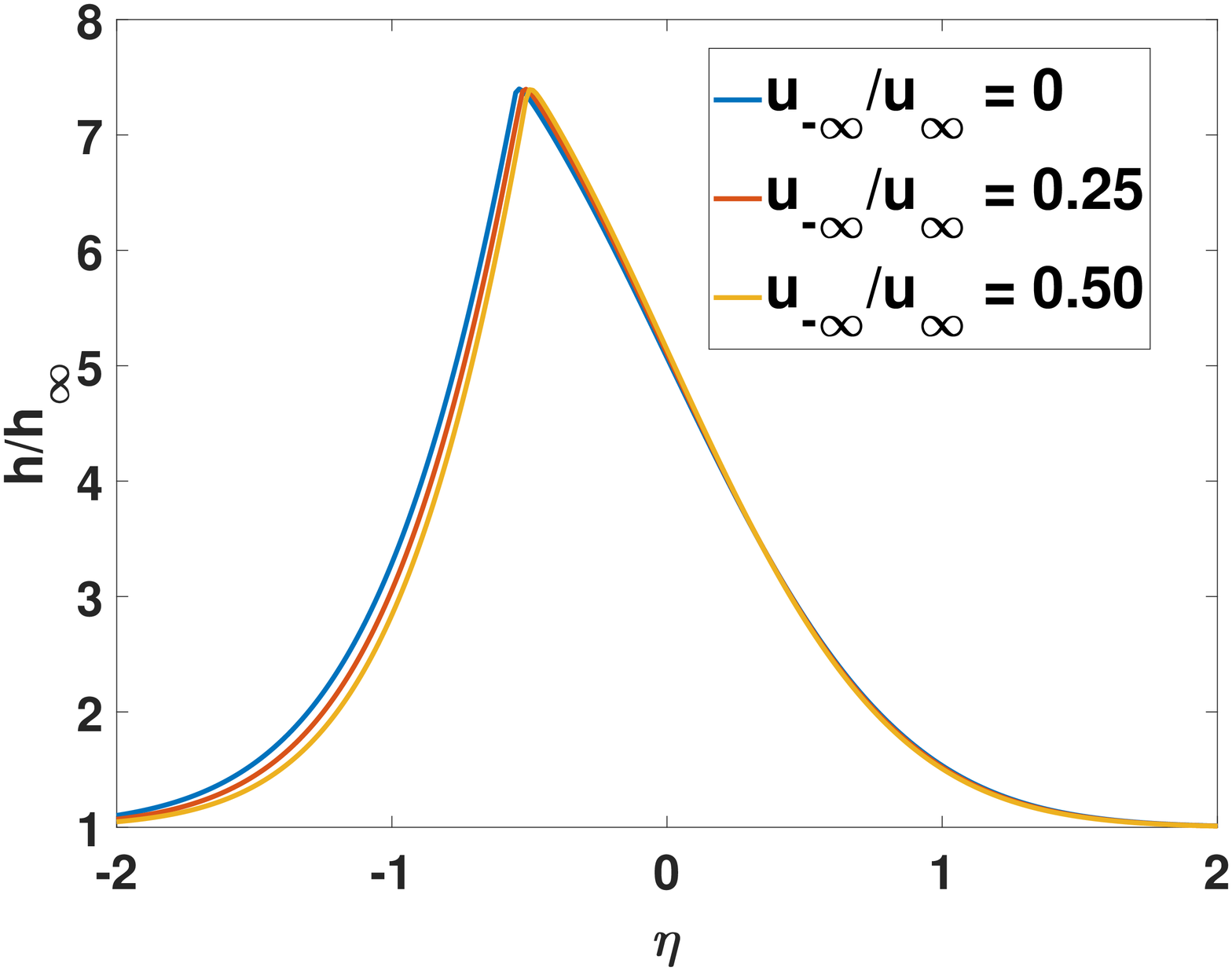}}
		\subfigure[$f$]{
			\includegraphics[height =5.3cm, width=0.48\linewidth]{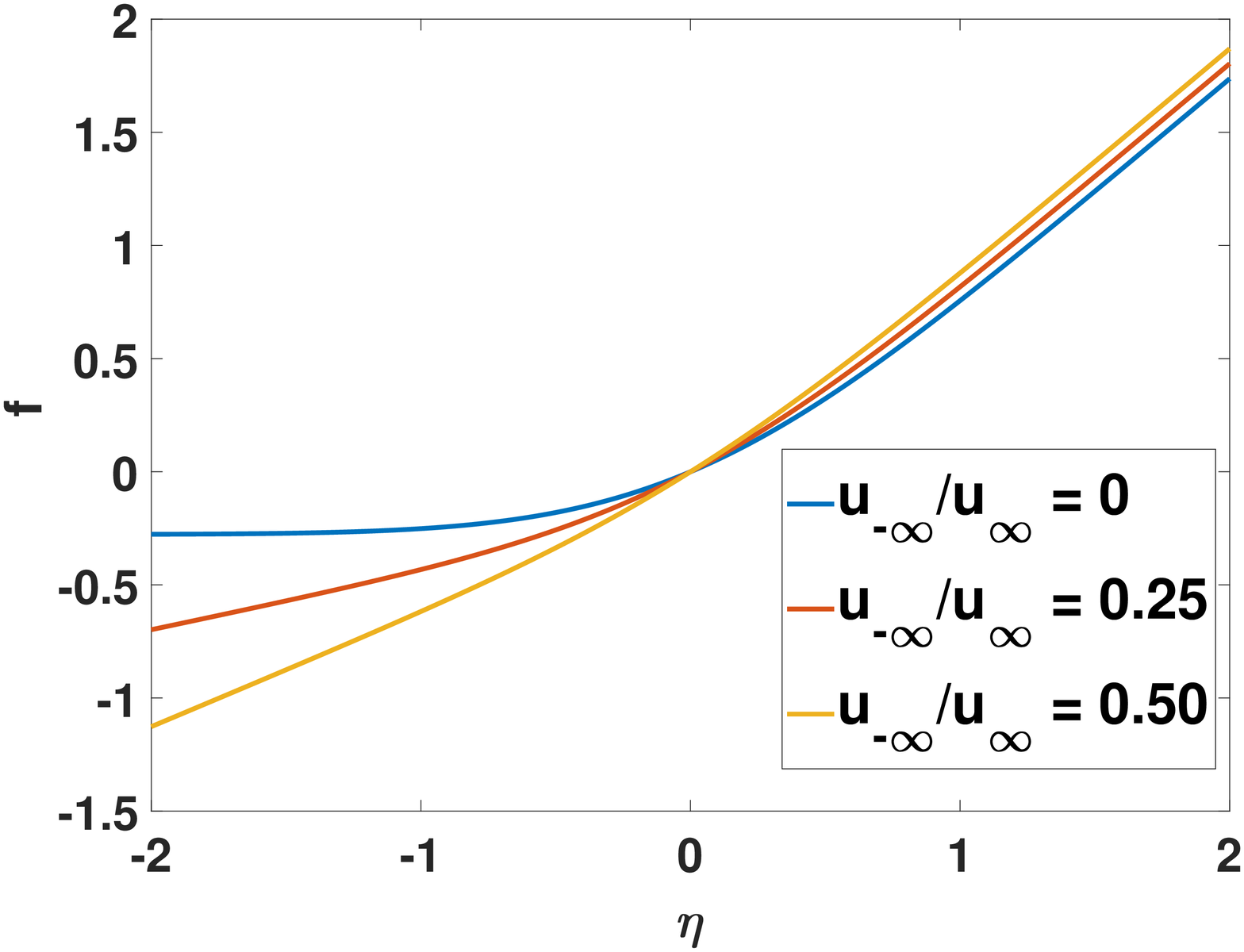}}
		\vspace{-0.1cm}
         \subfigure[$u/u_{\infty}$]{
			\includegraphics[height =5.3cm, width=0.48\linewidth]{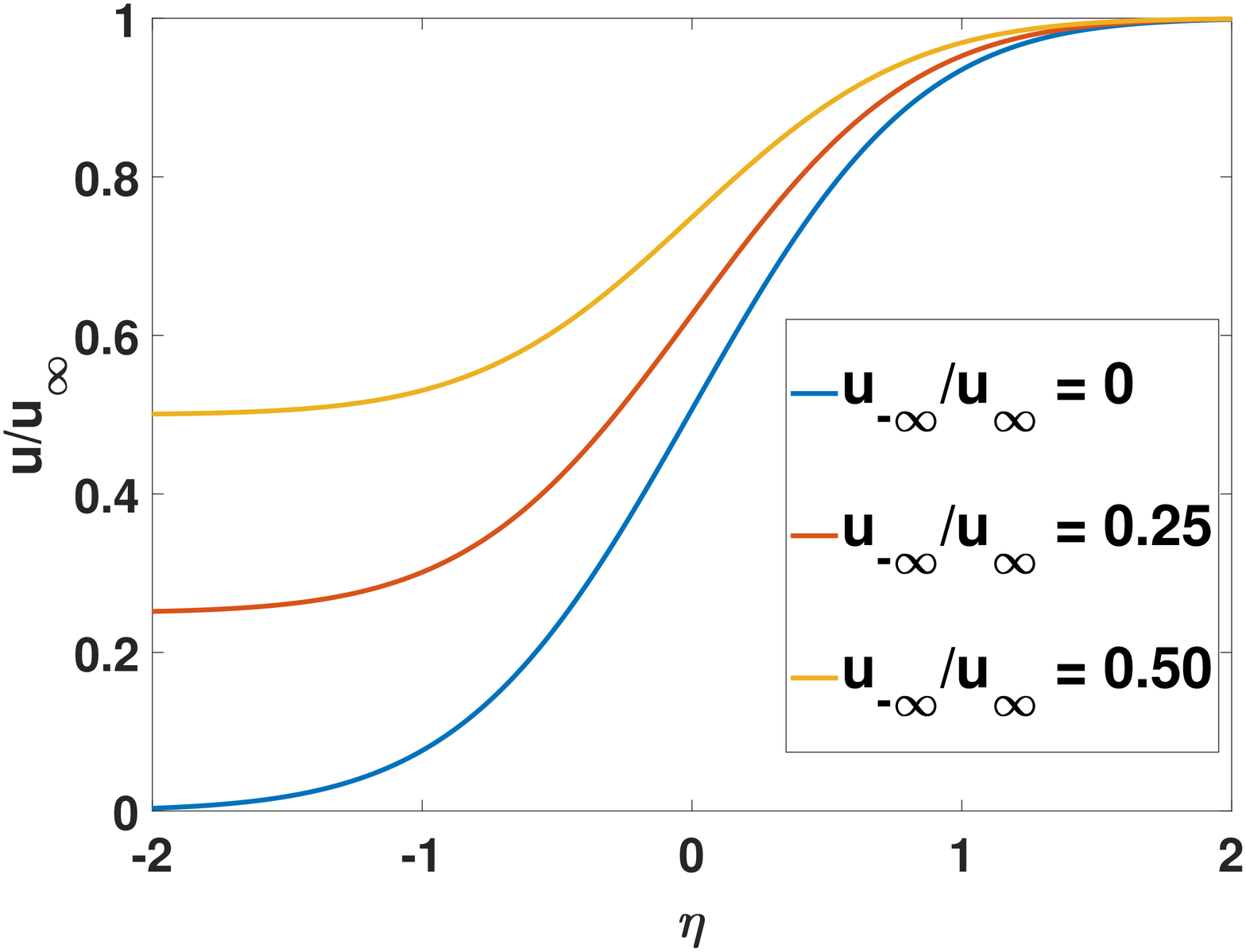}}
        \subfigure[$G$]{
			\includegraphics[height =5.3cm, width=0.48\linewidth]{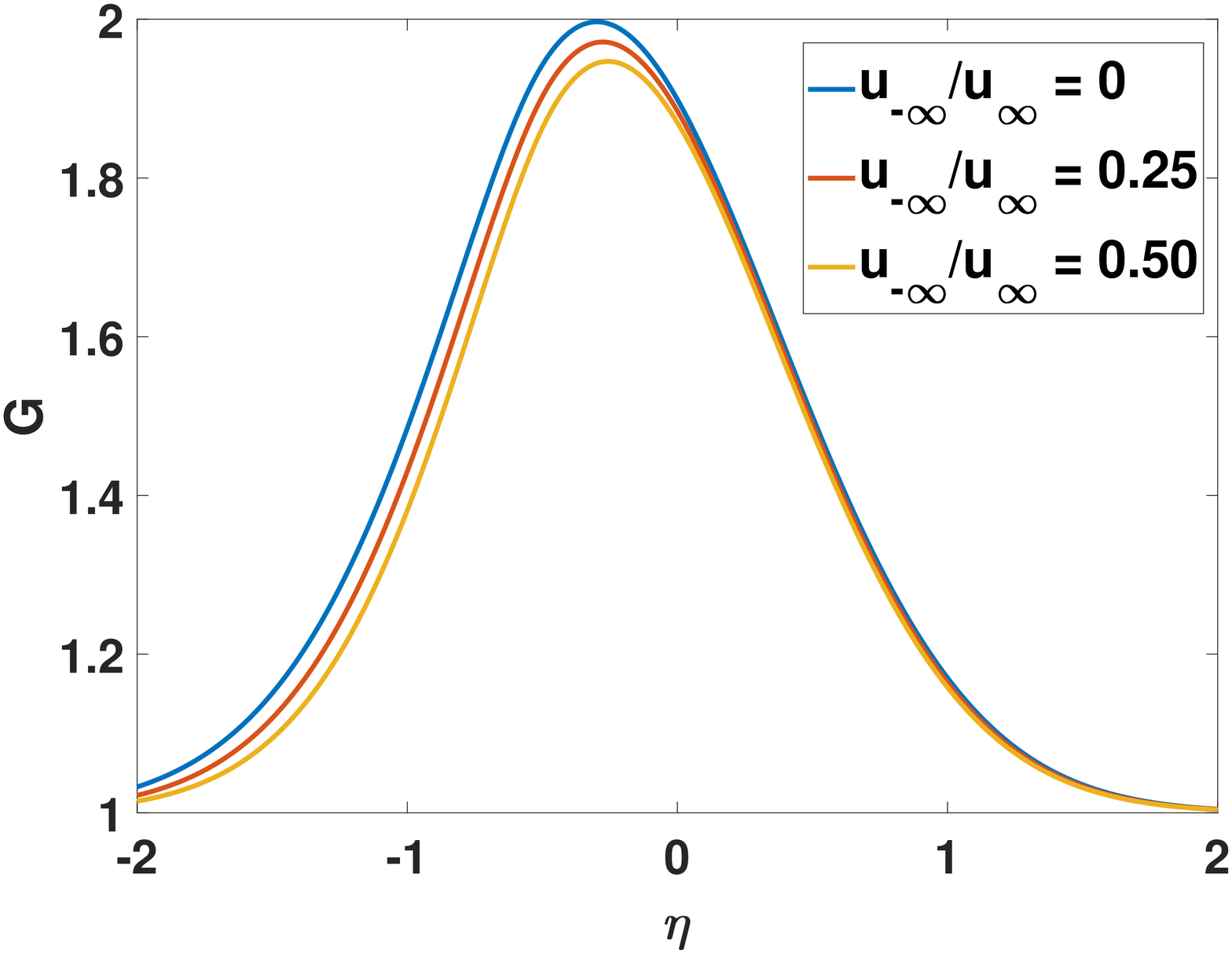}}
        \vspace{-0.1cm}
        \subfigure[$E$]{
			\includegraphics[height =5.0cm, width=0.48\linewidth]{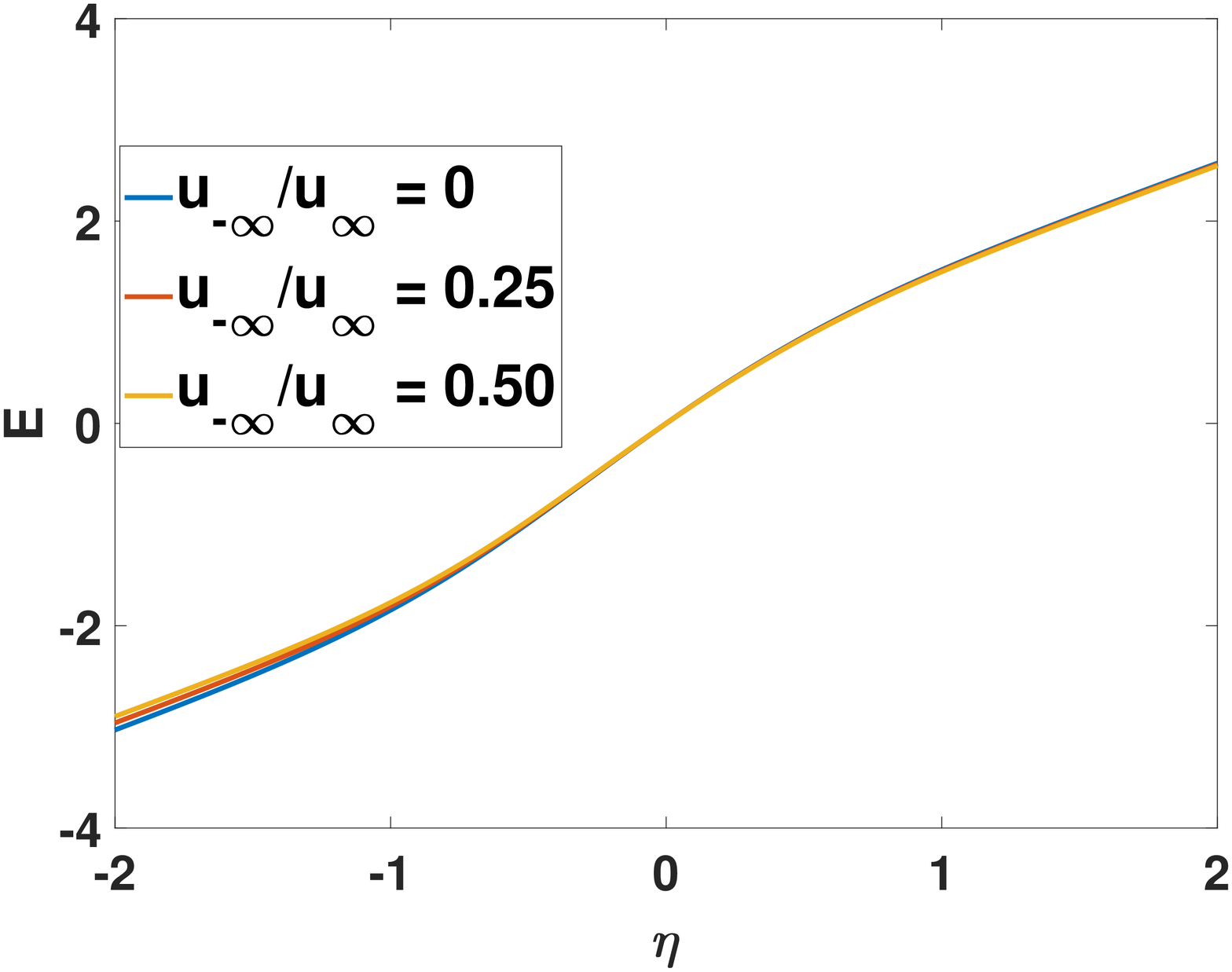}}
        \vspace{-0.1cm}
		\caption{Effect of velocity ratio on single diffusion flame solutions for enthalpy and dynamic field variables: $h, f,  u/u_{\infty},  G, E.$
			$ K =1.0 ;  Pr =1.0  ;  G_{\infty} = 1.0; u_{-\infty} / u_{\infty}= 0, 0.25, 0.50 $.}
		\label{DiffusionFlame9}
	\end{figure}

The velocity ratio $u_{-\infty}/u_{\infty}$ does have some impact on the velocity field as shown in Figure \ref{DiffusionFlame9}. As velocity ratio increases, the shear-strain rate decreases. While the imposed normal-strain rate remains fixed in value, it gains some importance relative to the shear; in particular, some narrowing of the mixing layer occurs due to the reduction of the flow displacement effect.

\begin{figure}[thbp]
		\centering
		\subfigure[$h/h_{\infty}$]{
			\includegraphics[height =5.0cm, width=0.48\linewidth]{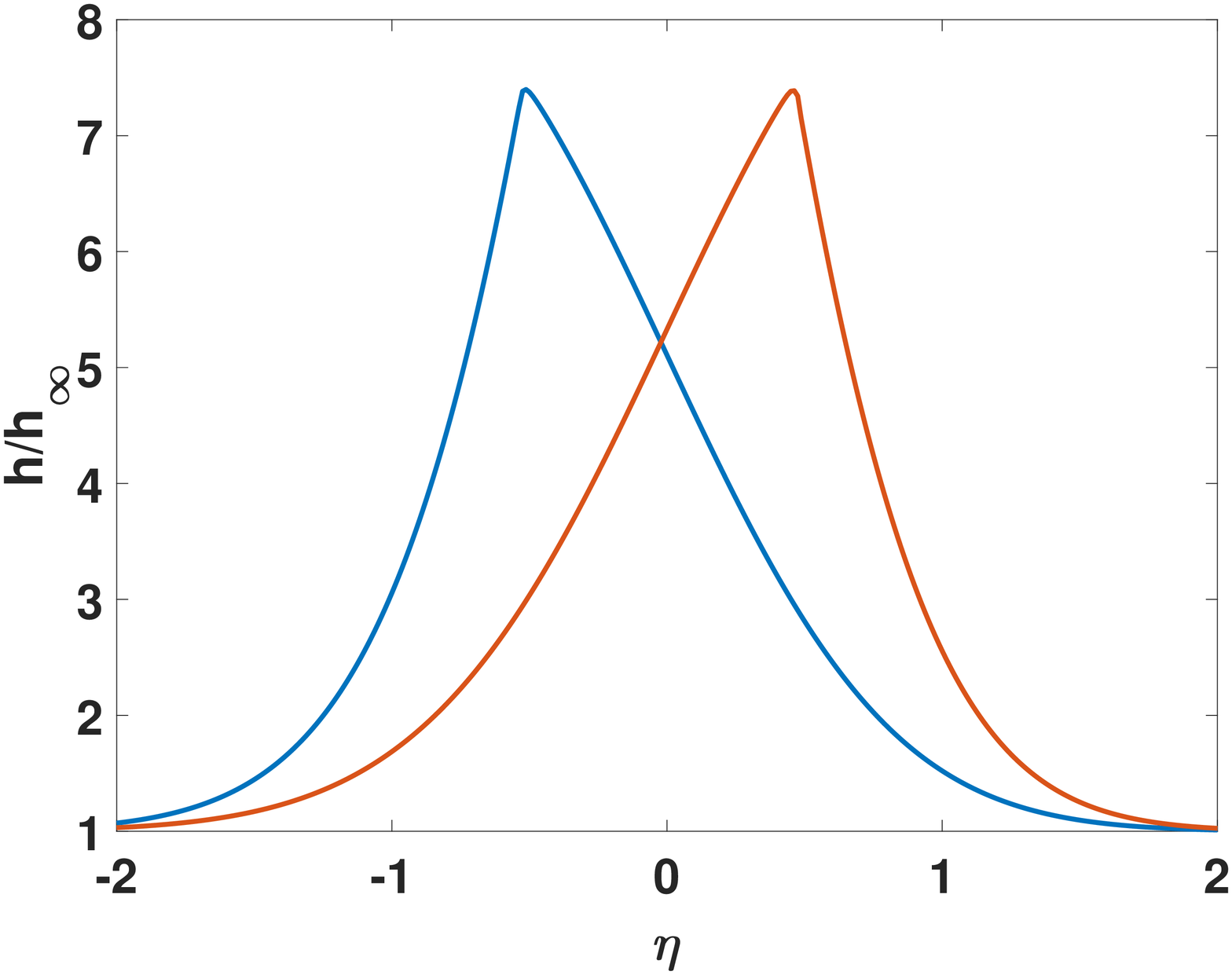}}
		\subfigure[$Y_F$]{
			\includegraphics[height =5.0cm, width=0.48\linewidth]{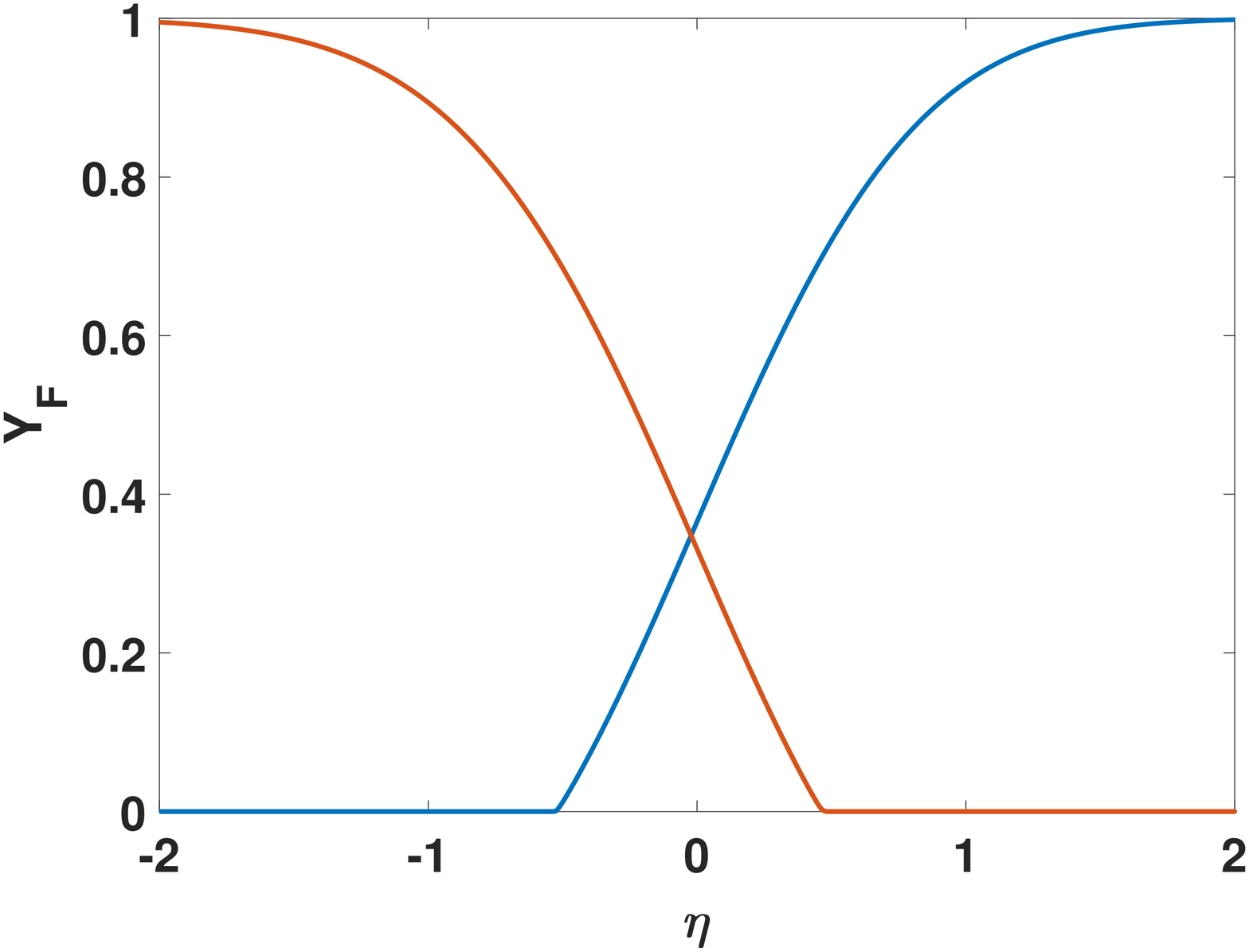}}
		\vspace{-0.1cm}
         \subfigure[$\nu Y_O$]{
			\includegraphics[height =5.0cm, width=0.48\linewidth]{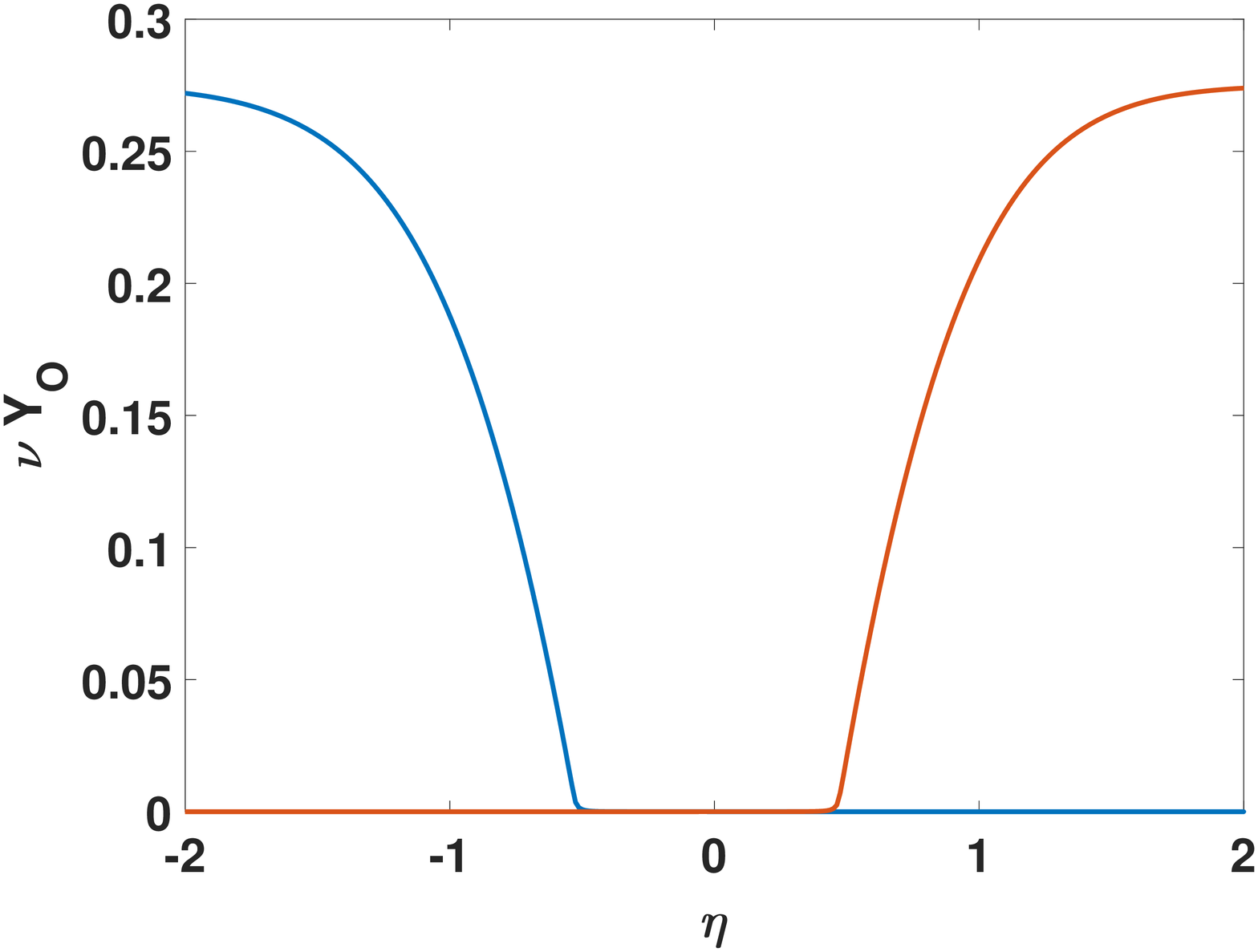}}
        \subfigure[$G$]{
			\includegraphics[height =5.0cm, width=0.48\linewidth]{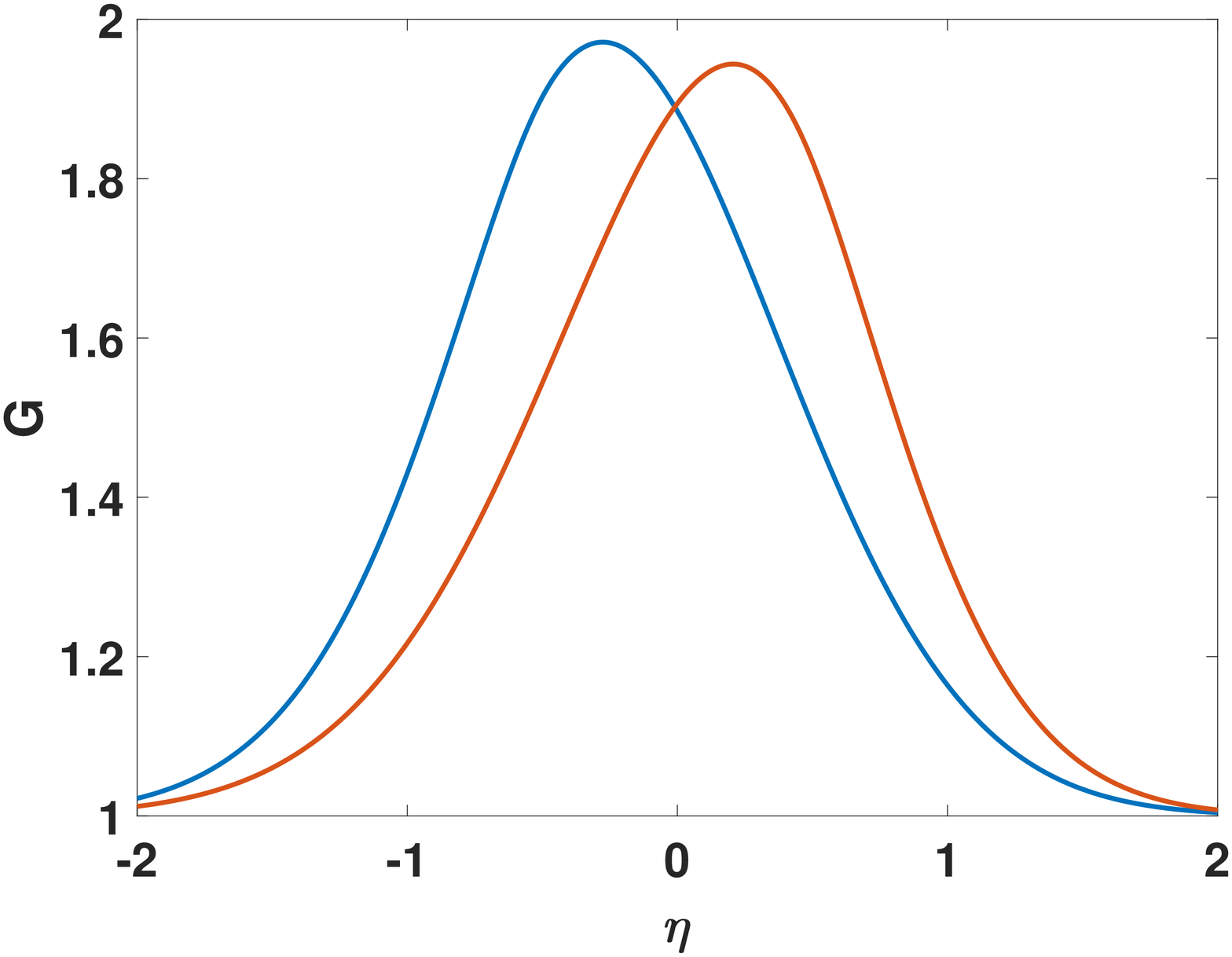}}
        \vspace{-0.1cm}
        \subfigure[$E$]{
			\includegraphics[height =5.0cm, width=0.48\linewidth]{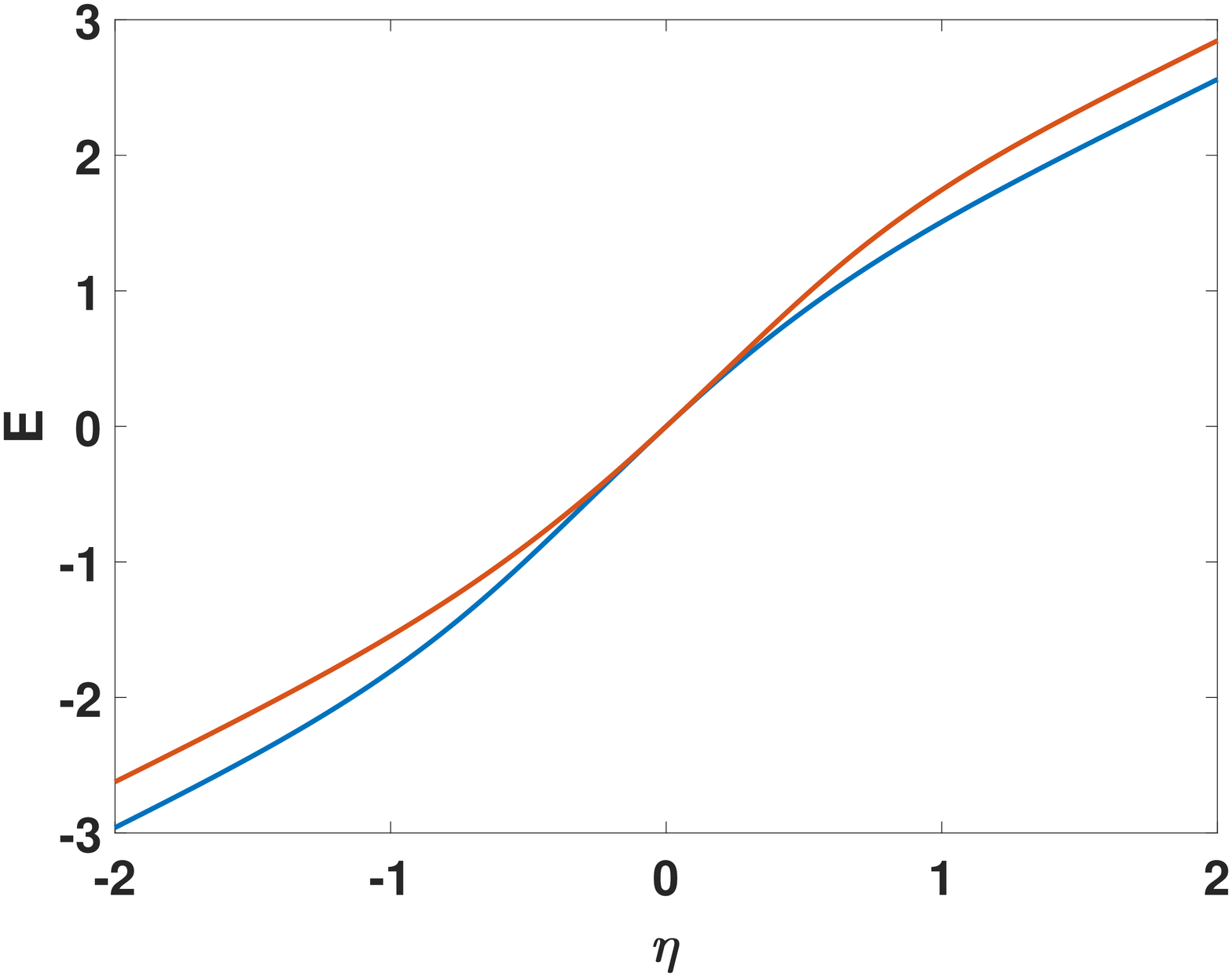}}
        \vspace{-0.1cm}
		\caption{Effect of reactant-stream reversal on single diffusion flame solutions for enthalpy and dynamic field variables: $h, Y_F,  \nu Y_O,  G, E.$
			$ K =1.0 ;  Pr =1.0  ;  G_{\infty} = 1.0; u_{-\infty} / u_{\infty}= 0.25 $.
Fuel in higher-speed stream and oxygen in lower-speed stream: blue line. Oxygen in higher-speed stream and fuel in lower-speed stream: red line.}
\label{DiffusionFlame10}
	\end{figure}

There are small differences displayed in Figure \ref{DiffusionFlame10} concerning the orientation of shear strain relative to the scalar gradient when fuel is moved from the higher-speed stream to the lower-speed stream with the opposite movement for the oxygen. The reaction zone always locates towards the oxygen side but slightly more so when oxygen flows in the slower stream. The gradients are always greater on the oxygen side but a little more so when the oxygen stream is the slower but accelerating stream with more convergence of the streamlines. Negligible effect of the reversal is seen on the $u$ velocity or $f$ or on the conserved scalars. Some effect is shown on the $G$ and $E$ variables.

\newpage
\subsection{Multiple Flames}\label{multiple}

\begin{figure}[thbp]
		\centering
		\subfigure[$h/h_{\infty}$]{
			\includegraphics[height =5.0cm, width=0.48\linewidth]{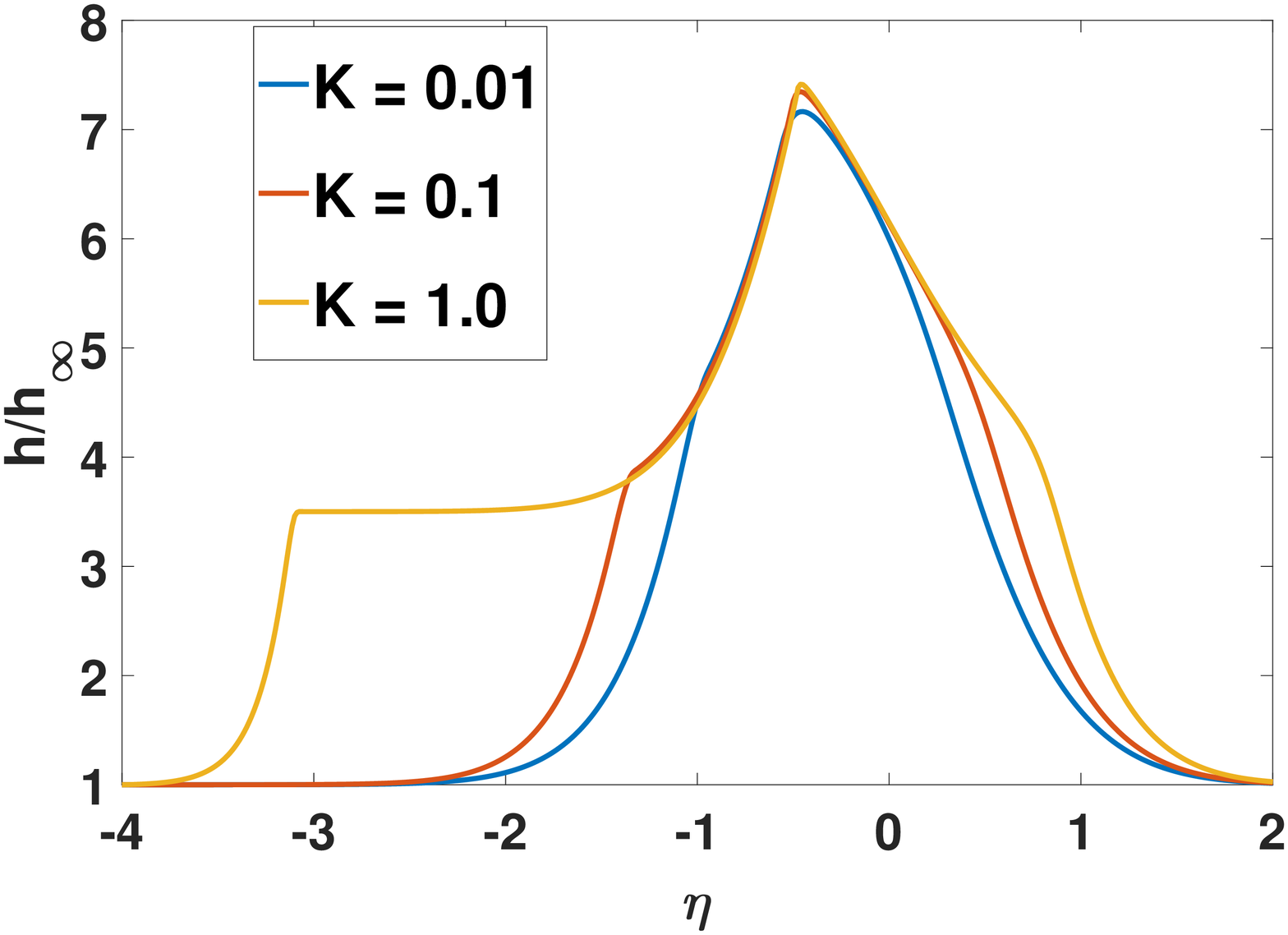}}
		\subfigure[$Y_F$]{
			\includegraphics[height =5.0cm, width=0.48\linewidth]{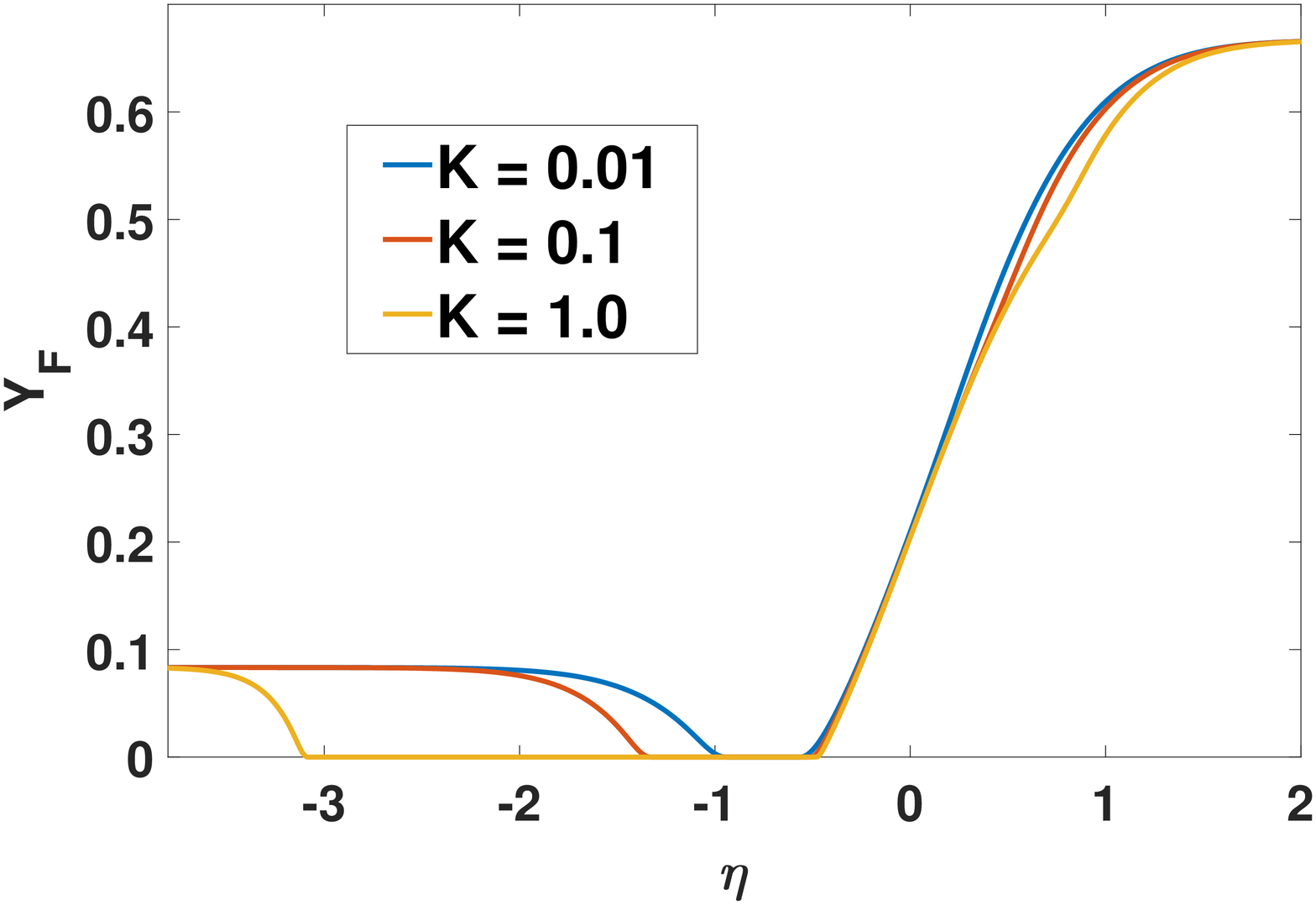}}
		\vspace{-0.1cm}
         \subfigure[$\nu Y_O$]{
			\includegraphics[height =5.0cm, width=0.48\linewidth]{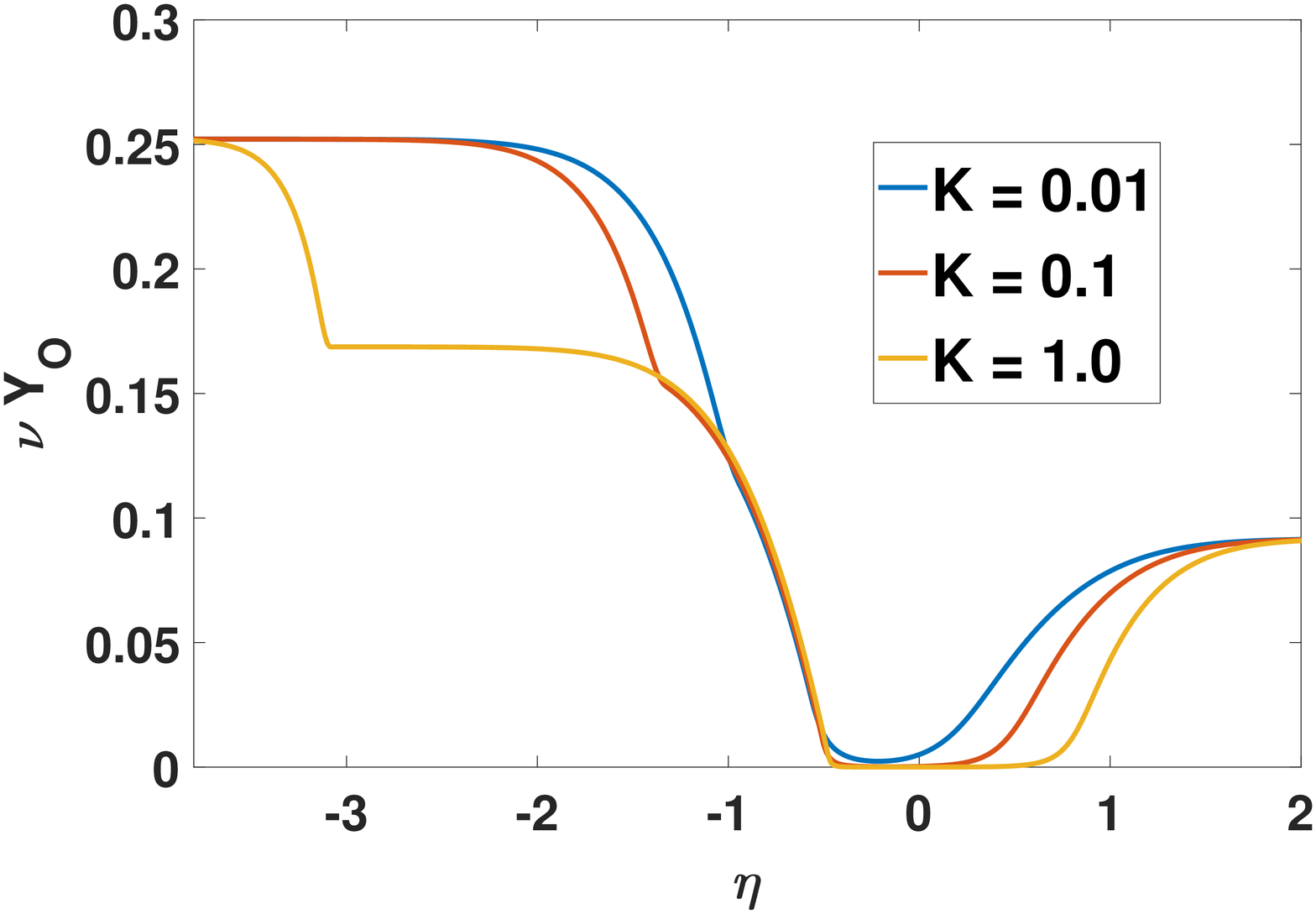}}
        \subfigure[$v g / \mu_{\infty}$]{
			\includegraphics[height =5.0cm, width=0.48\linewidth]{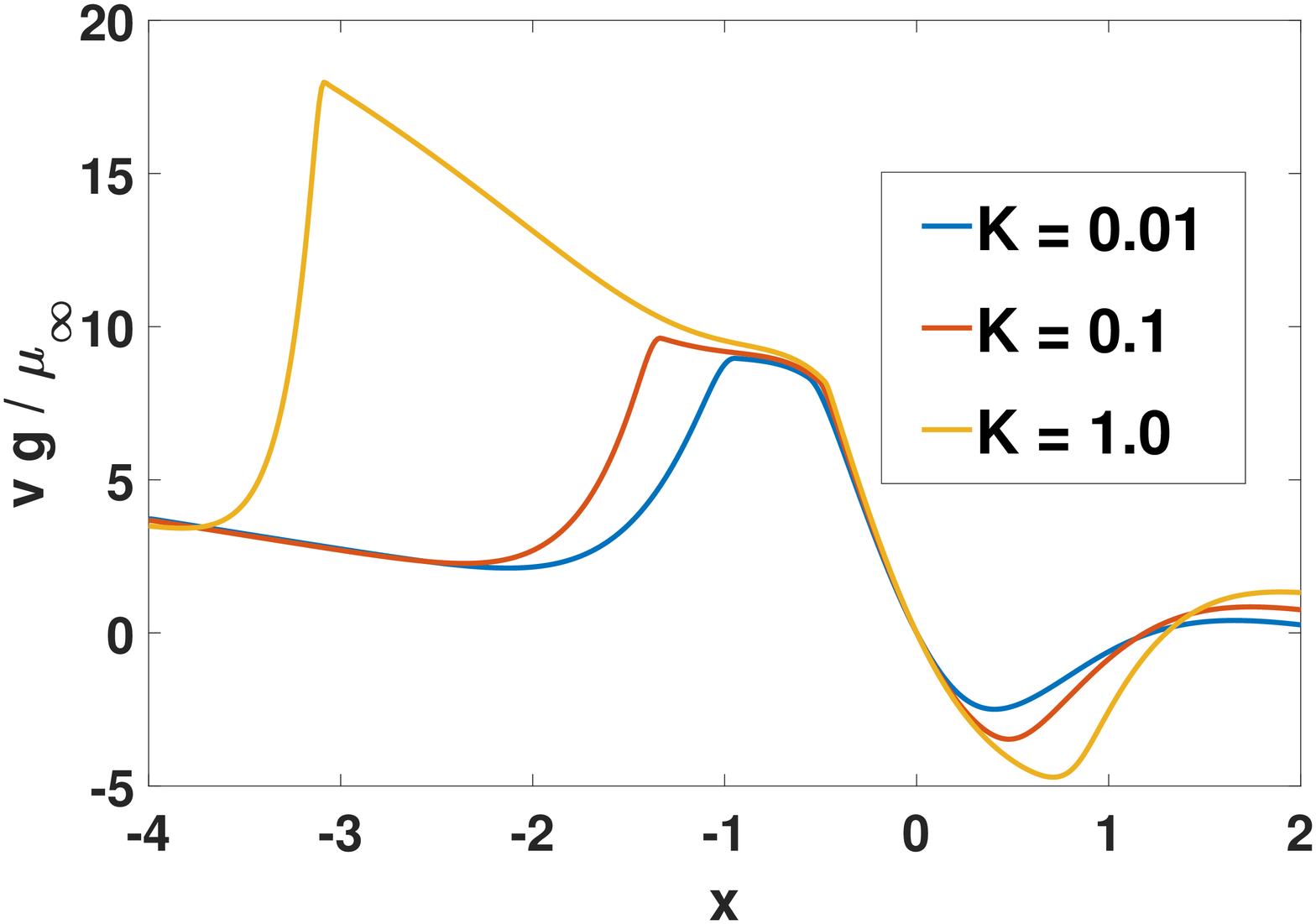}}
        \vspace{-0.1cm}
		\caption{Multiple flame solutions for enthalpy, mass fractions, and transverse velocity: $h, Y_F,\nu Y_O,  vg/\mu_{\infty}.$
		$ K =0.01, 0.1, 1.0;  Pr =1.0  ;  G_{\infty} = 1.0; u_{-\infty} / u_{\infty}= 0.25 $.}
\label{ThreeFlame}
	\end{figure}

If there is a combustible mixture in at least one of the two free streams and at least one reactant in the other stream, it is possible to have multiple flames: a diffusion flame with one or two partially premixed flames. The well known triple flame exhibits that character. This is well known for triple flames where several flame branches can appear \citep{Jorda2014, Rajamanickam}.  It also is found in counterflows \citep{Sirignano2019b, Lopez2019, Lopez2020}. This understanding leads to the consideration here of situations where a fuel-rich combustible mixture exists in the faster stream at $\eta = \infty$ and a fuel-lean combustible mixture exists in the slower stream at $\eta = -\infty$. Figure \ref{ThreeFlame} shows results with $Da$ varying over several orders of magnitude; specifically $K = 0.01, 0.1,$ and $1.0$.

Results show three flames: a fuel lean premixed flame to the left, a diffusion flame in the center, and a fuel-rich premixed flame to the right. It is noteworthy that the diffusion flame is to the left of the $v = 0$ surface because of asymmetry in the reaction-rate law. Excess oxygen from the left diffuses and advects through the fuel-lean premixed flame to reach the diffusion flame. Excess fuel from the right diffuses (but against advection) to reach the diffusion flame. Heat diffuses from the diffusion flame towards both premixed flames.

If the premixed flames behaves as classical isolated flames, the speed of propagation should increase with $\sqrt{K}$ which does not occur here. Also, propagation out of the mixing layer would occur. In fact, the propagation speed here is rather low and there is evidence of some diffusion control, especially for the fuel-rich flame and somewhat for the fuel-lean flame at low $Da$. The dependence on $Da$ and the re-location of the fuel-lean flame disallows rigor in the  use of the similar solution for this case. Thus, the results in Figure \ref{ThreeFlame} are self-contradictory and cannot be regarded as quantitatively accurate. However, they provide qualitative guidance about the behavior and the ability of three flames to co-exist. To obtain trustworthy quantitative information, the two-dimensional system of Equations (\ref{cont2}) through (\ref{species2}) should be solved.

\begin{figure}[thbp]
		\centering
		\subfigure[$h/h_{\infty}$]{
			\includegraphics[height =5.0cm, width=0.48\linewidth]{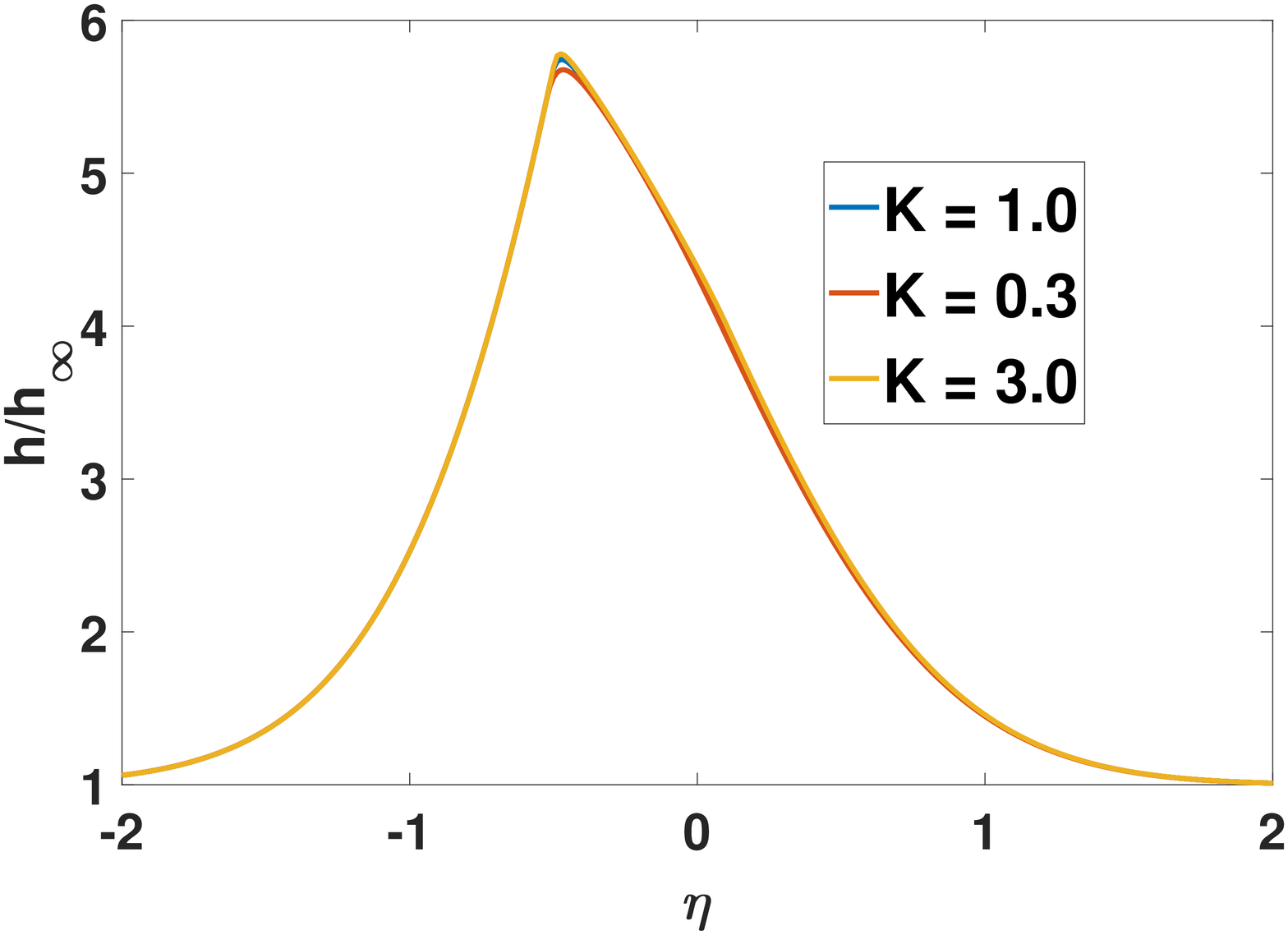}}
		\subfigure[$Y_F$]{
			\includegraphics[height =5.0cm, width=0.48\linewidth]{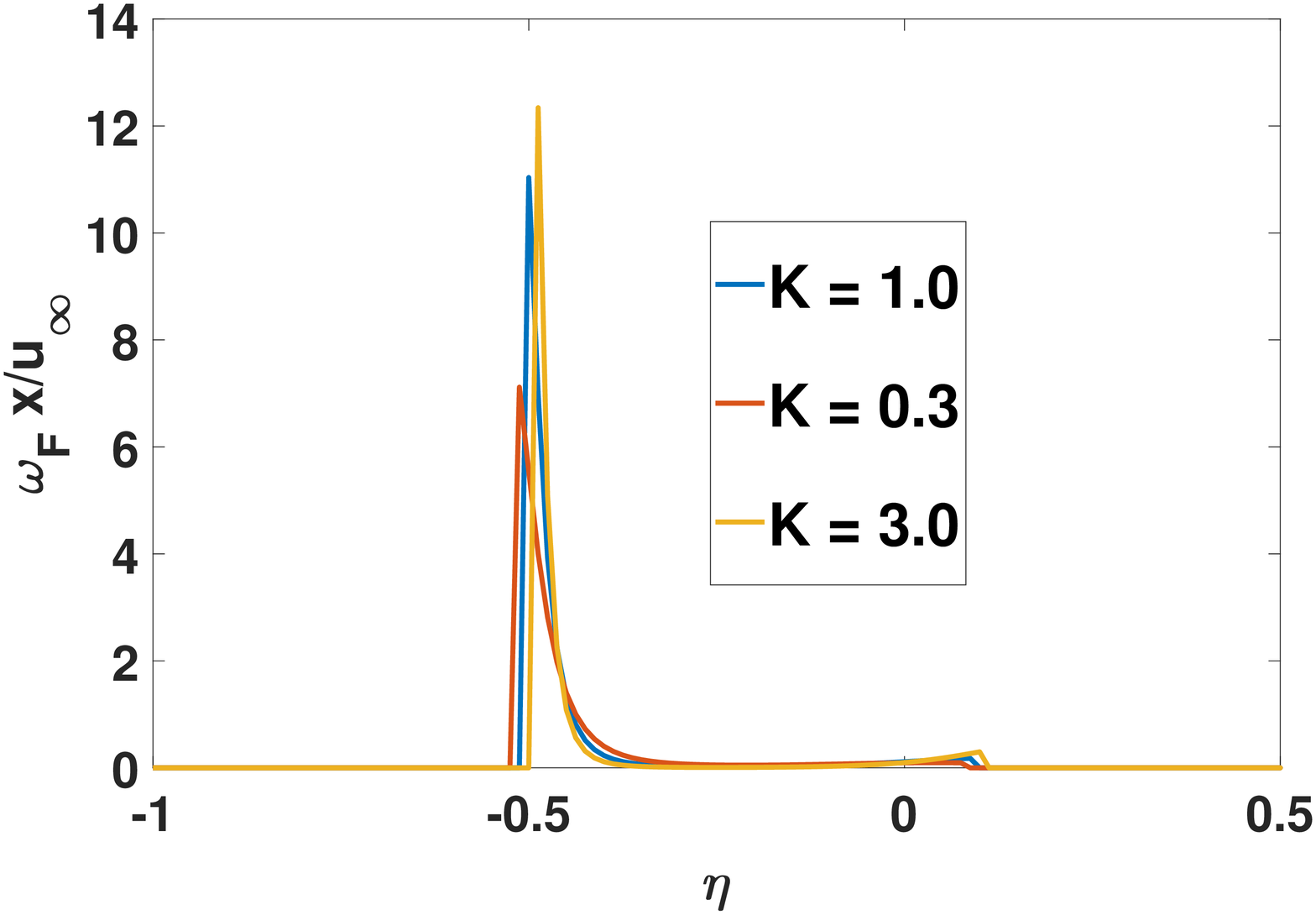}}
		\vspace{-0.1cm}
         \subfigure[$\nu Y_O$]{
			\includegraphics[height =5.0cm, width=0.48\linewidth]{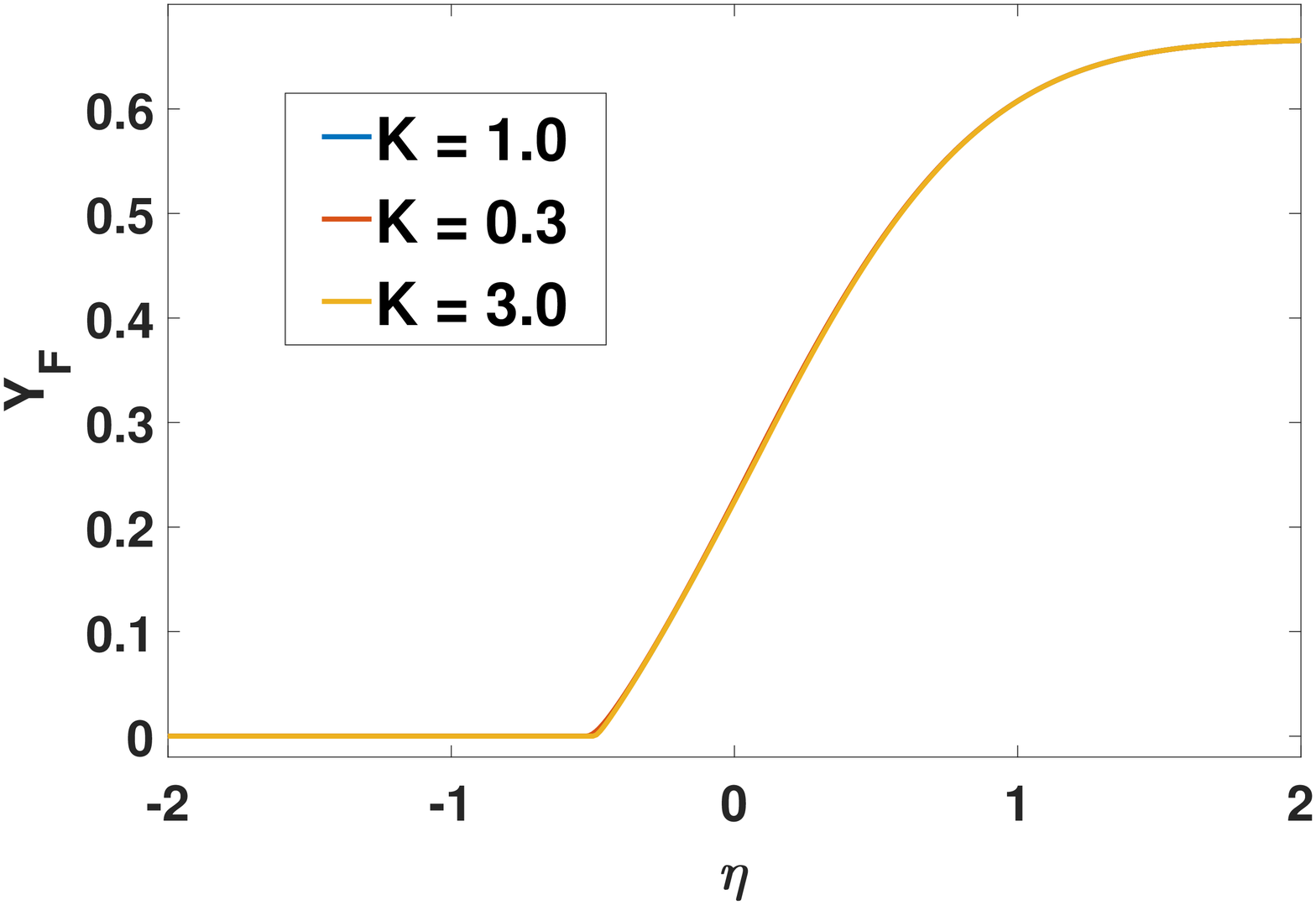}}
        \subfigure[$G$]{
			\includegraphics[height =5.0cm, width=0.48\linewidth]{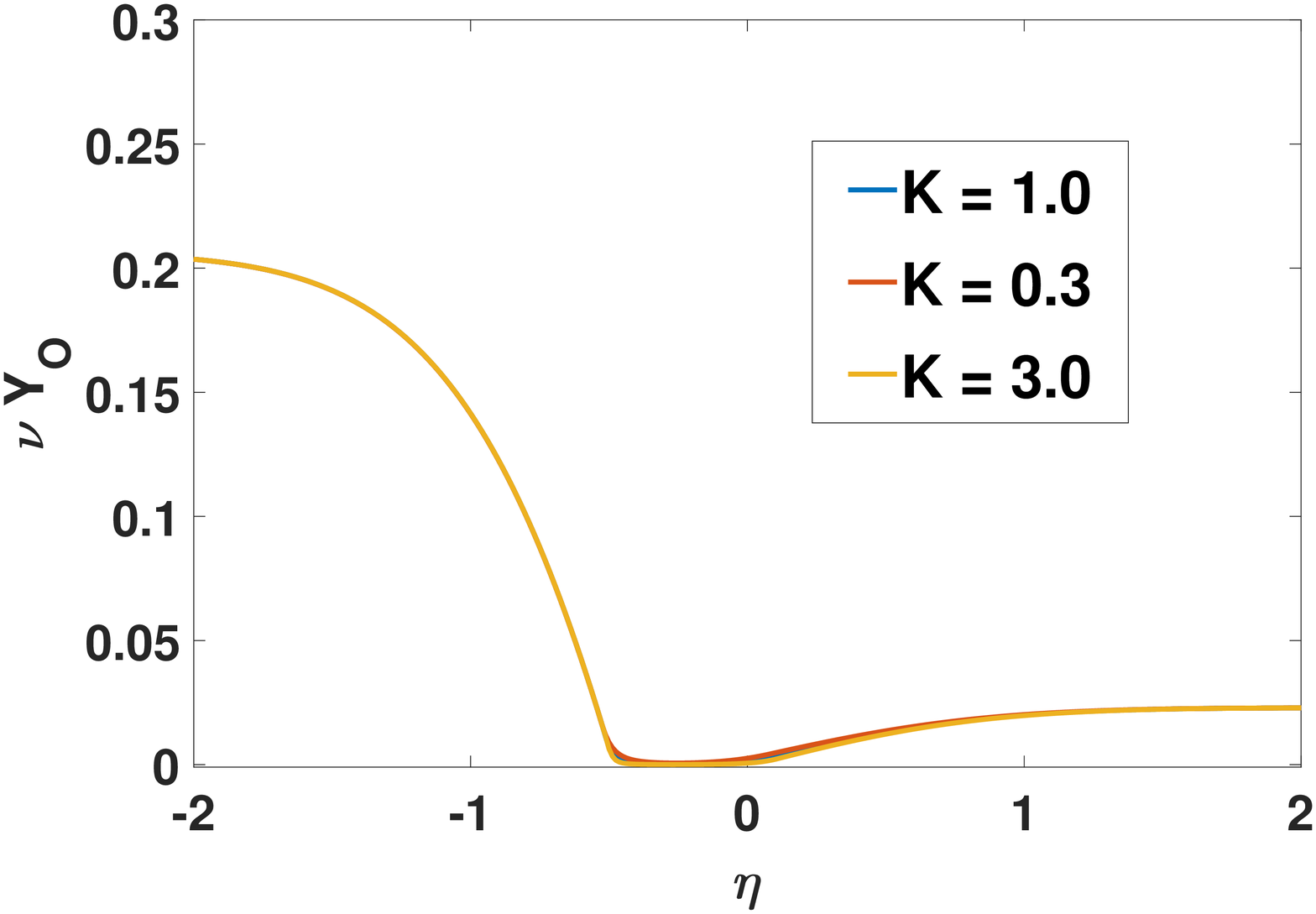}}
        \vspace{-0.1cm}
        \subfigure[$E$]{
			\includegraphics[height =5.0cm, width=0.48\linewidth]{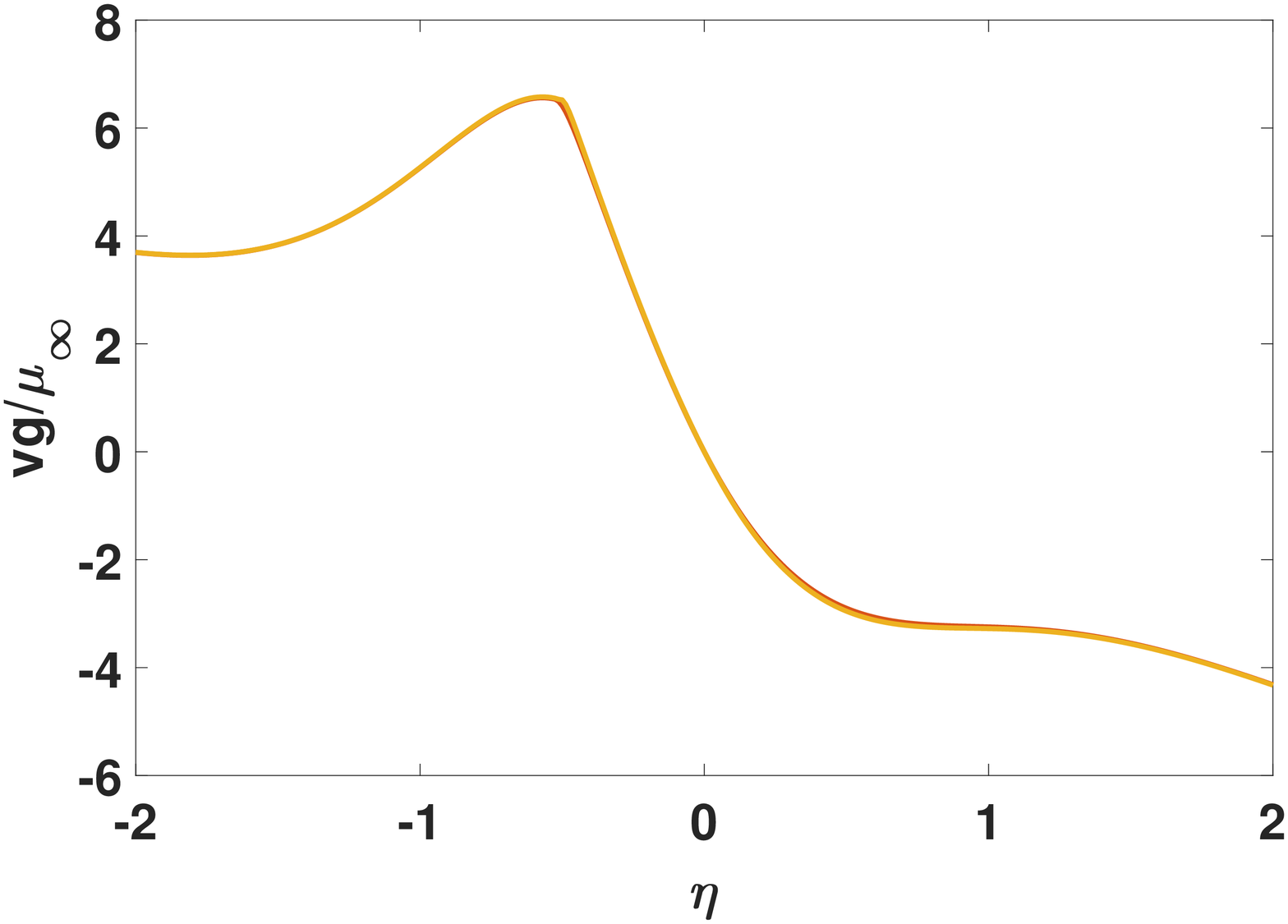}}
        \vspace{-0.1cm}
		\caption{Two-flame solutions for enthalpy and dynamic field variables: $h, Y_F,  \nu Y_O,  \omega_F x/u_{\infty} , vg/\mu_{\infty} , G.$
			$ K =1.0 ;  Pr =1.0  ;  G_{\infty} = 1.0; u_{-\infty} / u_{\infty}= 0.25 $.
}
\label{TwoFlame}
	\end{figure}
The above results with two combustible free streams leads  to the examination of  a configuration with pure oxygen in the slower stream at $\eta = -\infty$ and  a mixture in the faster stream at $\eta = \infty$ which is still more fuel-rich and therefore less reactive than the configuration in Figure \ref{ThreeFlame}. These results are shown in Figure \ref{TwoFlame}. The reaction-zone locations for both the diffusion flame and the weaker fuel-rich premixed flame do not change as $Da$ is varied over one order of magnitude. The transverse velocity $v$ also remains unchanged at the flame locations. The fuel-rich premixed flame here exhibits no wave-like character and is clearly diffusion-controlled.

Note that multiple-flame results here agree qualitatively with the counterflow-flame results of Sirignano \citep{Sirignano2019b} and Lopez et al. \citep{Lopez2019, Lopez2020}. Sirignano used one-step kinetics for propane and oxygen while Lopez et al. used detailed kinetics with methane-air chemistry.

\section{Concluding Remarks}   \label{conclusions}

 A three-dimensional configuration for a mixing layer with imposed normal strain (i.e., counterflow) in the transverse plane has been analyzed through one-dimensional similar solution for both  non-reacting and reacting configurations.
 A similar system of the Navier-Stokes equations coupled with equations for scalar transport is developed and solved. Variable  density, temperature, and composition are considered. One free stream is pure oxygen or sometimes a fuel-lean combustible mixture while the other stream is pure propane or sometimes a fuel-rich combustible mixture.

The enhancement of the mixing and combustion rates by imposed normal strain on a shear layer can be very substantial. Also, the imposition of shear strain and thereby vorticity on the counterflow can be substantial indicating the need for flamelet models with both shear strain and normal strain.

The  Damk\"{o}hler number $Da$ is an important parameter in the reacting case which has been varied over a range of one order of magnitude here. In the definition of $Da$ here, the reaction rate is normalized using a residence time $x/u_{\infty}$.  The diffusion flame becomes thinner with increasing $Da$ but its position does not change in either the single-flame or multi-flame configurations. The premixed flames show some diffusion control, especially at lower values of $Da$ and less flammable inflowing mixture ratios. In some cases, we can expect at least the fuel-rich premixed flame to fit the similar solution format with no change in location as $Da$ varies.

The impact of the magnitude of the dimensional shear-strain rate is made solely through the velocity ratio $u_{-\infty}/u_{\infty}$ .  The imposed normal strain rate is described through the parameter $G$. The increase in velocity ratio causes a decrease in the shear strain rate while an increase in $G$ results in an increased normal strain rate. The two strains have opposing effects on mixing layer thickness and thereby on scalar gradients, mixing rates, and burning rates.
Increased strain rate increases slightly both the displacement of streamlines and the width of the mixing layer. Increases in the normal strain rate can profoundly decrease the mixing layer width and increase transport rates.

The velocity profiles and the scalar profiles are shown to depend on the Prandtl number as well as the strain rates; higher $Pr$ caused steeper gradients and more narrow thermal and compositional layers.  A generalization has been developed for the classical Crocco integral to address non-unitary Prandtl number.

A closed form for the transverse velocity is found to match the similar solution formulation.

For the non-reacting case, the effect of Mach number $M$ is examined. Viscous dissipation is shown to have a modest role in the thermal behavior.

Flamelet theory as a closure model for turbulent combustion has been based on the tracking of two variables: a normalized conserved scalar and the strain rate; the latter may be given either directly or through a progress variable. Mixture fraction has traditionally been used for the conserved scalar.   A new normalized scalar is presented; the variable $\Sigma$  can be built on the Shvab-Zel'dovich conserved scalars or it can be a functional of the velocity field.

  \section*{Acknowledgements}
This research was supported by  the Air Force Office of Scientific Research under Grant FA9550-18-1-0392 with Dr. Mitat Birkan as the scientific officer.

\bibliographystyle{elsart-harv}
\bibliography {3DCompress}

\end{document}